\documentclass[aps,prd,reprint,11pt,amsmath,amssymb,nofootinbib,groupedaddress,superscriptaddress,floatfix,onecolumn,tightenlines,eqsecnum]{revtex4-2}

\usepackage{macros}
\graphicspath{{./figs}}

\usepackage{fontawesome5}
\makeatletter
\newcommand{\amarki}{\color{Blue}{\faCoffee}}
\newcommand{\amarkii}{\color{Blue}{\faBeer}}
\newcommand{\amarkiii}{\color{Blue}{\faRebel}}
\newcommand{\amarkiv}{\color{Blue}{\faPaperPlane[regular]}}
\def\@fnsymbol#1{{\ifcase#1\or \amarki\or \amarkii\or \amarkiii\or \amarkiv \else\@ctrerr\fi}}
\makeatother

\begin{document}

\title{Experimental targets for dark photon dark matter}
\author{David Cyncynates}
\email{davidcyn@uw.edu}
\affiliation{Department of Physics, University of Washington, Seattle, Washington, 98195, USA}
\author{Zachary J. Weiner}\email[]{zweiner@perimeterinstitute.ca}
\affiliation{Department of Physics, University of Washington, Seattle, Washington, 98195, USA}
\affiliation{Perimeter Institute for Theoretical Physics, Waterloo, Ontario N2L 2Y5, Canada}

\date{\today}

\begin{abstract}
Ultralight dark photon dark matter features distinctive cosmological and astrophysical signatures
and is also supported by a burgeoning direct-detection program searching for its kinetic mixing with
the ordinary photon over a wide mass range.
Dark photons, however, cannot necessarily constitute the dark matter in all of this parameter space.
In minimal models where the dark photon mass arises from a dark Higgs mechanism, early-Universe
dynamics can easily breach the regime of validity of the low-energy effective theory for a massive
vector field.
In the process, the dark sector can collapse into a cosmic string network, precluding dark photons
as viable dark matter.
We establish the general conditions under which dark photon production avoids significant
backreaction on the dark Higgs and identify regions of parameter space that naturally circumvent
these constraints.
After surveying implications for known dark photon production mechanisms, we propose novel models
that set well-motivated experimental targets across much of the accessible parameter space.
We also discuss complementary cosmological and astrophysical signatures that can probe the dark
sector physics responsible for dark photon production.
\end{abstract}

\maketitle
\makeatletter
\def\l@subsubsection#1#2{}
\def\l@subsection#1#2{}
\makeatother

\begin{spacing}{0.25}
\tableofcontents
\end{spacing}

\section{Introduction}

The cosmological and astrophysical evidence for cold, nonbaryonic dark matter provides some of the
most compelling motivation for physics beyond the Standard Model~\cite{Bertone:2004pz,
Bertone:2016nfn, Buckley:2017ijx}.
These observations, however, only evince dark matter through its gravitational effects, allowing a
broad range of new particles to effectively reproduce the cold dark matter paradigm in its observed
regimes.
From a bottom-up perspective, dark matter candidates span dozens of orders of magnitude in mass and
all possible spins; the only general requirement of their nongravitational interactions with
Standard Model (SM) particles is that they are sufficiently weak to have evaded detection thus far.

Confronted with so vast a landscape of possibilities, top-down theoretical approaches can provide
essential guidance by motivating concrete models and identifying consistent scenarios.
The canonical example, a weakly interacting massive particle~\cite{Bertone:2004pz}, sets
specific experimental targets by determining the dark matter relic abundance via its interaction
strength with SM particles through thermal freeze-out~\cite{zel1965magnetic, zel1965quarks,
Chiu:1966kg}.
Dark matter as a thermal relic would be too warm at late times if its mass were below
$\sim \mathrm{keV}$~\cite{Viel:2013fqw, Irsic:2017ixq, Garzilli:2019qki, Villasenor:2022aiy,
Hsueh:2019ynk, Gilman:2019nap, Enzi:2020ieg, Nadler:2021dft}; in this mass range, the dark matter
must also be bosonic~\cite{Tremaine:1979we, Boyarsky:2008ju, Hogan:2000bv, Dalcanton:2000hn,
Boyanovsky:2007ay, Gorbunov:2008ka, Shao:2012cg, Alvey:2020xsk, Domcke:2014kla, Randall:2016bqw,
DiPaolo:2017geq, Giraud:2018gxl, Savchenko:2019qnn}.
Axions and dark photons dominate theoretical and experimental efforts in this ultralight regime,
since symmetries can protect their masses from large quantum corrections.
Notably, the quantum chromodynamics (QCD) axion, initially proposed to resolve the strong \textit{CP}
problem~\cite{Peccei:1977hh, Weinberg:1977ma, Wilczek:1977pj}, is an excellent dark matter
candidate~\cite{Preskill:1982cy, Abbott:1982af, Dine:1982ah, Duffy:2009ig} and has inspired a broad
experimental program targeting its couplings to SM particles in the
Kim-Shifman-Vainshtein-Zakharov/Dine-Fischler-Srednicki-Zhitnisky range~\cite{Kim:1979if,
Shifman:1979if, Zhitnitsky:1980tq, Dine:1981rt}.

New, massive gauge bosons may not solve specific problems in the SM like the axion does, but they
are well motivated (if not expected) in extensions of the SM~\cite{Beasley:2008dc, Abel:2008ai,
Donagi:2008kj, Blumenhagen:2008zz, Arvanitaki:2009fg, Blumenhagen:2009gk, Goodsell:2009xc,
Bullimore:2010aj} as the force carriers of new interactions.
Indeed, massive gauge bosons are a feature of the SM itself.
Moreover, the dark photon inherits the axion's broad experimental prospects~\cite{Caputo:2021eaa}: while
detecting the axion's mixing with the SM photon requires catalysis by a background magnetic field,
the dark photon's kinetic mixing with the SM photon takes place in vacuum.
Every axion experiment is therefore also a dark photon experiment.

In this work, we investigate how theoretical considerations motivate particular parts of the
extensive dark photon parameter space accessible to upcoming searches.
We focus on a fundamental requirement for any dark matter candidate: a consistent mechanism for
generating its relic abundance.
In particular, the low-energy effective theory of a massive, kinetically mixed vector appropriate to
describe its phenomenology for direct detection may well be inadequate to describe its cosmological
production at early times (i.e., at higher energy).
In minimal models where the ultraviolet (UV) completion of the dark photon's mass is a Higgs
mechanism, the cutoff scale is directly tied to the kinetic mixing strength.
Reference~\cite{East:2022rsi} demonstrated that breaching this cutoff has severe consequences:
topological defects (Nielsen-Olesen strings~\cite{Nielsen:1973cs}) form, converting any coherent
dark photon background into string kinetic and potential energy.
The resulting string network is neither a viable dark matter candidate nor detectable by haloscopes.
In the minimal scenario of vector production from inflationary fluctuations~\cite{Graham:2015rva},
the entire experimentally accessible parameter space is excluded on these
grounds~\cite{East:2022rsi}.
Here we extend this analysis to constrain additional dark photon production mechanisms, propose
extensions thereof and novel scenarios with enhanced experimental prospects---including an in-depth
discussion and generalization of the models of scalar-mediated dark photon production introduced in
Refs.~\cite{Adshead:2023qiw, Cyncynates:2023zwj}---and identify theoretically motivated targets
within the dark photon parameter space.

The first loophole one might consider is that the dark photon's mass instead originates from the St\"uckelberg mechanism~\cite{Ruegg:2003ps}.
Typically, studies of dark photon dark matter that assume St\"uckelberg masses for simplicity in fact make a stronger assumption: that the theory is well described by Proca theory [see \cref{{eqn:proca}}], i.e., an Abelian gauge field with a bare mass term and nothing more.
In known examples from string theory and supersymmetry, however, St\"uckelberg masses are accompanied by radial degrees of freedom analogous to a Higgs~\cite{Reece:2018zvv}, and the validity of the Proca theory is similarly restricted in energy.
If this regime is breached at early times, at the very least we expect modifications to the dynamics of dark photon production that have not been studied.
Here we consider only consider Higgs masses, since the relevant dynamics have been established~\cite{East:2022rsi}, and comment further on St\"uckelberg masses in \cref{sec:model-agnostic}.

In \cref{sec:defect-bounds-on-existing-scenarios}, we derive model-independent bounds on kinetic
mixing as a function of wave number and redshift of production, applying these bounds to various
models proposed in prior literature.
We then consider the clockwork mechanisms as a production-agnostic means to relax these bounds in
\cref{sec:production-independent}, demonstrating that they introduce no additional fine-tuning
issues.
In \cref{sec:axion-enhance}, we revisit dark photon production from oscillating axions and describe
mechanisms that unify enhanced production efficiency with dynamics that avoid string formation.
Finally, in \cref{sec:scalar} we consider a class of models coupling the dark photon to a new
singlet scalar, discussing the scalar's potential to alleviate backreaction constraints by
modulating the fundamental parameters of the Abelian-Higgs theory.
We show that the Damour-Polyakov mechanism~\cite{Damour:1994zq} can weaken bounds on inflationary
production, allowing accessible kinetic mixing for dark photons heavier than an $\eV$.
We further discuss two qualitatively distinct postinflationary production mechanisms driven by the
scalar~\cite{Adshead:2023qiw, Cyncynates:2023zwj} that, with some fine-tuning, may open all kinetic
mixing parameter space accessible to upcoming haloscopes.
In \cref{sec:discussion}, we summarize and discuss the implications of our findings and conclude in
\cref{sec:conclusion}.
A number of appendices provide additional details on scalar--Abelian-Higgs models (\cref{app:SAH}),
quantum corrections to the scalar effective potential in these models
(\cref{app:quantum-corrections}), the strong-coupling (\cref{app:axionOscillations}) and
narrow-resonance (\cref{app:Delayed-Axion-Oscillations}) regimes of dark photon production from
axions, and the effect of the Standard Model plasma on the narrow scalar resonance discussed in
\cref{sec:narrow} (\cref{app:plasma}).

Throughout this paper, we use natural units in which $\hbar = c = 1$, define the reduced Planck mass
$\Mpl = 1/\sqrt{8\pi G}$, fix a cosmic-time Friedmann-Lema\^itre-Robertson-Walker (FLRW) metric
$\ud s^2 = \ud t^2 - a(t)^2 \delta_{i j} \ud x^i \ud x^j$ with $a(t)$ the scale factor, and employ
the Einstein summation convention for spacetime indices.
Dots denote derivatives with respect to cosmic time $t$, and the Hubble rate is
$H \equiv \dot{a} / a$.
We use boldface for spatial three vectors.

\section{Backreaction, defect formation, and a survey of production scenarios}\label{sec:defect-bounds-on-existing-scenarios}

We begin by reviewing Abelian-Higgs theory and the thresholds associated with backreaction onto the
dark Higgs and defect formation~\cite{East:2022rsi}.
We then identify generic conditions for postinflationary cosmological production of dark photons
that evade these thresholds.
We discuss defect bounds on inflationary production of dark photons in \cref{sec:inflation},
considering the effects of modified postinflationary expansion histories
(\cref{sec:postinflationary-histories}) and possible nonminimal couplings to gravity and the inflaton
itself (\cref{sec:nonminimal-couplings}).
In \cref{sec:thermal} we derive model-independent bounds from a dark photon's kinetic mixing with
the hot Standard Model plasma, and in \cref{sec:axion} we discuss dark photon production via an
oscillating axion.
We briefly consider vector misalignment in \cref{sec:vector-misalignment}.

In this work, we consider a dark sector described by various extensions of the Abelian-Higgs model.
We denote the dark photon with $\Ap$, not to be confused with the Standard Model photon $\Asm$.
The dark Higgs field $\Phi$ that provides the dark photon its mass is a complex scalar charged under
$\Ap$ with gauge coupling $\gD$ and a vacuum expectation value (VEV) $\vev$.
We refer to the dark Higgs simply as the Higgs, since the SM Higgs is not relevant to our
discussion.
The action of Abelian-Higgs model is $S = \int \ud^4 x \, \sqrt{-g} \mathcal{L}$ with Lagrangian
\begin{align}
    \mathcal{L}
    &= - \frac{1}{4} \Fp_{\mu\nu} {\Fp}^{\mu\nu}
        + \frac{1}{2} D_\mu \Phi \left( D^\mu \Phi \right)^\ast
        - \frac{\lambda}{4} \left( \abs{\Phi}^2 - v^2 \right)^2,
    \label{eqn:abelian-higgs-lagrangian}
\end{align}
where $\Fp_{\mu\nu} = \partial_\mu \Ap_\nu - \partial_\nu \Ap_\mu$ is the dark photon field
strength, $D_\mu = \partial_\mu - \I \gD \Ap_\mu$ is the gauge-covariant derivative, and $\lambda$
is the Higgs self-coupling.

In general, we seek to identify the conditions under which a cosmological abundance of dark photons
in the early Universe---typically, their maximum density which they reach when first
produced---invalidates the low-energy effective theory given by the Proca Lagrangian,
\begin{align}
    \mathcal{L}
    &= - \frac{1}{4} \Fp_{\mu\nu} {\Fp}^{\mu\nu}
        + \frac{1}{2} \mA^2 \Ap_\mu \Ap^\mu,
    \label{eqn:proca}
\end{align}
and whatever description of dark photon production that relies upon it.
The Proca Lagrangian derives from the low-energy limit of \cref{eqn:abelian-higgs-lagrangian} when
the dark photon's U(1) symmetry is broken.
In the broken phase, the dark photon and Higgs have masses $\mA = \gD \vev$ and
$m_\higgs = \sqrt{2\lambda} v$, where the Higgs is decomposed as
$\Phi \equiv (\vev + \higgs) e^{\I \goldstone/ \vev}$.
In unitary gauge, which sets $\Pi = 0$, the dark photon acquires a longitudinal mode as it eats the
Higgs phase, i.e., $A_\mu - \partial_\mu \goldstone / \mA \to \Ap_\mu$.
The theory is then described by the action
\begin{align}\label{eqn:abelian-higgs-action-expanded}
    \mathcal{L}
    &= - \frac{1}{4} \Fp_{\mu\nu} \Fp^{\mu\nu}
        + \frac{1}{2} \partial_\mu \higgs \partial^\mu \higgs
        - V_\mathrm{eff}(\higgs,\Ap),
\end{align}
where the effective potential for the Higgs and dark photon is
\begin{align}
    V_\mathrm{eff}(\higgs,\Ap)
    &= \frac{\lambda \vev^4}{4}
        \left[ 1 + \left( \frac{h}{\vev} + 1 \right)^4 \right]
        - \frac{1}{2} \left( \frac{h}{\vev} + 1 \right)^2
        \left( \lambda \vev^4 + \mA^2 \Ap_\mu \Ap^\mu \right).
    \label{eqn:higgs-effective-potential}
\end{align}
The first term in the effective potential is a pure quartic with a unique minimum at the
symmetric point $\higgs = -\vev$.
The latter term generates the symmetry-breaking minimum at $\higgs = 0$ provided that the dark
photon amplitude is not too large---that is, the dark photon backreacts strongly onto the Higgs if
\begin{align}\label{eqn:symmetry-restoration-condition}
    \left\vert \mA^2 \Ap_\mu \Ap^\mu \right\vert
    &\gtrsim \lambda \vev^4.
\end{align}
As pointed out in Ref.~\cite{Agrawal:2018vin}, beyond this threshold the dark photon is no longer
well described by the Proca Lagrangian \cref{eqn:proca}, and whatever dynamics are responsible for
producing the dark photon are (presumably) substantially modified.

More recently, Ref.~\cite{East:2022rsi} showed that the consequences of violating
\cref{eqn:symmetry-restoration-condition} can be severe.
The vacuum state of the Abelian-Higgs theory is characterized by a ring of degenerate vacua,
allowing for topological line defects---namely, Nielsen-Olisen strings~\cite{Nielsen:1973cs}.
Each string has a quantized magnetic flux $\Phi_0 = 2\pi/\gD$; the strings therefore interact with
the background dark electromagnetic field, converting the would-be dark matter into string
energy.\footnote{
    Though strings only have an intrinsic magnetic moment per unit length, moving strings also
    possess an electric dipole moment; they therefore interact with both electric and magnetic
    fields of the dark photon.
}
For the purpose of understanding string formation, the dark photon, which otherwise would have
constituted the cold dark matter, can often be approximated as a constant, homogeneous magnetic
field (see Ref.~\cite{East:2022rsi} for a more comprehensive discussion).
Magnetic flux in excess of one unit ($2\pi/\gD$) prefers to be confined to a topological defect by a
phenomenon known as flux trapping~\cite{Hindmarsh:2000kd}.
That is, strings are the energetically preferable field configuration when the magnetic field
exceeds the first critical field $B_{c1} \sim \gD\vev^2$, corresponding to a magnetic flux
$2\pi/\gD$ inside an area $\sim 1/\mA^2$.
However, there is an energy barrier to string formation (due to the differing spacetime symmetry and
topology before and after string formation), which only vanishes when the magnetic field exceeds the
parametrically larger ``superheating'' field $B_\mathrm{sh} = \sqrt{\lambda}\vev^2$.\footnote{
    The phase transition is entirely nonthermal, making ``superheating'' a misnomer.
}
In this case, the magnetic flux confines to strings with interstring spacing much larger than the
string core, $1/\sqrt{\gD B_\mathrm{sh}}\gg m_h^{-1}$, so $\mathrm{U}(1)$ symmetry is restored
\emph{locally}.
Only once the external magnetic field exceeds the second critical field
$B_{c2}\sim (\lambda/\gD^2)^{1/2} m_h^2$ do the string cores overlap, restoring symmetry globally.

The idealized limit of a purely magnetic, homogeneous background field is not precisely applicable
to all scenarios for dark photon production, some of which produce semirelativistic or
nonrelativistic modes (with wave numbers $k \lesssim \mA$) that are coherent on scales $1/\mA$ (or
larger) but also contain electric fields.
Reference~\cite{East:2022rsi} demonstrated with numerical simulations that defects nucleate
nonetheless for tachyonic dark photon production (see \cref{sec:axion}) at energy densities above
the superheating threshold.
The superheating threshold also turns out to match that for strong backreaction onto the Higgs
identified in \cref{eqn:symmetry-restoration-condition}.
We therefore take \cref{eqn:symmetry-restoration-condition} as generic upper limit on the dark
photon field amplitude: as Proca theory is an invalid description of the Abelian-Higgs model in this
regime either way, and the dynamics intended to produce massive dark photons are in all likelihood
disrupted.
In the rest of this section we discuss exceptions to this argument as needed.

\begin{figure}[t!]
    \centering
    \includegraphics[width=\textwidth]{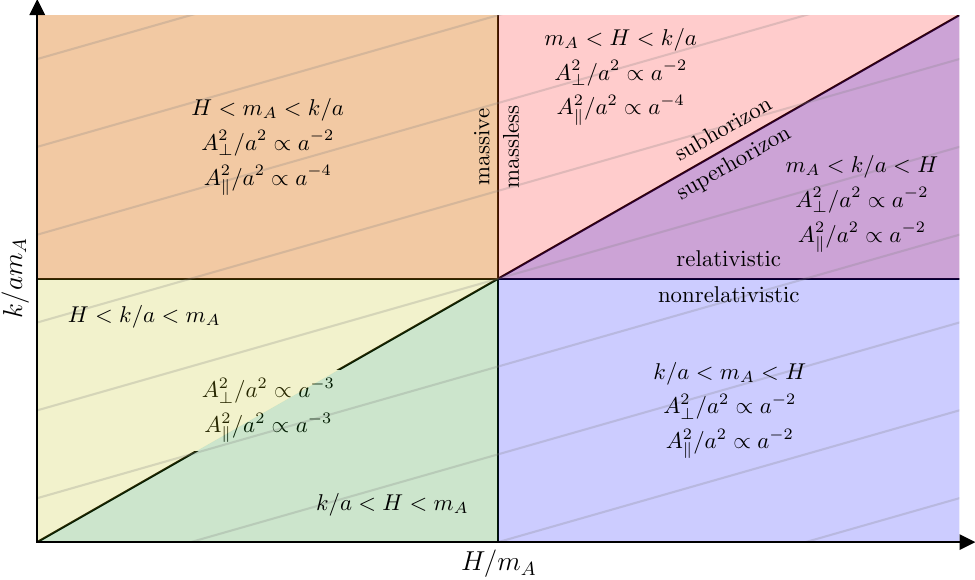}
    \caption{
        Redshifting of the dark photon's contribution to the Higgs's effective potential,
        $\propto \mA^2 A_\mu A^\mu$, in various regimes of free evolution.
        The horizontal and vertical axes are the physical Hubble scale and the physical momentum of
        a given mode.
        Gray lines depict the trajectory of modes of a fixed comoving wave number in time
        (from the top right to the bottom left).
        The relationship between $H$ and $k/a$ fixes a radiation-dominated Universe with
        $H \propto a^{-2}$; modifications to the expansion history would merely alter the line
        separating the sub- and superhorizon regimes.
    }
    \label{fig:free-evolution-backreaction-redshifting}
\end{figure}

To study the implications of \cref{eqn:symmetry-restoration-condition} as a consistency limit on
dark photon cosmologies, we first note that in almost all the cases we consider, dark photons are
produced with a peaked spectrum.
Namely, the relic abundance is dominated by dark photons over a narrow range of wave numbers
centered about the peak $k_\star$, which enables rephrasing the backreaction bound in terms of the
dark photon energy density $\rho_{A'}$.
Heuristically (see \cref{app:SAH} for more formal expressions),
\begin{align}
    \label{eqn:coupling-strength}
    \lambda v^4
    \gg -\mA^2 g^{\mu \nu} \Ap_\mu \Ap_\nu
    &\sim \frac{\mA^2}{\mA^2 + k_\star^2 / a^2} \rho_{\Ap, \perp}
        + \rho_{\Ap, \parallel},
\end{align}
where $\rho_{\Ap, \perp}$ and $\rho_{\Ap, \parallel}$ denote the contributions from transverse and
longitudinal modes, respectively.
\Cref{eqn:coupling-strength} shows that for equal energy densities, relativistic, transverse dark
photons couple more weakly (at fixed energy density) to the Higgs than nonrelativistic ones by a
factor of $(a \mA / k_\star)^2$.
But the energy density in relativistic modes redshifts as $a^{-4}$, one factor of $a$ faster than
nonrelativistic modes, and must therefore be produced with a larger initial energy density to match
the observed dark matter density.
The backreaction constraint is therefore relaxed, albeit modestly, if transverse dark photons
dominate $\rho_\Ap$ and become nonrelativistic as late as possible; \cref{eqn:coupling-strength} is
relatively more constraining if relativistic longitudinal modes dominate.
\Cref{fig:free-evolution-backreaction-redshifting} depicts the scaling of
\cref{eqn:coupling-strength} for freely evolving dark photons in all possible regimes of wave number
$k / a \mA$ and Hubble rate $H / \mA$.

\begin{figure}
    \centering
    \includegraphics[width=4.318in]{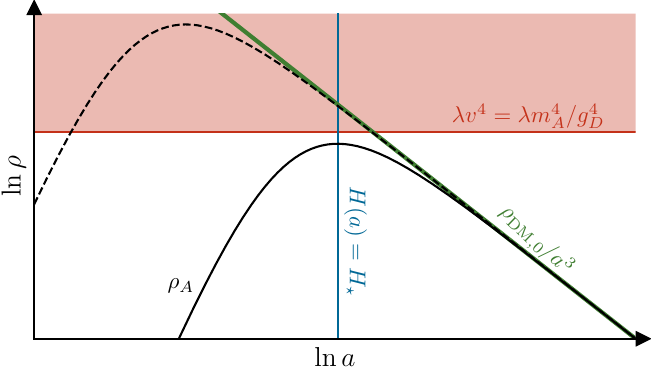}
    \caption{
        Illustration of dark photon production that avoids backreaction onto the Higgs by delaying
        the time of production (see also Ref.~\cite{Cyncynates:2023zwj}).
        After it is produced and becomes nonrelativistic, the dark matter has a known energy density at any scale factor $a$ (indicated by the green line)
        extrapolated from its present-day value.
        For any choice of model parameters, its energy density would exceed
        the threshold for backreaction (in the red shaded region) at some early time.
        By sufficiently delaying the production of dark photons, i.e., until some critical Hubble rate $H_\star$ (indicated by the blue line), they never backreact onto the Higgs (per the solid black curve).
        Dark photons produced too early (as in the dashed black curve) backreact onto the Higgs, possibly cutting off production before the total dark matter energy density is produced, and possibly collapsing into a string network that is not viable cold dark matter.
    }
    \label{fig:schematic}
\end{figure}

We derive a concrete bound on $\gD$ from \cref{eqn:coupling-strength} by taking dark photons to be
produced at a time $t_\star$ in the radiation era (when the Hubble rate is
$H_{\star} = 1 / 2 t_\star$) with typical wave number $k_\star$ and energy density
$\rho_{\Ap}(t_\star)$.
The energy density of the dominant modes scales like $a^{-4}$ until it becomes nonrelativistic when
$\sqrt{H_\mathrm{NR}/H_{\star}} k_{\star}/a_\star = \mA$ (where $H \propto a^{-2}$ during radiation
domination).
Afterward, the dark photon's energy density dilutes like matter ($\propto a^{-3}$).
For a present-day abundance of dark photons $\Omega_\Ap = \rho_A(t_0) / \rho(t_0)$ with $\rho(t_0)$
the critical density, and $\Omega_m$ defined likewise for the total matter density,
\begin{align}
    \frac{\Omega_A}{\Omega_m} \frac{3 \Mpl^2 H_\mathrm{eq}^2}{2}
    &\approx \left( \frac{H_\mathrm{eq}}{H_\mathrm{NR}} \right)^{3/2}
        \left( \frac{H_\mathrm{NR}}{H_\star} \right)^2
        \rho_{\Ap}(t_\star),
\end{align}
where $H_\mathrm{eq}$ is the Hubble rate at matter-radiation equality, equal to
$2.26 \times 10^{-28}~\eV$ in \textit{Planck}'s best-fit cosmology~\cite{Planck:2018vyg}.
Solving for $\rho_{\Ap}(t_\star)$ and plugging into \cref{eqn:coupling-strength} gives
\begin{align}
    \gD
    &\lesssim
        10^{-14} \,
        \lambda^{1/4}
        \left( \frac{\mA}{\mu\eV} \right)^{5/8}
        \left( \frac{\mA}{H_\star} \right)^{3/8}
        \left( \frac{\Omega_\Ap}{\Omega_m} \right)^{-1/4}
        \begin{dcases}
            \left( \frac{H_\mathrm{NR}}{H_\star} \right)^{-1/8},
            & \rho_{A, \perp} \gg \rho_{A, \parallel},
            \\
            \left( \frac{H_\mathrm{NR}}{H_\star} \right)^{1/8},
            & \rho_{A, \perp} \ll \rho_{A, \parallel}.
        \end{dcases}
    \label{eqn:defectBound}
\end{align}
Independent of polarization, the bounds on $\gD$ are least severe when $H_{\star}$ is as small as
possible, as illustrated by \cref{fig:schematic}.
That is, the later dark photons are produced, the lower their peak energy density.
The case with purely transverse dark photons benefits from remaining relativistic until as late as
is allowed, while longitudinal-mode production is least constrained if nonrelativistic at the start;
the dependence on $H_\mathrm{NR} / H_\star$ is modest in either case.
As we show in the following, models of nonthermal production typically take place when
$H_\star \gtrsim \mA$, setting a useful benchmark.

Crucially, reducing the gauge coupling $\gD$ to avoid backreaction comes at the cost of also
decreasing the strength of kinetic mixing between the dark and SM photons.
In the simplest scenarios, kinetic mixing is generated by loops of fermions charged under both the
visible and dark photons and is therefore proportional to their charge under both
groups~\cite{Holdom:1985ag}:
\begin{align}\label{eqn:epsilon-Benchmark}
    \varepsilon
    &\sim \frac{e \gD}{16 \pi^2},
\end{align}
with loop factor $16 \pi^2$ chosen to match Ref.~\cite{East:2022rsi}.
On the other hand, kinetic mixing is a dimension-four operator, and from the perspective of
effective field theory there is no fundamental obstruction to simply choosing $\varepsilon \gg g_D$.
However, in the basis with diagonal kinetic and mass terms ($\Asm \to \Asm - \varepsilon \Ap$), the SM
fermions have a charge $Q \sim \varepsilon / e \gD$ under the dark gauge group---i.e., choosing an
arbitrary value of $\varepsilon$ is equivalent taking the SM fermions to have an arbitrarily large
charge measured in units of the dark Higgs charge ($g_D$).
A prototypical mechanism to mechanize a large hierarchy in charges from a theory with order-unity
parameters is clockwork~\cite{Giudice:2016yja}, which we consider in detail in
\cref{sec:production-independent}.

\begin{figure}
    \centering
    \includegraphics[width=\columnwidth]{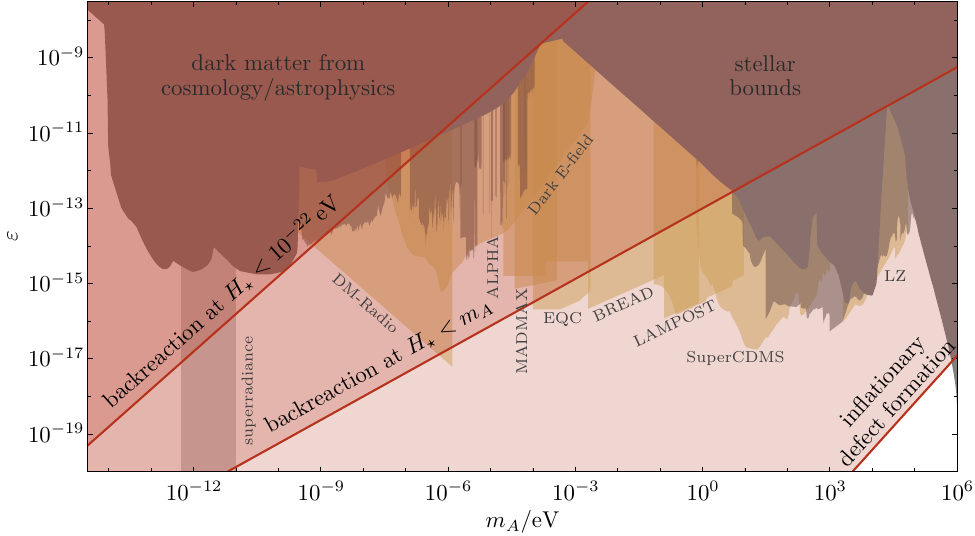}
    \caption{
        Parameter space for kinetically mixed dark photons.
        Current exclusions from astrophysical~\cite{Redondo:2008aa, Zechlin:2008tj, Li:2023vpv,
        Dubovsky:2015cca, Vinyoles:2015aba, Baryakhtar:2017ngi, Wadekar:2019mpc, Linden:2024fby,
        Hong:2020bxo, Bi:2020ths, Fedderke:2021aqo} and cosmological~\cite{Arias:2012az,
        McDermott:2019lch, Caputo:2020rnx, Caputo:2020bdy, Witte:2020rvb} probes are depicted in
        dark gray and those from haloscope and other laboratory searches~\cite{Suzuki:2015sza,
        Knirck:2018ojz, Brun:2019kak, Hochberg:2019cyy, Nguyen:2019xuh, Phipps:2019cqy,
        SuperCDMS:2019jxx, XENON:2019gfn, DAMIC:2019dcn, An:2020bxd, Dixit:2020ymh,
        FUNKExperiment:2020ofv, SENSEI:2020dpa, Tomita:2020usq, XENON:2020rca, Chiles:2021gxk,
        Fedderke:2021rrm, Manenti:2021whp, XENON:2021qze, XENON:2022ltv, An:2022hhb,
        Cervantes:2022yzp, DarkSide:2022knj, DOSUE-RR:2022ise, Fan:2022uwu, Ramanathan:2022egk,
        Bajjali:2023uis, An:2023wij, BREAD:2023xhc, An:2024wmc, Levine:2024noa} in light gray.
        The projected reaches~\cite{Caputo:2021eaa} of DMRadio~\cite{DMRadio:2022pkf}, Dark
        E-field~\cite{Godfrey:2021tvs}, ALPHA~\cite{ALPHA:2022rxj}, MADMAX~\cite{Gelmini:2020kcu},
        BREAD~\cite{BREAD:2021tpx} and an extension using a highly excited
        cyclotron~\cite{Fan:2024mhm}, LAMPOST~\cite{Baryakhtar:2018doz},
        SuperCDMS~\cite{Bloch:2016sjj}, and LZ~\cite{LZ:2021xov} are depicted in yellow.
        The red regions depict various benchmark thresholds for which the dark photon substantially
        backreacts onto the dark Higgs's effective potential [\cref{eqn:defectBound}] and can
        collapse into a string network, taking the fiducial kinetic mixing of
        \cref{eqn:epsilon-Benchmark}.
        These benchmark scenarios include inflationary production
        [\cref{eqn:minimal-gD-inflation-bound}]; nonrelativistic, postinflationary production
        taking place when the Hubble $H_\star < \mA$; and a maximally conservative bound that only
        requires avoiding backreaction only around the time when scales relevant to CMB anisotropies
        enter the horizon.
    }
    \label{fig:defect-limits-existing-models}
\end{figure}

Taking \cref{eqn:defectBound} as a fiducial benchmark, \cref{fig:defect-limits-existing-models}
depicts bounds for several benchmark scenarios in the $\varepsilon$-$\mA$ parameter space,
superimposed with current limits~\cite{Redondo:2008aa, Zechlin:2008tj, Li:2023vpv, Dubovsky:2015cca,
Vinyoles:2015aba, Baryakhtar:2017ngi, Wadekar:2019mpc, Linden:2024fby, Hong:2020bxo, Bi:2020ths,
Fedderke:2021aqo, Arias:2012az, McDermott:2019lch, Caputo:2020rnx, Caputo:2020bdy, Witte:2020rvb,
Chluba:2024wui, Arsenadze:2024ywr, Suzuki:2015sza, Knirck:2018ojz, Brun:2019kak, Hochberg:2019cyy,
Nguyen:2019xuh, Phipps:2019cqy, SuperCDMS:2019jxx, XENON:2019gfn, DAMIC:2019dcn, An:2020bxd,
Dixit:2020ymh, FUNKExperiment:2020ofv, SENSEI:2020dpa, Tomita:2020usq, XENON:2020rca,
Chiles:2021gxk, Fedderke:2021rrm, Manenti:2021whp, XENON:2021qze, XENON:2022ltv, An:2022hhb,
Cervantes:2022yzp, DarkSide:2022knj, DOSUE-RR:2022ise, Fan:2022uwu, Ramanathan:2022egk,
Bajjali:2023uis, An:2023wij, BREAD:2023xhc, An:2024wmc, Levine:2024noa} and the prospective reach of
DMRadio~\cite{DMRadio:2022pkf}, Dark E-field~\cite{Godfrey:2021tvs}, ALPHA~\cite{ALPHA:2022rxj},
MADMAX~\cite{Gelmini:2020kcu}, BREAD~\cite{BREAD:2021tpx} and an extension using a highly excited
cyclotron~\cite{Fan:2024mhm}, LAMPOST~\cite{Baryakhtar:2018doz}, SuperCDMS~\cite{Bloch:2016sjj}, and
LZ~\cite{LZ:2021xov}.
Much of the open, experimentally accessible parameter space lies in the range
$10^{-16} \lesssim \varepsilon \lesssim 10^{-10}$; only the lower end of this range, however, is
within the reach of future searches for scenarios with $H_\star \gtrsim \mA$.

A maximally conservative bound might assume that the dark matter existed no earlier than directly
evidenced by data---namely, no earlier than required by observations of the cosmic microwave
background (CMB) anisotropies.
The visible CMB is sensitive to scales that enter the horizon in the few decades of expansion around
matter-radiation equality; we take $H_\star \approx 10^{-22}~\eV$ as a reasonable threshold.
\Cref{fig:defect-limits-existing-models} shows that a hypothetical scenario delaying production to
so late a time is viable in nearly all the parameter space of future searches.
Interestingly, bounds on dark photon dark matter from cosmological
observations~\cite{Arias:2012az,McDermott:2019lch,Caputo:2020rnx,Caputo:2020bdy,Witte:2020rvb}
(labeled ``dark matter from cosmology/astrophysics'' in \cref{fig:defect-limits-existing-models})
reside within this conservative exclusion region---that is, in the parameter space excluded by
Refs.~\cite{Arias:2012az,McDermott:2019lch,Caputo:2020rnx,Caputo:2020bdy,Witte:2020rvb}, dark
photons with kinetic mixing \cref{eqn:epsilon-Benchmark} and a standard Higgs mass could not be the
dark matter anyway.

\subsection{Inflationary production}\label{sec:inflation}

The most minimal mechanism for cosmological particle production is
gravitational~\cite{Kolb:2023ydq}.
During inflation, minimally coupled degrees of freedom are excited as their wave number exits the
horizon, so long as their action breaks conformal symmetry.
Though the transverse polarizations of a vector are conformally symmetric up to their small mass
$\mA \ll H_I$, the longitudinal mode of a massive vector is equivalent to a minimally coupled scalar
field; its action thus violates conformal symmetry no matter how small its mass.
Reference~\cite{Graham:2015rva} showed that longitudinal modes exit the horizon during inflation
with amplitude $k H_I / 2 \pi \mA$; per \cref{fig:free-evolution-backreaction-redshifting}, they
remain frozen until they either become massive (i.e., when $H \leq \mA$) or enter the horizon.
A mode with wave number $k$ reenters the horizon at $a_\times = k / H(a_\times)$ and becomes
nonrelativistic at $a_\mathrm{NR} = k / \mA$.
Assuming that the radiation era began immediately after inflation (so that $H \propto a^2$) and
using \cref{eqn:rho-perp-parallel}, the energy density per logarithmic wave number thus scales as
\begin{align}
    \dd{\rho_\parallel}{\ln k}
    \propto \frac{\mA^2}{a^2} \dd{\left\vert A_\parallel \right\vert^2}{\ln k}
    &\sim
        \mA^2 \left( \frac{H_I}{2 \pi} \right)^2
        \left( \frac{H(a)}{\mA} \right)^{3/2}
        \cdot \begin{dcases}
            \left( \frac{k}{a_\star \mA} \right)^2,
            & k < a_\star \mA,
            \\
            \left( \frac{k}{a_\star \mA} \right)^{-1},
            & k > a_\star \mA
        \end{dcases}
    \label{eqn:inflation-energy-spectrum}
\end{align}
at late times $a > a_\star$ and $a > a_\times$.
The present abundance of longitudinal dark photons produced gravitationally is thus dominated by the
mode that enters the horizon when $H = \mA$ and takes the form~\cite{Graham:2015rva}
\begin{align}\label{eqn:inflationaryEnergyDensity}
    \Omega_{\Ap}
    &= \Omega_\mathrm{DM} \left( \frac{\mA}{0.1~\meV} \right)^{1/2}
        \left( \frac{H_I}{5 \times 10^{13}~\GeV} \right)^2.
\end{align}
Dark photons with mass $\mA$ below $0.1~\meV$ cannot be sufficiently produced during
inflation due to bounds on the tensor-to-scalar ratio from recent \textit{Planck} and
BICEP/\textit{Keck} observations~\cite{BICEP:2021xfz}.\footnote{
    Namely, taking the amplitude of the scalar power spectrum
    $A_s = 2.1 \times 10^{-9}$~\cite{Planck:2018vyg} and the tensor-to-scalar ratio
    $r < 0.036$~\cite{BICEP:2021xfz} bounds
    $H_I = \sqrt{ \pi^2 r A_s / 2} \Mpl < 4.7 \times 10^{13}~\mathrm{GeV}$.
    \label{footnote:H-inf}
}

During inflation, unsuppressed formation of topological defects takes place if $H_I^2\gtrsim \mu =
\pi\vev^2\ln(\min[\lambda/\gD^2,\sqrt{\lambda}v/H_I])$~\cite{Basu:1991ig,East:2022rsi}, where $\mu$
is the string tension.\footnote{
    References~\cite{Redi:2022zkt, Sato:2022jya} extended the study of inflationary dark photon
    production to account for a Higgs mass, though without considering the implications for direct
    detection.
}
After inflation, these strings approach a scaling solution and absorb any remaining coherent dark
photon field.
Thus, if inflation is to account for vector dark matter, then the scale of inflation $H_I$ must be
below the Higgs VEV $\vev$, in turn bounding the dark gauge coupling by~\cite{East:2022rsi}
\begin{align}
    \label{eqn:minimal-gD-inflation-bound}
    \gD
    \lesssim \frac{\mA}{H_I}
    &\lesssim 2 \times 10^{-22}
        \left( \frac{\mA}{\eV} \right)^{5/4}
        \left( \frac{\Omega_A}{\Omega_\mathrm{DM}} \right)^{-1/2}.
\end{align}
The corresponding kinetic mixing [\cref{eqn:epsilon-Benchmark}] is well out of range of any foreseeable experiment (\cref{fig:defect-limits-existing-models}).\footnote{
    Reference~\cite{East:2022rsi} points out that this bound may in fact be conservative, as even if
    string production is exponentially suppressed ($H_I \lesssim v$) strings may continue to grow in
    the background of the dark electromagnetic field to reach a scaling solution.
    In fact, this concern is also relevant for \cref{eqn:defectBound}: large fluctuations, though
    exponentially unlikely, can produce strings.
    The precise threshold value of $H_I$ or $\rho_A(t_\star)$ for which strings would be
    cosmologically relevant today requires further study.
}

The abundance in \cref{eqn:inflationaryEnergyDensity}, as computed in Ref.~\cite{Graham:2015rva},
assumes instantaneous reheating, i.e., that the radiation era begins immediately following
inflation.
There is no guarantee that a prolonged era dominated by matter (or something more exotic) did not
occur first after inflation, which modifies the relationship between the inflationary scale $H_I$
and the relic abundance when the dark photon's mass $\mA$ exceeds the Hubble rate at reheating
$H_\mathrm{RH}$.
(Here by reheating we mean the moment the radiation era began, not necessarily that when the SM
itself is populated and thermalized.)
Modified expansion histories therefore shift the bound on $\gD$ in
\cref{eqn:minimal-gD-inflation-bound} (as well as the minimal allowed dark photon mass $\mA$ given
an upper bound on $H_I$ from CMB observations).

\subsubsection{Postinflationary expansion history}\label{sec:postinflationary-histories}

\begin{figure}
    \centering
    \includegraphics[width=\textwidth]{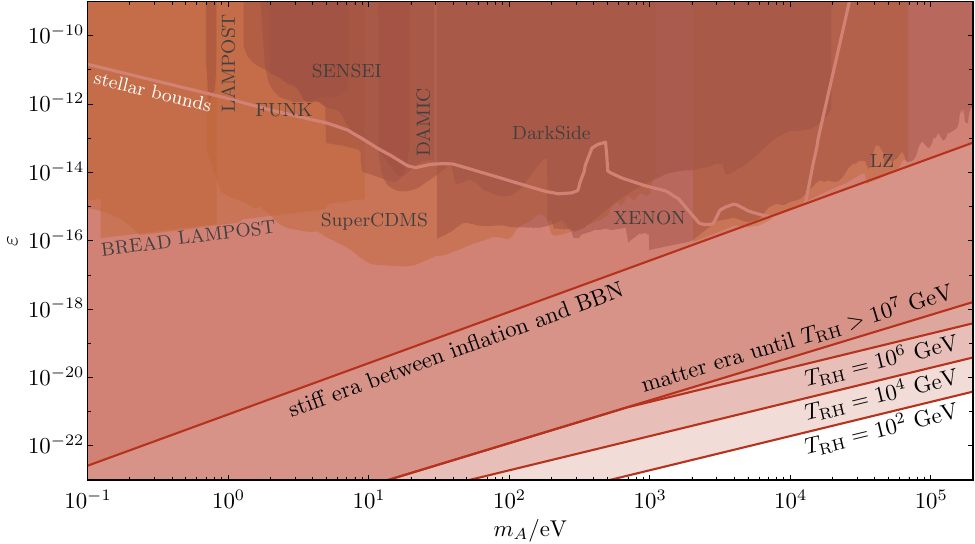}
    \caption{
        Kinetic mixing parameter space [\cref{eqn:epsilon-Benchmark}] of dark photon dark matter
        produced during inflation.
        Red regions depict parameter space that are constrained by defect formation under various
        assumptions on the expansion history between inflation and nucleosynthesis, as labeled on
        the figure: the standard case of instantaneous reheating, an extreme case in which the
        Universe is kination-dominated ($w = 1$) until BBN, and various extended epochs of matter
        domination with reheating temperatures indicated on the figure.
        Current limits from LAMPOST~\cite{Chiles:2021gxk}, FUNK~\cite{FUNKExperiment:2020ofv},
        SENSEI~\cite{SENSEI:2020dpa}, DAMIC~\cite{DAMIC:2019dcn}, DarkSide~\cite{DarkSide:2022knj},
        and XENON~\cite{XENON:2019gfn, XENON:2020rca, XENON:2021qze, XENON:2022ltv} are depicted in
        gray and the projected reach~\cite{Caputo:2021eaa} of BREAD~\cite{BREAD:2021tpx},
        LAMPOST~\cite{Baryakhtar:2018doz}, SuperCDMS~\cite{Bloch:2016sjj}, and LZ~\cite{LZ:2021xov}
        in yellow.
        Joint bounds from stellar probes are outlined in white.
    }
    \label{fig:inflation-parameter-space}
\end{figure}

The results of Ref.~\cite{Graham:2015rva} were extended to an arbitrary reheating temperature in
Refs.~\cite{Ema:2019yrd, Kolb:2020fwh} and to arbitrary expansion histories before reheating in
Ref.~\cite{Ahmed:2020fhc}.
In the simplest inflationary scenarios, once slow roll ends, the inflaton retains the entirety of
the Universe's energy as it oscillates about the minimum of its potential (which is typically taken
to be quadratic, such that the inflaton is matterlike in this period).
The Universe only reheats once the inflaton's dominant decay channel becomes efficient relative to
the expansion rate.
Since the inflaton's energy density redshifts more slowly than the SM radiation it eventually decays
into, the relative abundance of dark photons is lower for later reheating.
Achieving the full relic abundance of the dark matter at a fixed dark photon mass then requires a
larger inflationary scale $H_I$, such that avoiding defect formation imposes an even stronger limit
on $\gD$.
In fact, the relic abundance is independent of $\mA$ for
$\mA > H_\mathrm{RH}$~\cite{Ema:2019yrd, Kolb:2020fwh, Ahmed:2020fhc}, instead scaling with
$\sqrt{H_\mathrm{RH}} H_I^2$.
Requiring the dark photon to make up all of the dark matter then places a lower bound on the Hubble
rate when the radiation era began as a function of $H_I$.
The bound on the dark gauge coupling \cref{eqn:minimal-gD-inflation-bound} is then penalized by a
factor of $(\mA / H_\mathrm{RH})^{-1/4}$ for dark photons heavier than $H_\mathrm{RH}$ (\cref{fig:inflation-parameter-space}).

More generally, any early epoch with equation of state $w < 1/3$ further reduces the maximal gauge
coupling that avoids inflationary defect formation.
For stiffer equations of state $w > 1/3$, the Universe's energy density redshifts faster than
radiation, instead easing constraints on $\gD$.
The net enhancement in abundance in the $w \to 1$ limit is
$\sim \sqrt{\mA / H_\mathrm{RH}}$~\cite{Ahmed:2020fhc}, requiring a lower inflationary scale to
match the dark matter relic abundance and alleviating bounds on $\gD$ by a factor
$\sim \sqrt[4]{\mA / H_\mathrm{RH}}$.
The minimum reheat temperature allowed by big bang nucleosynthesis (BBN) is of order
$10~\meV$~\cite{Giudice:2000ex, Kawasaki:2000en, Kawasaki:2004qu, Hannestad:2004px,
Ichikawa:2005vw, deSalas:2015glj, Hasegawa:2019jsa}, corresponding to
$H_\mathrm{RH} \approx 10^{-14}~\eV$ and an enhancement in allowed $\gD$ of about three orders of
magnitude for $\mA \sim \eV$.
\Cref{fig:inflation-parameter-space} shows that, while no viable parameter space is accessible to
current or future searches assuming an instantaneous onset of radiation domination,
a stiff era indeed marginally allows for kinetic mixing within reach of LZ~\cite{LZ:2021xov}
and only just below current exclusions from XENON~\cite{XENON:2019gfn, XENON:2020rca, XENON:2021qze,
XENON:2022ltv}.
On the other hand, if the Universe were instead matter-dominated for some or all of the epoch
between inflation and BBN (a possibility one has little reason to exclude \textit{a priori}), viable
kinetic mixings are up to $\sim 2.5$ decades further out of reach of these experiments, depending on
how late reheating occurs.\footnote{
    Note that $T_\mathrm{RH} = 444~\mathrm{GeV}$ is the lowest possible reheating temperature
    for which dark photons (of any mass) produced gravitationally during inflation may make up the
    entire dark matter abundance when taking $H_I < 4.7 \times 10^{13}~\mathrm{GeV}$
    (see \cref{footnote:H-inf}).
}

\subsubsection{Nonminimal couplings}\label{sec:nonminimal-couplings}

Nonminimal couplings to gravity may affect the relic abundance of dark photons generated during
inflation~\cite{Cembranos:2023qph, Ozsoy:2023gnl}.
However, nonminimally coupled Proca theories of the form
\begin{align}
    \mathcal{L}
    &= - \frac{1}{4} \Fp_{\mu\nu} {\Fp}^{\mu\nu}
        + \frac{1}{2} \mA^2 \Ap_\mu \Ap^\mu
        - \frac{1}{2} \xi_1 R \Ap_\mu \Ap^\mu
        - \frac{1}{2} \xi_2 R^{\mu \nu} \Ap_\mu \Ap_\nu
    \label{eqn:nonminimal-proca}
\end{align}
exhibit ghost instabilities~\cite{Himmetoglu:2008zp, Himmetoglu:2009qi, Karciauskas:2010as}.
References~\cite{Capanelli:2024pzd, Capanelli:2024rlk} more recently discuss these runaway
instabilities during inflation, arguing that they are present if the nonminimal couplings $\xi_1$
and $\xi_2$ are larger than $\sim (\mA / H_I)^2$ in magnitude; the dynamics have only been solved
numerically, preventing a simple analysis of the effect on defect formation bounds.
References~\cite{Capanelli:2024pzd, Capanelli:2024rlk} claim that radiative corrections in
Abelian-Higgs theories generate nonminimal couplings of order
$(\mA / m_\higgs)^6 = \gD^6 / (2 \lambda)^3$.
The runaway instability is then absent if $\gD / \sqrt{2 \lambda} < \sqrt[3]{\mA / H_I}$, which is
a dramatically weaker requirement than that to evade inflationary defect formation
[\cref{eqn:minimal-gD-inflation-bound}].

Moreover, the nonminimal couplings in \cref{eqn:nonminimal-proca} must ultimately descend from
gauge-invariant operators.
(Otherwise, technical naturalness would not protect the dark photon's mass from large radiative
corrections.)
The only gauge-invariant operator linear in $\Ap_\mu$ is $D_\mu \Phi$; the nonminimally coupled
Abelian-Higgs theory therefore must take the form
\begin{align}
    \mathcal{L}
    &= - \frac{1}{4} \Fp_{\mu\nu} {\Fp}^{\mu\nu}
        + \frac{1}{2}
        \left(
            g^{\mu \nu}
            - \frac{\xi_1}{\mA^2} R g^{\mu \nu}
            - \frac{\xi_2}{\mA^2} R^{\mu \nu}
        \right)
        D_\mu \Phi \left( D_\nu \Phi \right)^\ast
        - \frac{\lambda}{4} \left( \abs{\Phi}^2 - v^2 \right)^2
    .
    \label{eqn:nonminimal-proca-gauge-invariant}
\end{align}
As mentioned by Ref.~\cite{East:2022rsi}, such a nonminimal coupling for the dark Higgs drives its
VEV to larger values during inflation, which would mitigate defect formation if an increase in the
inflationary scale were not required to compensate for a suppression incurred in the dark photon
abundance (as derived in \cref{sec:inflation-varying-mass}).
Furthermore, the effective Higgs mass squared is also suppressed by a factor $\sim \mA^2 / \xi_i R$,
not to mention the additional derivative couplings induced by the time dependence of the coefficient
of the Higgs's kinetic term.
Unless $m_\higgs / H_I$ is even larger than $\sqrt{\vert \xi_i \vert} H_I / \mA$, the Higgs is thus
light during inflation and would be produced gravitationally to a comparable degree; since the Higgs
is much heavier at late times, it would dominate the dark sector density.
When requiring the Higgs to remain heavy during inflation (i.e.,
$\mA^2 m_\higgs^2 / \vert \xi_i \vert H_I^4 > 1$), if $\vert \xi_i \vert \sim (\mA / m_\higgs)^6$
as expected from radiative corrections then $\vert \xi_i \vert < (\mA / H_I)^3$ and the runaway
instability is absent.

Though \cref{eqn:nonminimal-proca-gauge-invariant} is written in terms of $\xi_i / \mA^2$ to match
the form of \cref{eqn:nonminimal-proca}, in reality we expect the coefficient of these dimension-6
operators to be $1 / \Lambda^2$ where $\Lambda$ is the energy scale of the physics that generates
the nonminimal couplings.
(Written in terms of $\Lambda$, the $\mA \to 0$ limit explicitly restores gauge symmetry and the
dark photon mass's technical naturalness remains manifest.)
In order that the low-energy theory remain a valid description of gravity during inflation, $H_I$
(if not the energy scale $\sqrt{H_I \Mpl} \gg H_I$) should be below $\Lambda$; one would thus expect
that $\xi_i \sim (\mA / \Lambda)^2 < (\mA / H_I)^2$, even discounting powers of the dark gauge
coupling expected from vertex factors.
It is therefore not even clear whether the radiatively generated $\xi_i$ should be large enough to
trigger the runaway instability in the first place without breaching the validity of the effective
field theory.
Because it is unclear whether nonminimal couplings can have a significant impact on dark photon dark
matter production without overproducing the dark Higgs, triggering vortex formation, or violating
perturbativity, we do not consider them further (except when referencing their invocation in past
literature on vector misalignment in \cref{sec:vector-misalignment}).

\subsubsection{Direct inflaton couplings}\label{sec:direct-inflaton-couplings}

The dark photon (or one of the particles it interacts with) could also directly couple to the
inflaton, enabling more efficient production of dark photon fluctuations and therefore a lower scale
of inflation to achieve the dark matter relic abundance~\cite{Bastero-Gil:2018uel, Salehian:2020asa,
Firouzjahi:2020whk, Nakai:2020cfw, Bastero-Gil:2021wsf, Nakai:2022dni, Barrie:2022mma}.
These mechanisms are nonetheless constrained at least to the same extent as any late-time production
mechanism, i.e., \cref{eqn:defectBound} with $H_\star = \mA$; however, the maximal energy density is
necessarily higher than that assumed in \cref{eqn:defectBound} given that the dark photons were
produced before the end of inflation.
Referring to \cref{fig:free-evolution-backreaction-redshifting}, $\Ap_\mu \Ap^\mu$ redshifts no more
slowly than $a^{-2}$, and while $H > \mA$ all modes redshift as such except for subhorizon,
longitudinal ones for which $\Ap_\mu \Ap^\mu \propto a^{-4}$.
Taking a general equation of state $w$ in this epoch such that $H \propto a^{-3 (1 + w) / 2}$,
\cref{eqn:defectBound} is penalized by an additional factor $(\mA / H_I)^{1/3(1+w)}$.
If longitudinal modes enter the horizon at a scale factor $a_\times$ before $H = \mA$, a further
penalty of $(a_\times / a_\star)^{1/2}$ is incurred.
These factors only account for the dark photons existing at the end of inflation; production earlier
in inflation yields yet stronger bounds (but with model dependence that is difficult to capture on
general grounds).

Similar scenarios where the dark photon instead couples to a rolling spectator field during
inflation~\cite{Nakai:2022dni} are subject to the same bounds.
However, Refs.~\cite{Salehian:2020asa, Firouzjahi:2020whk, Nakai:2020cfw, Nakai:2022dni} considered
only phenomenological parametrizations of the time dependence of the coupling function to the
inflaton or spectator, neglecting the production that occurs (in concrete models) as the coupled
scalar eventually oscillates about the minimum of its potential---i.e., preheating.
This contribution most likely substantially modifies the results in these models, given that
preheating would produce dark photons on scales as small as $k \sim a_e H_I$ where $a_e$ is the
scale factor at the end of inflation.
On the other hand, the results of Refs.~\cite{Bastero-Gil:2018uel, Bastero-Gil:2021wsf,
Barrie:2022mma} are limited to the linear regime in which backreaction effects are negligible; full
nonlinear dynamics of such models were treated in Refs.~\cite{Deskins:2013dwa, Adshead:2015pva,
Adshead:2016iae, Adshead:2017xll, Adshead:2018doq, Cuissa:2018oiw, Lozanov:2019jff, Adshead:2019lbr,
Adshead:2019igv, Adshead:2023mvt}, albeit not specialized to dark photons specifically as the dark
matter.

\subsection{Irreducible thermal fluctuations}\label{sec:thermal}

Dark photons that kinetically mix with the SM photon are inevitably thermally excited to some
degree; depending on the thermal history of the Universe, these fluctuations can be large enough to
restore symmetry to the dark Higgs such that topological defects form via the Kibble-Zurek
mechanism~\cite{Kibble:1976sj}.
The dynamics of the photon--dark-photon system are described by the Lagrangian
\begin{align}
    \mathcal{L}
    &= - \frac{1}{4} \Fp_{\mu \nu} \Fp^{\mu \nu}
        - \frac{1}{4} \Fsm_{\mu \nu} \Fsm^{\mu \nu}
        + \frac{1}{2} \mA^2 \Ap_\mu \Ap^\mu
        - \frac{\varepsilon}{2} \Fp_{\mu \nu} \Fsm^{\mu \nu}
        - e J_\mathrm{SM}^\mu \Asm_\mu,
\end{align}
where $J_\mathrm{SM}$ is the current of electrically charged SM particles.
Charged SM particles are sterile with respect to the combination $S = \Ap + \varepsilon \Asm$,
making the $\{\Asm, S\}$ basis a convenient choice to discuss the propagation of (dark) photons in
the SM plasma~\cite{Redondo:2008aa}.
The dispersion relation of the active state $A_\mathrm{SM}$ receives corrections in the background of the Standard
Model plasma, resulting in the well-known effective mixing angle between $A_\mathrm{SM}$ and
$S$~\cite{Redondo:2008aa},
\begin{align}\label{eqn:in-medium-mixing-angle}
    \epsilon_{\mathrm{eff},\lambda}^2
    &= \frac{\varepsilon^2\mA^4}{(m_{\lambda}^2 - \mA^2)^2 + \omega^2\Gamma_\lambda^2}
        + \mathcal{O}(\varepsilon^4),
\end{align}
where $m_{\lambda}$ and $\Gamma_\lambda$ are the plasma frequency and damping rate of SM photons
with polarization state $\lambda$ and $\omega$ is the frequency of the photons.

At sufficiently early times, the plasma frequency of transverse modes far exceeds the bare mass of
the dark photon, i.e., $m_\perp^2 \propto e^2 T^2\gg\mA^2$; the effective mixing angle is then
suppressed relative to its vacuum value as
$\varepsilon_{\mathrm{eff}, \perp} \sim \varepsilon \mA^2 / m_\perp^2$.
In other words, the transverse photon $A_\mathrm{SM,\perp}$ is not only an interaction eigenstate but
also nearly a propagation eigenstate.
The amplitude of the sterile state $S$ is therefore suppressed relative to the SM photon by a factor
$\varepsilon \mA^2/m_\perp^2$ and hence
\begin{align}
   \Ap_\perp
   &\approx - \varepsilon \left[ 1 + \mathcal{O}(\mA^2 / m_\perp^2) \right] A_\mathrm{SM,\perp}
   .
\end{align}
The dark photon fluctuations of order $\varepsilon A_\mathrm{SM,\perp}$ do not set the late-time relic
dark matter abundance---they are strongly coupled to the plasma and continually reabsorbed---rather,
the sterile state itself adiabatically transitions into the dark photon as the Universe cools and
the plasma frequency drops.
These $\Ap_\perp$ fluctuations do, however, backreact onto the dark Higgs, and at sufficiently large
temperature and dark gauge coupling the small, thermal dark photon fluctuations are enough to
restore symmetry globally to the Higgs.
That is, the active state $A_\mathrm{SM}$ receives a small effective mass from the Higgs field
$\varepsilon \gD \vert \Phi \vert$, leading to a thermal effective potential for the
Higgs~\cite{Kapusta:1989tk}
$V_\mathrm{eff}(\vert \Phi \vert^2)\supset \varepsilon^2 \gD^2 \vert \Phi \vert^2 T^2/24$.
This effective potential restores symmetry to the Higgs when Standard Model temperature is
sufficiently large, and consequently defects form by the Kibble-Zurek mechanism, precluding the
formation of dark photon dark matter at some later epoch.
Requiring that symmetry remains broken when SM temperature is its largest ($T_\mathrm{max}$) sets a
bound
\begin{align}\label{eqn:irreducible-thermal-abundance-bound}
    \varepsilon
    &\lesssim
        1.1 \times 10^{-6} \lambda^{1/6}
        \left( \frac{\mA}{\ueV} \right)^{1/3}
        \left( \frac{T_\mathrm{max}}{10~\MeV} \right)^{-1/3}
        \left( \frac{16 \pi^2 \varepsilon}{e \gD} \right)^{2/3}.
\end{align}
This limit is depicted in \cref{fig:final-parameter-space} (red lines labeled by values of
$T_\mathrm{max}$) and discussed in \cref{sec:discussion}.

\subsection{Axion oscillations and tachyonic resonance}\label{sec:axion}

While inflationary and thermal dark photon production are both minimal scenarios in their physical
content and assumptions, neither can produce dark photon dark matter at small masses and large
couplings---nor can either produce viable dark photon dark matter below $0.1~\meV$ at all.
Detectably large kinetic mixing in the sub-meV mass range requires a nonthermal, postinflationary,
and nonminimal production mechanism (or an \textit{ad hoc} mechanism to boost $\varepsilon$ relative
to $\gD$; see \cref{sec:production-independent}).
One possibility is that the dark photon inherits the relic abundance of a misaligned scalar field
via resonant particle production.
A canonical example is a coupling to an axion $\phi_a = \fa \theta$ with mass $\ma$ and decay
constant $\fa$ through the axial term
$\tilde{\Fp}_{\mu\nu}\Fp^{\mu\nu} / 4 = \three{E} \cdot \three{B}$~\cite{Agrawal:2018vin,
Co:2018lka, Co:2021rhi, Kitajima:2023pby},
\begin{align}
    \mathcal{L}
    \supset \frac{1}{2} f_a^2 \partial_\mu \theta \partial^\mu \theta
        - V(\theta)
        + \frac{\beta}{4} \theta \tilde{F}_{\mu\nu} F^{\mu\nu},
\end{align}
where $\tilde{F}_{\mu\nu} \equiv \epsilon_{\mu\nu\rho\sigma} F^{\rho\sigma}$ is the dual field
strength tensor.
The axion lives in a compact field space, and its potential often takes the simple periodic form
\begin{align}
    V
    &= \ma^2 \fa^2 \left( 1 - \cos \theta \right).
\end{align}
After a period of inflation with $\ma \ll H_I \ll \fa$, the axion field assumes a nearly homogeneous
initial condition $\theta(0, \three{x}) \equiv \theta_0$ selected from the interval $[0, 2\pi)$
with approximately uniform probability.
To linear order in spatial fluctuations, the equations of motion for the circular polarizations of
the dark photon $\Ap_\pm$ [\cref{eqn:Ai-polarization-fourier-expansion}] and the spatially averaged
axion $\bar{\theta}(t)$ are
\begin{subequations}\label{eqn:axion-photon-eoms}
\begin{align}
    0
    &= \ddot{\theta} + 3 H \dot{\theta} + \ma^2 \theta
        - \frac{\beta}{\fa^2} \gen{\three{E} \cdot \three{B}}
    \label{eqn:axion-eom}
    \\
    0
    &= \ddot{\Ap}_\pm + H \dot{\Ap}_\pm
        + \left(
            \frac{k^2}{a^2}
            + \mA^2
            \pm \beta \dot{\bar{\theta}} \frac{k}{a}
        \right)
        \Ap_\pm
    \label{eqn:dark-photon-axion-eom}
    ,
\end{align}
\end{subequations}
neglecting nonlinearities in the axion potential.

Provided $\beta$ is large enough that $\beta \dot{\bar{\theta}} \geq 2 \mA$, the effective
frequency of one of the two transverse polarization states is negative, leading to the exponential
growth of $A_\pm$ due to a tachyonic resonance~\cite{Agrawal:2018vin, Co:2018lka}.
In the simplest scenarios, the axion's damped and (nearly) harmonic oscillations begin around the
time when $H = \ma$.
The maximum velocity of an axion oscillating in a quadratic potential is set by its mass, i.e.
$\vert \dot{\bar{\theta}} \vert \leq \ma \theta_0$; tachyonic resonance then occurs if
$\beta \theta_0 > 2 \mA / \ma$.
A more precise determination of the conditions for efficient resonance requires solving the
equations of motion (see \cref{app:Delayed-Axion-Oscillations}); the so-called ``broad resonance''
regime requires
\begin{align}
    \label{eqn:axion-DP-Production-Threshold-Broad}
    \beta \theta_0
    &\gg \max \left\{ 1, 4 \mA / \ma \right\}.
\end{align}
In the broad resonance regime, the dark photon mode that grows the most has wave number
$\mathcal{O}(\beta \ma)$.\footnote{
    The dynamics of broad parametric resonance via an oscillating, parity-even
    scalar~\cite{Adshead:2023qiw} (or the dark Higgs itself~\cite{Dror:2018pdh})
    are qualitatively similar.
    Production via scalars does allow for a number of qualitatively distinct regimes,
    which we discuss in detail in \cref{sec:scalar}.
}

Assuming $\theta_0 \sim 1$, efficient dark photon production requires $\beta$ larger than
unity to some degree.
If $\beta$ is just large enough for production to be efficient
[\cref{eqn:axion-DP-Production-Threshold-Broad}], then by kinematics the dark photon cannot be
heavier than the axion and production must occur at $H_{\star} \gtrsim \mA$.
This regime is well studied~\cite{Agrawal:2018vin, Co:2018lka, Co:2021rhi, Kitajima:2023pby}.
The resulting limit on the kinetic mixing that avoids backreaction onto the Higgs is displayed in
\cref{fig:defect-limits-existing-models} [assuming the fiducial relationship between $\gD$ and
$\varepsilon$, \cref{eqn:epsilon-Benchmark}] and coincides with the constraint taken in
Ref.~\cite{Agrawal:2018vin} to ensure backreaction onto the Higgs's mass is negligible.
Production at $H_\star \gtrsim \mA$ does offer substantially improved detection prospects compared
to minimal inflationary production.
Notably, however, most of the experimental prospects are only for masses above an $\meV$---a range
already accessible to inflationary production (i.e., given an \textit{ad hoc} enhancement of the
kinetic mixing in the latter case; see \cref{sec:production-independent}).
Thus, the sub-$\meV$ mass range (as probed by axion haloscopes) remains out of reach without
additional machinery.

When the dark photon's energy density becomes comparable to that of the axion, it backreacts onto
the axion via nonlinear processes, altering the rate and direction of energy transfer.
For resonant production mechanisms from massive, oscillating (pseudo)scalars in general, the final
energy partition between the dark photon and the scalar ultimately must be quantified by numerical
simulations.
Those presented in Ref.~\cite{Agrawal:2018vin} show that whether the majority of the axion's energy
density ends up in the dark photon depends on the mass ratio and becomes less efficient as
$m_{a}/\mA$ becomes large, though the range of ratios that have been simulated is limited.
The nonlinear dynamics of backreaction do not generically yield a large hierarchy in abundance;
Ref.~\cite{Co:2018lka} discusses a number of model-dependent remedies to dilute any remaining axion
energy density.

In standard constructions where the axion-photon coupling is generated via Peccei-Quinn (PQ) fermions,
order-unity couplings $\beta$ are not the natural expectation.
The anomaly coefficient to the PQ charge of the $\mathrm{U}(1)_D$-charged fermions
is~\cite{Agrawal:2018mkd, Agrawal:2018vin}
\begin{align}\label{eqn:anomaly-Decomposition}
    C_{\theta\Ap\Ap}
    &\sim \sum_{i = 1}^{N_f} Q_i^2
    \sim N_f Q_f^2,
\end{align}
where $Q_i$ is the PQ charge of the $i$th fermion, and $N_f$ is the total number of fermions.
In terms of $C_{\theta\Ap\Ap}$ and the dark fine-structure constant $\alpha_D = \gD^2 / 4 \pi$,
the axial coupling is
\begin{align}\label{eqn:beta-Decomposition}
    \beta
    &= C_{\theta\Ap\Ap} \frac{\alpha_{D}}{2\pi}.
\end{align}
For modest charges and number of PQ fermions, $\beta \sim \gD^2 \ll 1$, in extreme tension with the
requirement for tachyonic resonance [\cref{eqn:axion-DP-Production-Threshold-Broad}].
Namely, for gauge couplings that evade backreaction onto the Higgs [\cref{eqn:defectBound}],
\begin{align}
    C_{\theta\Ap\Ap}
    &\gtrsim
        10^{31} \frac{\beta}{10}
        \sqrt{\frac{\Omega_A}{\Omega_\mathrm{DM}}}
        \left( \frac{\mA}{\mu\eV} \right)^{-5/4}
        \left( \frac{\mA}{m_a} \right)^{-1/4}.
    \label{eqn:axial-coupling-requirement}
\end{align}
References~\cite{Agrawal:2017cmd,Agrawal:2018mkd,Agrawal:2018vin} discuss several model-building
avenues to enhance the axial coupling relative to the nominal expectation, including taking $N_f$
and $Q_f$ large as well as the clockwork mechanism~\cite{Giudice:2016yja}.
We discuss clockwork and its consequences for backreaction onto the Higgs in more detail in
\cref{sec:production-independent}.
Abelian clockwork may also simultaneously account for large hierarchies between $\varepsilon$ and
$\gD$ and between $\beta$ and $\alpha_D$~\cite{Agrawal:2018vin}.

\subsection{Vector misalignment}\label{sec:vector-misalignment}

Much like scalar fields, the energy density of a homogeneous vector field dilutes as $a^{-3}$ when
$H\lesssim \mA$, and the resulting condensate behaves as cold dark matter~\cite{Nelson:2011sf}.
But in contrast to a scalar, a vector's energy density dilutes as $a^{-2}$ at early times
($H \gtrsim \mA$) when it is not oscillating, as can be inferred from the $k / a \ll m_A$ limit in
\cref{fig:free-evolution-backreaction-redshifting}.
Successful vector misalignment after inflation therefore requires a nonminimal coupling to gravity
of the form \cref{eqn:nonminimal-proca} with $\xi_1 = -1/6$ and $\xi_2 = 0$ to avoid this additional
$a^{-2}$ dilution~\cite{Arias:2012az}.
Setting aside the theoretical issues with nonminimal couplings discussed in
\cref{sec:nonminimal-couplings}, vector misalignment provides a useful benchmark for nonthermal
production of dark photons---in particular, one whose initial conditions lack magnetic fields.

In the absence of spatial gradients, the temporal component $\Ap_0$ vanishes and the dark photon is
purely electric.
Defining the rescaled field $\tilde{\Ap}_i = \bar{A}_i / a$, each component of $\tilde{\Ap}_i$
satisfies the equation for the evolution of a homogeneous scalar with additional terms proportional
to $1 + 6 \xi_1 \equiv 1 - \kappa$:
\begin{align}
    0
    &= \ddot{\tilde{A}}_i
        + 3 H \dot{\tilde{A}}_i
        + \left[ \mA^2 + (1 - \kappa) \left( \dot{H} + 2 H^2 \right) \right] \tilde{A}_i.
\end{align}
During inflation, the solutions to this equation all decay exponentially if $\kappa < 1$, while one
instead grows exponentially if $\kappa > 1$; the vector field $\tilde{A}_i$ is only constant if
$\kappa$ is tuned very close to one.
Fixing $\kappa = 1$, vector misalignment and scalar misalignment are mathematically identical in
vacuum.

A homogeneous dark photon field contains no magnetic flux, offering misalignment a possible loophole
to avoid the defect bound \cref{eqn:symmetry-restoration-condition}.
On the other hand, if $\mA^2\tilde{A}_i(0)^2\gtrsim\lambda \vev^4$, the dark photon condensate
causes the Higgs field to oscillate about the symmetric point $\Phi = 0$; the dark photon itself
then slowly rolls due to its oscillating (and decreasing) mass.
Perturbations in one of $\Phi$ or $\Phi^\ast$ experience a tachyonic instability that ultimately
fragments the Higgs, although whether misalignment would seed topological defects is a question that
can only be addressed by numerical simulations (see Appendix B
of~\cite{East:2022rsi}).

\section{Clockworked couplings and backreaction}\label{sec:production-independent}

The fiducial kinetic mixing \cref{eqn:epsilon-Benchmark} is comparable to the dark gauge coupling, as prescribed by the simple, canonical model that kinetic mixing is generated by $N_f$ heavy fermions charged under
both the SM and dark gauge groups~\cite{Holdom:1985ag, Rizzo:2018vlb}:
\begin{align}
    \varepsilon
    &= \frac{e \gD}{12 \pi^2}
        \sum_{i = 1}^{N_f} Q_{e,i} Q_{D,i} \ln\frac{m_i^2}{\mu^2},
\end{align}
where $m_i$ are the masses of the so-called ``portal matter'' fermions, each with charges $Q_{e,i}$
and $Q_{D,i}$ under electromagnetism and the dark $\mathrm{U}(1)$, respectively.
The additional constraint
\begin{align}
    \sum_{i = 1}^{N_f} Q_{e,i} Q_{D,i}
    &= 0
\end{align}
is often imposed so that $\varepsilon$ is independent of the renormalization scale $\mu$.
Assuming that the portal matter masses are comparable and their charges are order unity leads to the
parametric estimate of \cref{eqn:epsilon-Benchmark}.
Clearly, taking a large number of fermions and/or their charges large enhances $\varepsilon / \gD$.
A less contrived possibility, perhaps, is the clockwork mechanism~\cite{Giudice:2016yja,
East:2022rsi}.
In this section we consider the viability of clockwork as a means to evade backreaction onto the
Higgs and possible defect formation when invoked both to enhance the kinetic mixing itself
(\cref{sec:clockwork-mixing}) and to generate the hierarchically large axion couplings required for
successful tachyonic production (\cref{sec:clockwork-axion}), as discussed in \cref{sec:axion}.

The Abelian clockwork mechanism employs a chain of $N + 1$ $\mathrm{U}(1)$ gauge symmetries that are
broken to a single $\mathrm{U}(1)$ when $N$ charged scalars (Higgses) acquire VEVs.
Each gauge field $A_i$ is labeled by an integer $i \in \{1, \ldots, N+1\}$, and each Higgs
$\Phi_{i,i+1}$ is charged under two neighboring $\mathrm{U}(1)$s on the chain with charges $1$ and
$-Q$.
We denote the additional dark Higgs that couples only to the dark photon $\Ap$ (and generates its
mass) by $\Phi$ without subscripts, and for clarity we never refer to the clockwork gauge fields
without an index subscript (leaving $\Ap$ to denote only ``the'' dark photon).
The clockwork Lagrangian is explicitly
\begin{align}\label{eqn:clockwork-lagrangian}
\begin{split}
    \mathcal{L}
    &= \frac{1}{2} \left\vert \left( \partial^\mu - \I g A_{N + 1}^\mu \right) \Phi \right\vert^2
        + \frac{\lambda}{4} \left( \vert \Phi \vert^2 - v^2 \right)^2
        + g J_{\mu} A_1^\mu
    \\ &\hphantom{ {}={} }
        - \frac{1}{4} \sum_{i = 1}^{N+1} F^{\mu\nu}_i F^i_{\mu\nu}
        + \sum_{i = 1}^{N} \left[
            \frac{1}{2} \left\vert D_{i,i + 1} \Phi_{i,i+1} \right\vert^2
            - \frac{\lambda_{i,i+1}}{4}
            \left( \left\vert \Phi_{i,i + 1} \right\vert^2 - v_{i,i+1}^2 \right)^2
        \right],
\end{split}
\end{align}
where the gauge-covariant derivative is
\begin{align}
    D_{i,i + 1}^{\mu}
    &= \partial^\mu + \I g A_i^{\mu} - \I Q g A_{i+1}^{\mu}
\end{align}
and $J_\mu$ is the current of particles charged under $A_1$ (e.g. the portal fermions).
We also take all gauge couplings $g$ and charges $Q$ independent of $i$ for simplicity.
We take the $N$ Higgses $\Phi_{i,i + 1}$ to have VEVs $v_{i,i+1}$ larger than the VEV $\vev$ of the
Higgs $\Phi$ that ultimately breaks the last $\mathrm{U}(1)$ and gives the dark photon its mass.

The effective potential for the gauge fields is a sum of independent quadratics
$g^2 \sum_i \vev_{i,i+1}^2 (A_i - Q A_{i+1})^2$ that is minimized when each term in the sum is
independently zero.
At low enough energies, integrating out the heavy Higgses and heavy gauge fields therefore sets
$\vert \Phi_{i,i+1} \vert = v_{i,i+1}$ and $A_i = Q A_{i + 1}$ for all $i < N + 1$.
The resulting low-energy effective theory is
\begin{align}\label{eqn:low-enegry-clockworked-lagrangian}
    \mathcal{L}
    &= \frac{1}{2} \left\vert \left( \partial_\mu - \I g_D \Ap_\mu \right) \Phi \right\vert^2
        + \frac{\lambda}{4} \left( \vert \Phi \vert^2 - v^2 \right)^2
        - \frac{1}{4} F^{\mu\nu} F_{\mu\nu}
        + \gD Q^{N} J_{\mu} \Ap^\mu,
\end{align}
where the canonically normalized dark photon field is identified as
\begin{align}
    \Ap^\mu
    &= \sqrt{ \sum_{i = 0}^{N} Q^{2i} }
        A_{N + 1}^\mu
    = \sqrt{ \frac{Q^{2 (N + 1)} - 1}{Q^{2} - 1} }
        A_{N + 1}^\mu
\end{align}
with a dark gauge coupling
\begin{align}
    \gD
    &= g \sqrt{ \frac{Q^{2} - 1}{Q^{2 (N + 1)} - 1} }.
\end{align}
Particles charged under the first clockwork site have an effective charge under the dark photon
$Q^N$ times larger than their charge under $A_1$ itself, exponentially enhancing the kinetic mixing
relative to the Higgs coupling $\gD$:
\begin{align}
    \varepsilon
    &\sim Q^N \frac{e \gD}{16 \pi^2}.
\end{align}
Implementing clockwork to explain large kinetic mixing with a small effective charge for the dark
Higgs $\Phi$ is the reverse of the traditional invocation---to account for a small kinetic mixing
with order-unity gauge couplings.

While \cref{eqn:low-enegry-clockworked-lagrangian} takes the form of the standard Abelian-Higgs
model \cref{eqn:abelian-higgs-lagrangian} (plus a current), it is only strictly correct while the final gauge symmetry remains unbroken.
After $\Phi$ acquires its VEV, the ground state configuration of the clockwork gauge fields is
modified from that above ($A_i = Q A_{i + 1}$ for all $i < N + 1$) because the effective potential
has an additional contribution $g^2 v^2 A_{N+1}^2 / 2$.
Whereas in \cref{eqn:low-enegry-clockworked-lagrangian} the massless mode couples to no clockwork
Higgses in the low-energy theory, the lightest mass eigenstate [after the final
$\mathrm{U}(1)$ is broken] couples to them all---and, with a large abundance, can in principle
backreact strongly on any of the clockwork Higgses.
In \cref{sec:clockwork-mixing} we assess whether the dark photon backreacts on any of the clockwork
Higgses at a lower threshold than it does onto the dark Higgs $\Phi$, i.e., whether clockwork
modifies the backreaction condition \cref{eqn:symmetry-restoration-condition}.
In \cref{sec:clockwork-axion} we then consider whether the heavier clockwork gauge fields themselves
may be produced (as a byproduct of the dynamics that produce the lightest mode, the dark photon) and
themselves trigger backreaction onto any of the clockwork scalar fields.

\subsection{Large kinetic mixing}\label{sec:clockwork-mixing}

In the absence of the additional dark Higgs $\Phi$, the massless mode of the clockwork fields
is precisely that which couples to none of the clockwork Higgses $\Phi_{i, i+1}$.
Breaking the last gauge symmetry (with $\Phi$) only generates effective couplings to the clockwork
Higgses by perturbing the lightest mode in the clockwork spectrum away from the would-be massless
mode identified in \cref{eqn:low-enegry-clockworked-lagrangian}.
The size of the dark photon's couplings to the clockwork Higgses must therefore be controlled by
that of the dark Higgs, $\gD$~\cite{Craig:2018yld}.
Since $v_{i,i + 1} \gtrsim \vev$, the threshold dark photon density at which any of the other
clockwork gauge symmetries are restored is manifestly higher than that of the dark photon and dark
Higgs, i.e., the dark Higgs still provides the strongest condition to evade backreaction and defect
formation.

To verify this expectation, we explicitly compute the dark photon's couplings to the clockwork
Higgses by expanding in small $\vev / v_{i, i+1}$.
The gauge fields have an unperturbed mass matrix
\begin{align}\label{eqn:GeneralMassMatrix}
    M^2 &= g^2
        \begin{pmatrix}
            v_{1, 2}^2 & - Q v_{1, 2}^2
            \\
            - Q v_{1, 2}^2 & Q^2 v_{1, 2}^2 + v_{2, 3}^2 & - Q v_{2, 3}^2
            \\
            & \ddots & \ddots & \ddots
            \\
            & & - Q v_{N-1, N}^2 & Q^2 v_{N-1, N}^2 + v_{N, N+1}^2 & - Q v_{N, N+1}^2
            \\
            & & & - Q v_{N, N+1}^2 & Q^2 v_{N, N+1}^2
        \end{pmatrix}.
\end{align}
Independent of the VEVs of the $\Phi_{i,i+1}$, by inspection this matrix always has a null
eigenvector
\begin{align}
    \three{e}_0
    &=
        \sqrt{ \frac{1 - Q^{-2}}{1 - Q^{-2(N + 1)}} }
        \left[ 1, Q^{-1}, \ldots, Q^{-N} \right],
\end{align}
which, for $\vev = 0$, is the (massless) dark photon field $\Ap$.
However, specifying the other eigenvectors requires specifying the VEVs of the clockwork Higgses.
To simplify the calculation, we assume all the VEVs are the same, $v_{i,i+1} = \mathrm{v}$, in which
case the remaining eigenvectors have the form~\cite{Giudice:2016yja}
\begin{align}
    [\three{e}_{k}]_j
    &= \sqrt{\frac{2 g^2\mathrm{v}^2}{(N + 1)m_k^2}}
        \left( Q \sin\frac{\pi jk}{N+1} - \sin{\frac{\pi(j + 1)k}{N+1}} \right),
\end{align}
where the mass-squared eigenvalues are
\begin{align}
    m_k^2
    &= g^2\mathrm{v}^2 \left[ Q^2 + 1 - 2 Q \cos\frac{\pi k}{N + 1} \right].
    \label{eqn:clockwork-heavy-mass-eigenstates}
\end{align}
Note that all the masses of the broken $\mathrm{U}(1)$s are parametrically $g^2Q^2\mathrm{v}^2$.

When $v \ll \mathrm{v}$, breaking the final gauge symmetry modifies the mass matrix
\cref{eqn:GeneralMassMatrix} by a small perturbation
\begin{align}
    \delta M^2
    &= g^2 \diag\left( 0, \ldots, 0, v^2 \right).
\end{align}
The perturbation to the eigenvectors can be computed at first order in $v^2/\mathrm{v}^2$ as
\begin{align}
    \delta {\three{e}}_k
    &= \sum_{j\neq k}\frac{\three{e}_k^T\delta M^2\three{e}_j}{m_k^2 - m_j^2}\three{e}_j.
\end{align}
The lightest of these states, which we identify as the dark photon, is then
\begin{align}
    \left[ \delta \three{e}_0 \right]_i
    &= -
        \left( \frac{\vev}{\mathrm{v}} \right)^2
        \frac{\gD}{g}
        \frac{2 Q}{N + 1}
        \sum_{j = 1}^{N}
        \sin\frac{\pi N j}{N+1}
        \frac{
            Q \sin \frac{\pi i j}{N+1} - \sin{\frac{\pi(i + 1) j}{N+1}}
        }{
            \left[ Q^2 + 1 - 2 Q \cos\frac{\pi j}{N + 1} \right]^2
        }
    .
\end{align}
We may straightforwardly infer the dark photon's couplings to all the clockwork Higgses via their
kinetic terms, projecting each clockwork photon onto the dark photon (i.e., the mass eigenstate
$\three{e}_0 + \delta \three{e}_0$):
\begin{align}
    \mathcal{L}
    &\supset \frac{1}{2}
        g^2
        A_\mu A^\mu
        \sum_{i = 1}^N
        \left(
            \left[ \delta \three{e}_0 \right]_i
            - Q
            \left[ \delta \three{e}_0 \right]_{i+1}
        \right)^2
        \vert \Phi_{i, i+1} \vert^2,
\end{align}
which we empirically observe to evaluate to
\begin{align}
    \mathcal{L}
    &\supset \frac{1}{2}
        \gD^2
        \left( \frac{v}{\mathrm{v}} \right)^4
        A_\mu A^\mu
        \sum_{i = 1}^{N - 1}
        \left(
            Q^{N - i} \frac{Q^{2 (i + 1)} - 1}{Q^{2 (N + 1)} - 1}
        \right)^2
        \vert \Phi_{i, i+1} \vert^2.
\end{align}
In the limit of large $Q^2$,
\begin{align}\label{eqn:simplified-dark-photon-gear-coupling}
    \mathcal{L}
    &\supset \frac{1}{2}
        \gD^2
        \left( \frac{v}{\mathrm{v}} \right)^4
        A_\mu A^\mu
        \sum_{i = 1}^{N - 1}
        \frac{\vert \Phi_{i, i+1} \vert^2}{Q^{2 (N - i)}}
    = \frac{1}{2}
        \mA^2
        A_\mu A^\mu
        \left( \frac{v}{\mathrm{v}} \right)^2
        \sum_{i = 1}^{N - 1}
        \frac{\vert \Phi_{i, i+1} / \mathrm{v} \vert^2}{Q^{2 (N - i)}}.
\end{align}
The dark photon $\Ap$ backreacts significantly onto to $\Phi_{i,i+1}$ when the corresponding terms
in \cref{eqn:simplified-dark-photon-gear-coupling} exceed $\lambda \mathrm{v}^4$.
Not only is the dark photon's contribution to the clockwork Higgses' effective potentials in
\cref{eqn:simplified-dark-photon-gear-coupling} suppressed relative to that for the dark Higgs
$\Phi$ by $(v / \mathrm{v})^2$ and at least two powers of $Q$, their backreaction thresholds are
parametrically higher by a factor of $(\mathrm{v} / v)^4$.
If only the $\Ap$ field is produced and no other eigenstate, clockwork therefore remains a viable
means to boost kinetic mixing without requiring a large hierarchy $\mathrm{v} / \vev \sim g / \gD$,
contrary to the claim made in Appendix C.1 of Ref.~\cite{East:2022rsi}.

\subsection{Large axion coupling}\label{sec:clockwork-axion}

While the preceding section establishes that the dark photon itself does not backreact strongly onto
any of the clockwork Higgses before it does so to the dark Higgs, any production channel via an
interaction-basis gauge field could, in principle, produce mass eigenstates other than the dark
photon.
(Production could simply occur via a coupling directly to the linear combination of clockwork fields
that make up the lightest mode, but this possibility seems rather contrived.)
So long as the clockwork gauge fields' masses lie above a kinematic threshold for production, these
heavier states are not directly produced.
In principle, however, the interactions directly responsible for production may also generate
interactions between the dark photon and other mass eigenstates.

We now show that such effects are negligible for production of dark photons from an oscillating
axion (\cref{sec:axion}) as a concrete example.
Efficient dark photon production from an oscillating axion typically requires couplings $\beta$ that
are parametrically large compared to that expected to arise by integrating out heavy, PQ-charged
fermions charged under the $\mathrm{U}(1)$ gauge group of $\Ap$, for which
$\beta \propto \gD^2 \ll 1$.
If these fermions couple to the more strongly coupled end of the clockwork chain [as in
\cref{eqn:low-enegry-clockworked-lagrangian}], then $\beta$ is enhanced to
$\beta \propto Q^{2N} \gD^2 \sim g^2$, after which only a moderate enhancement to $\beta$ is necessary
to reach $\beta \gtrsim 40$ (see the discussion in \cref{sec:axion-enhance} and
Ref.~\cite{Agrawal:2018vin}).

Following the notation of the previous section, the Chern-Simons term for the clockwork vector
$A_1^\mu$ expands to
\begin{align}
    F_1^{\mu \nu} \tilde{F}^1_{\mu \nu}
    &= \sum_{i, j = 0}^{N}
        [\three{e}_{i}]_1 [\three{e}_{j}]_1
        \mathsf{F}_i^{\mu \nu} \tilde{\mathsf{F}}^j_{\mu \nu},
\end{align}
where we use $\mathsf{A}_i^\mu$ to denote the $i$th mass-basis vector and $\mathsf{F}_i^{\mu \nu}$
its field strength.
The equations of motion for the mass-basis fields are then
\begin{align}
    \ddot{\mathsf{A}}_{i, \pm}
        + H \dot{\mathsf{A}}_{i, \pm}
        + \left[
            \frac{k^2}{a^{2}}
            + m_i^2
            \pm \beta \dot{\theta}
            \frac{k}{a}
            \left( [\three{e}_{i}]_1 \right)^2
        \right]
        \mathsf{A}_{i, \pm}
    &= \mp \beta \dot{\theta}
        \frac{k}{a}
        \sum_{j = 0,\, j \neq i}^{N}
        [\three{e}_{i}]_1 [\three{e}_{j}]_1
        \mathsf{A}_{j, \pm}.
    \label{eqn:clockwork-axion-coupling-eom}
\end{align}
Following \cref{sec:axion}, if the massive modes are sufficiently heavy
($m_i \sim g \mathrm{v} > \beta \dot{\theta} \sim \beta \ma \theta_0$), only the lightest mode
(the dark photon) is produced.
The mixing term on the right-hand side of \cref{eqn:clockwork-axion-coupling-eom} is therefore
dominated by the $j = 0$ term in the sum.
This source generates heavy modes with amplitude of order
\begin{align}
    \mathsf{A}_{i, \pm}
    &\sim
        \frac{\beta \dot{\theta} k / a}{k^2 / a^2 + m_i^2}
        [\three{e}_{i}]_1 [\three{e}_{0}]_1
        \mathsf{A}_{0, \pm}
    \propto \left( \ma / m_i \right)^2
        [\three{e}_{i}]_1 [\three{e}_{0}]_1
        \mathsf{A}_{0, \pm},
    \label{eqn:clockwork-heavy-driving}
\end{align}
noting that $\dot{\theta} \sim \ma \theta_0$ and $k / a \sim \ma$ in the broad resonance regime.
The eigenvector elements are each no larger in magnitude than $\sim 1 / Q$, so
the backreaction thresholds for the clockwork Higgses are order
$\lambda \mathrm{v}^4 \gtrsim m_i^2 A_i^\mu A^i_\mu
\sim \ma^4 \mathsf{A}_{0, \pm}^2 / m_i^2$.
Taking the dark photon's own kinematic threshold from
\cref{eqn:axion-DP-Production-Threshold-Broad} for order-unity axion misalignments,
$\ma \gtrsim \mA / \beta$, shows that avoiding backreaction onto the clockwork Higgses requires
$\lambda \mathrm{v}^4 \gtrsim (\mA / m_i)^2 / \beta^4 \cdot \mA^2 \mathsf{A}_{0, \pm}^2$.
Since $\beta > 1$ and $\mA < m_i$ (and $v < \mathrm{v}$) by construction, avoiding backreaction onto the dark Higgs for
the dark photon itself is a parametrically stronger condition than for the other clockwork gauge
symmetries; the backreaction bound \cref{eqn:symmetry-restoration-condition} is therefore not
modified by clockwork.

\section{Detectable dark photons from axions}\label{sec:axion-enhance}

As discussed in \cref{sec:axion}, the dimensionless axion-photon coupling is typically
$\beta \sim \alpha_D / 2 \pi \ll 1$, and without additional model building, tachyonic resonance from
axion oscillations does not occur.
Suitable model building can generate a hierarchy in $\beta / \alpha_D$, as achieved by Abelian
clockwork mechanism described in \cref{sec:production-independent}.
The enhancement from clockwork alone, however, is limited to $\beta \lesssim 1$ by perturbative
unitarity (i.e., of scatterings with the fermions with largest charge); an additional mechanism must
further enhance $\beta$ to the values $\gg 10$ required for efficient
resonance~\cite{Agrawal:2017cmd, Agrawal:2018mkd, Agrawal:2018vin}.
Even taking a viable origin of arbitrarily large $\beta$ for granted, backreaction constraints still
limit production in much of the experimentally accessible parameter space
(\cref{fig:defect-limits-existing-models}), since production typically occurs no later than when
$H \approx \mA$.
The same mechanisms that boost $\beta$ can also enhance $\varepsilon$ relative to $\gD$, and in
principle axions can produce dark photons across all experimentally accessible parameter
space~\cite[see also \cref{sec:clockwork-axion}]{Agrawal:2018vin}.
However, a $10^{30}$ enhancement to the axion-photon coupling is rather implausible \textit{a
priori} [see \cref{eqn:axial-coupling-requirement}].

\begin{figure}
    \centering
    \includegraphics[width=4.318in]{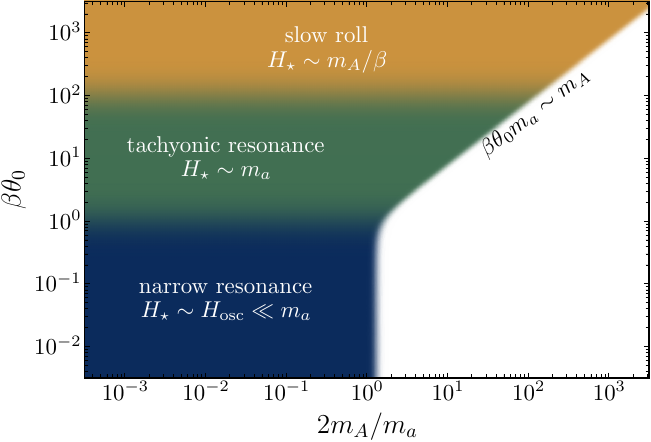}
    \caption{
        Illustration of the parameter space in which an axion may transfer its energy efficiently to
        a dark photon.
        The green region represents the usual range of moderate axion couplings considered, for
        example, in Refs.~\cite{Agrawal:2018vin, Co:2018lka, Dror:2018pdh}.
        The blue region corresponds to the narrow resonance discussed in
        \cref{sec:axion-enhance-narrow-resonance} and Ref.~\cite{Kitajima:2023pby}, in which delayed
        axion oscillations lead to a more efficient resonance and thus allow for $\beta \ll 1$.
        In the yellow region, the axion--dark-photon coupling is so large that the axion's
        oscillations are overdamped and energy transfer takes a parametrically long time.
        A rigorous understanding of this regime, however, requires numerical simulations (see
        \cref{sec:slow-roll}).
        Such large couplings could possibly be realized if the axion is coupled to the dark photon
        in the magnetic basis~\cite{Heidenreich:2023pbi, Ringwald:2023yni}, but this possibility
        also remains speculative and requires further model-building study.
    }
    \label{fig:axion-production-schematic}
\end{figure}

In this section, we explore scenarios that simultaneously facilitate efficient dark photon
production and weaken backreaction and defect formation limits.
We demonstrate that models that delay production (thereby relaxing these bounds, as described in
\cref{sec:defect-bounds-on-existing-scenarios}) can also enable efficient tachyonic resonance at
smaller values of $\beta$.
In this regime, both objectives work in tandem: delayed production allows for larger $\gD$, raising
the expected axion coupling $\alpha_D / 2 \pi \propto \gD^2$, and also reduces the degree to which
$\beta$ must be enhanced relative to this expectation.
We also consider how the opposite regime---strong coupling---can itself delay production and may
require less model-building machinery than the moderately large values of $\beta$ considered in
prior literature.
\Cref{fig:axion-production-schematic} illustrates these various dynamical regimes in parameter space.

\subsection{Narrow resonance}
\label{sec:axion-enhance-narrow-resonance}

In the standard scenario described in \cref{sec:axion}, only dark photon modes with
$\mA < k / a < \beta \vert \dot{\bar{\theta}} \vert \propto a^{-3/2}$ undergo tachyonic growth.
As the Universe expands, modes continuously exit the resonance band since their wave numbers
redshift more slowly than does the width of the resonance band itself.
Because the expansion rate $H$ is only just below the axion's frequency $\ma$, resonance is
efficient only during the axion's first few oscillations before unstable modes redshift out of the
band and are no longer tachyonic.
However, if axion oscillations begin when $H$ is parametrically lower than $\ma$, modes grow over a
proportionally larger number of oscillations, making resonance efficient at smaller
$\beta \theta_0$.

As derived in \cref{app:Delayed-Axion-Oscillations-Condition} and Ref.~\cite{Kitajima:2023pby},
when axion oscillations begin at an arbitrary Hubble rate $H_\mathrm{osc} < \ma$, the resonance
condition relaxes to
\begin{align}\label{eqn:axion-DP-Production-Threshold-Efficient-Delay}
    \beta \theta_0 \sqrt{\frac{\ma}{2 H_\mathrm{osc}}}
    \gg 1.
\end{align}
Delaying axion oscillations until $H \ll \ma$ thus not only reduces the required enhancement of the
axion coupling $\beta$, but also weakens constraints on the dark gauge coupling from backreaction
onto the Higgs [per \cref{eqn:defectBound}].
Because resonance is narrow in wave number about $a_\mathrm{osc} \ma / 2$, resonantly produced dark
photons become nonrelativistic when
$H_\mathrm{NR} / H_\star = (a_\star / a_\mathrm{NR})^2 = (2 \mA / m_a)^2$.
\Cref{fig:axion-parameter-space} shows that a sufficiently long delay allows for viable kinetic
mixing in reach of all future searches.

To quantify the joint impact on the required hierarchy in $\beta$ and $\alpha_D$, combine the
efficient resonance condition \cref{eqn:axion-DP-Production-Threshold-Efficient-Delay} with the
backreaction bound \cref{eqn:defectBound}:
\begin{align}
    \frac{\beta}{\alpha_D / 2 \pi}
    &\gg 10^{10}
        \theta_0^{-1}
        \lambda^{-1/2}
        \sqrt{\frac{\Omega_A}{\Omega_\mathrm{DM}}}
        \frac{2 \mA}{\ma}
        \left( \frac{\mA}{\ueV} \right)^{-5/2}
        \left( \frac{H_\mathrm{osc}}{10^{-22}~\eV} \right)^{5/4}.
    \label{eqn:delayedCouplingRequirement}
\end{align}
\Cref{eqn:delayedCouplingRequirement} represents a significant reduction in the required hierarchy
relative to the standard scenario, traded instead for a large hierarchy in
$\ma / H_\mathrm{osc}$.
The requirement is further reduced from the fiducial values in \cref{eqn:delayedCouplingRequirement}
by taking a heavier axion or dark photon, but at the expense of exacerbating the hierarchy in
$\ma / H_\mathrm{osc}$.
For instance, the fiducial hierarchy is order unity for dark photons with mass
$10~\meV$, for which $\ma / H_\mathrm{osc} > 10^{20}$.
Delayed oscillations require a dynamical hierarchy rather than a parameter hierarchy, but candidate
models may well require parameter hierarchies or tunings of their own.

There are numerous concrete models that delay the onset of axion oscillations~\cite{Arvanitaki:2019rax,
Huang:2020etx, Cyncynates:2021xzw, Cyncynates:2022wlq, Kitajima:2023pby}, although mechanisms relying on attractive self-interactions~\cite{Arvanitaki:2019rax, Huang:2020etx} result in axion
fragmentation, spoiling the dark photon's tachyonic resonance (see
\cref{app:axion-fragmentation-time}).
Repulsive self-interactions suppress axion perturbations, but constructing an axion potential with
repulsive self-interactions (while possible~\cite{Fan:2016rda}) is its own challenge because the
periodic axion field always has attractive self-interactions at large enough field excursions.
\Cref{eqn:delayedCouplingRequirement} tacitly assumes axion self-interactions do not disrupt dark
photon production.

Example models whose dynamics can plausibly delay axion oscillations while also avoiding axion
fragmentation may arise in axiverse scenarios~\cite{Arvanitaki:2009fg}.
Given a large number of axions logarithmically distributed in mass, there is a reasonable chance
that a pair of axions have masses within an order-unity factor of each
other~\cite{Cyncynates:2021xzw}; candidate pairs indeed appear in concrete string theory
compactifications~\cite{Gendler:2023kjt}.
Such ``friendly'' pairs can undergo nonlinear autoresonance, in which the axion with the smaller
decay constant $\fa$ oscillates with fixed amplitude until a parametrically late time, after which
it dilutes like matter~\cite{Cyncynates:2021xzw, Cyncynates:2022wlq}.
Importantly, autoresonance may occur whether axion self-interactions are attractive or
repulsive~\cite{Cyncynates:2021xzw}.
The time between $H = \ma$ and when the axion starts diluting is controlled by the ratio of decay
constants between the friendly axions, $\mathcal{F} \gg 1$, as
$H_\mathrm{osc} \sim \ma \mathcal{F}^{-4/3}$.
Quantified in decades, a hierarchy in $\ma / H_\mathrm{osc}$ translates to one $3/4$ as severe in
decay constants.

\subsection{Slow roll}
\label{sec:slow-roll}

Given that dark photon production from axion oscillations already requires a significant enhancement
of $\beta$ beyond its natural expectation, one may as well consider enhancements yet larger than the
bare minimum required for efficient resonance.
We now point out the possibility that such large $\beta$ (much greater than, say, $50$) may yield a
new dynamical regime in which dark photon production is delayed.
Moreover, because such delayed production allows for larger $\gD$ without backreaction or defect formation,
requiring larger $\beta$ itself does not necessarily exacerbate the hierarchy in $\beta / \alpha_D$,
a claim we quantify below.

The cosmological evolution of massive, noninteracting scalars is relatively simple: the
(homogeneous) field is frozen when Hubble friction dominates over the curvature of its potential ($H
\gg \ma$) and subsequently undergoes underdamped oscillations once $H$ drops below $\ma$.
The axion's interaction with dark photons, however, yields a source term
$\propto \beta \gen{\three{E} \cdot \three{B}}$ in its equation of motion \cref{eqn:axion-eom}.
The relative importance of dark photon backreaction may be parametrically estimated by comparing
energy densities, i.e., comparing $\beta \rho_A$ and $\rho_a$.
For the moderate values of $\beta$ considered in the standard scenario (\cref{sec:axion}), tachyonic
resonance of dark photons only becomes efficient in the oscillatory regime, and their backreaction
onto the homogeneous axion only becomes important as energy transfer completes.
However, for sufficiently large $\beta$, even the axion's parametrically small velocity at early
times,
\begin{align}
    \dot{\bar{\theta}}
    &= - \ma \theta_0
        \left[ \ma / 5 H + \mathcal{O}([\ma / H]^3) \right],
\label{eqn:overdamped-axion-velocity}
\end{align}
is enough to produce dark photons.
As derived in \cref{app:axionOscillations}, if
\begin{align}
    \theta_0 \beta
    &\gg 48.5 + 3.2 \ln\frac{\fa^4}{\ma^4 \theta_0^2\beta^6},
\end{align}
then not only does the slowly rolling axion efficiently produce dark photons, but the dark photons
that are produced induce significant backreaction---not just because larger $\beta$ increases the
magnitude of their backreaction at fixed energy density, but also because dark photons are produced
at earlier times $\ma t_\mathrm{BR} \sim (\beta \theta_0)^{-1/2}$.
Rather than proceeding to oscillate when $H \leq \ma$, the axion may instead undergo an extended phase
of slow roll.

Neglecting spatial axion perturbations (a strong assumption we comment on shortly), the structure of
\cref{eqn:axion-photon-eoms}---where the dark photon's backreaction grows with a rate proportional
to the axion velocity---suggests the possibility of a quasiequilibrium state where the rolling
axion produces a dark photon background that exerts friction on the axion, in turn regulating the
axion's velocity.
In particular, when $\dot{\bar{\theta}} = 2 \mA / \beta$, the axion rolls just fast enough
that one dark photon polarization is massless.
(The other is massive and may be ignored.)
If $\dot{\bar{\theta}}$ were to roll any faster, the massless mode would become tachyonic, the
energy in the rolling axion would be converted into dark photons, and $\theta$ would slow down.
Under these assumptions, the totality of the axion energy density would
gradually drain into the dark photon by the time $H_\star \approx \mA / \beta$, realizing both
efficient and delayed dark photon production at once.
Because $\dot{\bar{\theta}}$ is nearly constant, the axion rolls to its minimum almost entirely
during the last $e$-fold of production; the majority of the dark photon energy density is then
concentrated at wavelengths $k = \mA a_\star$.\footnote{
    At extremely large $\beta$ and $\ma/\mA$, the axion velocity at the onset of slow roll is
    parametrically larger than its quasiequilibrium value by an amount proportional to
    $\sqrt{\beta\theta_0}\ma/\mA$.
    It is therefore likely that much of the axion's initial energy converts to highly relativistic
    dark photons before it slows to its quasiequilibrium velocity; a large axion decay constant
    may thus be necessary to achieve the correct dark photon relic abundance.
}

In \cref{app:axionOscillations}, we show that the axion enters the overdamped regime just as its
perturbations become comparable to its homogeneous component.
If the axion indeed fully fragments, the tachyonic resonance is disrupted, meaning the slow-roll
solution is not realized.
On the other hand, this effective slow-roll regime is indeed realized in numerical simulations of
preheating into massless gauge bosons after axion inflation~\cite{Adshead:2015pva, Adshead:2016iae,
Figueroa:2017qmv, Adshead:2018doq, Cuissa:2018oiw, Adshead:2019lbr, Adshead:2019igv,
Adshead:2023mvt}, indicating that axion perturbations may be regulated at large axion-photon
coupling---in which case the axion would remain approximately homogeneous and overdamped as it rolls
to the bottom of its potential.
Suggestive as these results are, it remains to be shown whether they extend to the
radiation-dominated epoch relevant for dark photon dark matter production, a possibility only
testable with nonlinear simulations.
Given the magnitude of $\beta$ required, the initial resonance is quite broad and nonlinear
interactions between modes should be strong (even if they do not disrupt the homogeneous mode);
robust simulations may require significant computational resources.
We leave this investigation to future work and reiterate the uncertainty that any conclusions are
subject to.

\begin{figure}
    \centering
    \includegraphics[width=\textwidth]{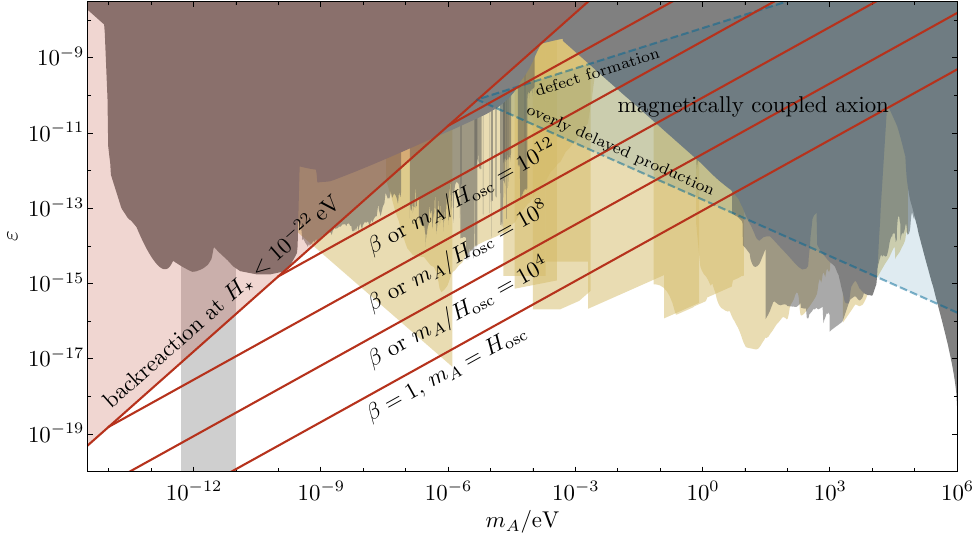}
    \caption{
        Kinetic mixing parameter space of dark photon dark matter produced from rolling axions.
        Red lines depict the maximum kinetic mixing [\cref{eqn:epsilon-Benchmark}] that evades
        backreaction onto the Higgs when either the axial coupling $\beta$ is large enough to realize
        slow roll (\cref{sec:slow-roll}) or when axion oscillations are delayed (relative
        to $\mA \lesssim m_a$) sufficiently long by other means
        (\cref{sec:axion-enhance-narrow-resonance}).
        The blue region depicts the parameter space available when the axion coupling is
        $\beta \propto 1 / \alpha_D$, as may be for an axion coupled in the magnetic basis.
        We note that the slow-roll scenario is subject to theoretical uncertainty regarding
        nonlinear dynamics and, for the magnetic axion, the consistency of large magnetic couplings
        (both discussed in \cref{sec:slow-roll}).
        The red region depicts parameter space for which backreaction onto the Higgs occurs when
        scales relevant to CMB anisotropies enter the horizon, i.e., when the dark photon must
        behave as dark matter; current bounds and projected search space are depicted in gray and
        yellow as labeled in \cref{fig:defect-limits-existing-models}.
    }
    \label{fig:axion-parameter-space}
\end{figure}

Slow roll, should it occur, requires larger couplings, but not necessarily in units of $\alpha_D$.
That is, the increase in the maximum $\gD$ that evades backreaction onto the Higgs and defect
formation can offset the required increase in $\beta$ itself.
Taking $H_\star = \mA / \beta$ and substituting \cref{eqn:defectBound},
\begin{align}
    \frac{\beta}{\alpha_D / 2 \pi}
    &\gtrsim 10^{34}
        \lambda^{-1/2}
        \sqrt{\frac{\Omega_A}{\Omega_\mathrm{DM}}}
        \left( \frac{\mA}{\ueV} \right)^{-1}
        \left( \frac{H_\star}{10^{-22}~\eV} \right)^{-1/4},
\end{align}
which is in the same ballpark as the standard case [\cref{eqn:axial-coupling-requirement}], though
with slightly different mass dependence.
However, axion--dark-photon couplings $\beta \gg 1$ may in fact require less model building than
moderate $\beta \sim \mathcal{O}(10) \gg \alpha_D$.
If the axion couples to PQ-charged fermions that have both dark electric and magnetic charges
(dyons), then $\beta$ may be proportional to (the square of) the dark magnetic charge $4\pi/\gD$
rather than the electric charge $\gD$~\cite{Sokolov:2021eaz}.
As a result, $\beta$ would naturally be of order $1/\alpha_D$, realizing the slow-roll regime.
Because $\beta$ is determined by the dark gauge coupling and itself sets the time of dark photon
production $H_\star \sim \mA/\beta$, the magnetically coupled axion is potentially viable only in a
narrow range of kinetic mixing bounded from above and below.
Substituting $H_\star = \alpha_D \mA / 2 \pi$ into \cref{eqn:defectBound} yields
\begin{align}
    \gD
    &\lesssim 2 \times 10^{-8} \lambda^{1/7}
        \left( \frac{\mA}{\ueV} \right)^{5/14}
        \left( \frac{\Omega_A}{\Omega_\mathrm{DM}} \right)^{-1/7}.
\end{align}
On the other hand, if the dark gauge coupling is too small, then (at a given $\mA$) $\beta$ would be
so large that dark photon production is delayed until after we know the dark matter must exist.
This places a \emph{lower} bound on the gauge coupling of
\begin{align}\label{eqn:gD-lower-bound-magnetic}
    \gD
    &\gtrsim 9 \times 10^{-8}
        \left( \frac{\mA}{\ueV} \right)^{-1/2}
        \left( \frac{H_{\star,\mathrm{min}}}{10^{-22}~\eV} \right)^{1/2}.
\end{align}
We plot the resulting parameter space in \cref{fig:axion-parameter-space}.
Intriguingly, production from a magnetically coupled axion may only both evade defect formation and
occur early enough for masses $\mA \gtrsim \mu\eV$.
Moreover, the available parameter space is already largely excluded, but numerous future axion
haloscopes, such as ALPHA~\cite{ALPHA:2022rxj}, MADMAX~\cite{Gelmini:2020kcu}, Dark
E-field~\cite{Levine:2024noa}, BREAD~\cite{BREAD:2021tpx}, LAMPOST~\cite{Baryakhtar:2018doz}, and
SuperCDMS~\cite{SuperCDMS:2019jxx,Bloch:2016sjj} could easily detect or exclude the parameter space
that remains below an $\eV$.
LZ~\cite{LZ:2021xov} could also potentially detect viable parameter space around $10~\mathrm{keV}$.

The viability of a magnetic-basis coupling as a means to boost the axion-photon coupling remains
unsettled at present~\cite{Heidenreich:2023pbi, Ringwald:2023yni}.
Although most of the objections raised by Ref.~\cite{Heidenreich:2023pbi} are specific to the case
of the QCD axion coupled to the Standard Model photon, questions remain even for the case of
axionlike particles coupled to a dark photon.
Because of the dual Witten effect~\cite{Witten:1979ey}, as the axion rolls from $0$ to $2\pi$, the
magnetic charge of every dyon changes by one unit, implying the existence of an infinite dyonic
tower.
Not only must the magnetically charged dyons be very heavy to avoid magnetic Schwinger pair
production~\cite{Affleck:1981bma,Ho:2021uem}, but Ref.~\cite{Heidenreich:2023pbi} also suggests that
integrating out such a tower generically drives the theory to weak coupling.
The coupling to heavy dyons would also generate a large mass for the axion, though because $H_\star$
is independent of $\ma$ once the axion has achieved its equilibrium, slow-roll velocity, production
may still proceed as described even if the axion were quite heavy.
The dark Higgs is one such electrically charged state that would form the basis of such a dyonic
tower~\cite{matt-private-comm}.
The possibility of axions strongly coupled to dark photons---perhaps with a St\"uckelberg rather
than a Higgs mass---remains an intriguing possibility, and further investigation is required to
establish its viability.
Finally, note that even if one ignores possible backreaction onto the radial mode associated with
the St\"uckelberg mass, the lower bound [\cref{eqn:gD-lower-bound-magnetic}] on $\varepsilon$
remains.

\section{Dark photon dark matter in scalar--Abelian-Higgs theories}\label{sec:scalar}

Unlike axions, whose couplings are restricted to those that perturbatively preserve their continuous
shift symmetry, scalars can couple to any operator and therefore offer more possible means to
produce dark photons and evade strong backreaction.
In particular, a scalar field effectively modulates the fundamental parameters of a theory
throughout spacetime~\cite{Baryakhtar:2024rky, Cyncynates:2024bxw}.
A homogeneous and dynamical scalar field coupled to the Abelian-Higgs model may therefore modulate
the backreaction threshold [\cref{eqn:symmetry-restoration-condition}] and the string production
threshold [\cref{eqn:minimal-gD-inflation-bound}] over time.
For example, the dark Higgs VEV could be larger when dark photons are first produced than it is
today, raising the backreaction and defect formation thresholds at early times when the dark photon
energy density is largest.

We begin this section by introducing a useful parametrization of a scalar's interactions with the
Abelian-Higgs Lagrangian in \cref{sec:SAH-EOM}.
In \cref{sec:universal-coupling}, we illustrate how ``universal'' scalar couplings can alleviate
defect bounds in existing scenarios, focusing on gravitational production during inflation; in
\cref{sec:damour-polyakov-dilaton}, we show that this scenario can be realized in well-motivated
physics models such as the string dilaton~\cite{Damour:1994zq}.
Scalar couplings also offer unique mechanisms for particle production: in
\cref{sec:transverse-kinetic-coupling,sec:narrow}, we explore two models that avoid backreaction and
defect formation through the same physics that drives dark photon
production~\cite{Cyncynates:2023zwj, Adshead:2023qiw}.

\subsection{Equations of motion}
\label{sec:SAH-EOM}

The classical Lagrangian describing an Abelian-Higgs theory coupled to a scalar $\phi$ may be
parametrized as\footnote{
    \Cref{eqn:scalar-abelian-higgs-action-unbroken} neglects a possible fourth coupling that
    modulates the quadratic term of the dark Higgs potential independent of the quartic.
    However, our choice of parametrization is sufficient to modulate all the Abelian-Higgs
    parameters independently and also includes all possible (two-)derivative interactions.
}
\begin{align}
    \mathcal{L}_\mathrm{SAH}
    &= \frac{1}{2} \partial_\mu \phi \partial^\mu \phi
        - V(\phi)
        - \frac{W(\phi)}{4} F_{\mu\nu} F^{\mu\nu}
        + \frac{X(\phi)}{2} D_\mu \Phi \left( D^\mu \Phi \right)^\ast
        - \frac{\lambda Y(\phi)}{4} \left( \abs{\Phi}^2 - v^2 \right)^2
    \label{eqn:scalar-abelian-higgs-action-unbroken}
    ,
\end{align}
generalizing \cref{eqn:abelian-higgs-lagrangian}.
As discussed in \cref{app:SAH}, written in terms of radial fluctuations $\higgs$ of the Higgs about
its VEV in unitary gauge ($\goldstone = 0$), \cref{eqn:scalar-abelian-higgs-action-unbroken} expands
to
\begin{align}
\begin{split}
    \mathcal{L}_\mathrm{SAH}
    &= \frac{1}{2} \partial_\mu \phi \partial^\mu \phi
        - V(\phi)
        - \frac{W(\phi)}{4} F_{\mu\nu} F^{\mu\nu}
        + \frac{X(\phi)}{2}
        \left( 1 + \frac{\higgs}{v} \right)^2
        \mA^2 A_\mu A^\mu
    \\ &\hphantom{{}={}}
        + \frac{X(\phi)}{2} \partial_\mu \higgs \partial^\mu \higgs
        - Y(\phi) \left(
            \frac{\lambda}{4} \higgs^4
             + \lambda v \higgs^3
             + \frac{1}{2} m_\higgs^2 \higgs^2
        \right).
    \label{eqn:scalar-abelian-higgs-action-expanded-unitary}
\end{split}
\end{align}
The coupling functions $W$, $X$, and $Y$ could in general depend on multiple scalar fields; for
simplicity we consider only a single scalar field, but the extension to multiple is straightforward.

To rewrite the theory with canonically normalized kinetic terms, we rescale the vector and Higgs
fields as $A_\mu = \mathcal{A}_\mu / \sqrt{W(\phi)}$ and $\higgs = \rhiggs / \sqrt{X(\phi)}$.
Then, suppressing the explicit $\phi$ dependence of $W$, $X$, and $Y$ for brevity,
\begin{align}
\begin{split}
    \mathcal{L}_\mathrm{SAH}
    &= \frac{1}{2} \partial_\mu \phi \partial^\mu \phi
        - V(\phi)
        - \frac{1}{4} \left(
            \mathcal{F}_{\mu \nu}
            + \mathcal{G}_{\mu \nu}
        \right)
        \left(
            \mathcal{F}^{\mu \nu}
            + \mathcal{G}^{\mu \nu}
        \right)
        + \frac{1}{2} \frac{X \mA^2}{W}
        \left( 1 + \frac{\rhiggs}{\sqrt{X} v} \right)^2
        \mathcal{A}_\mu \mathcal{A}^\mu
    \\ &\hphantom{{}={}}
        + \frac{1}{2} \partial_\mu \rhiggs \partial^\mu \rhiggs
        - \frac{1}{2} \rhiggs \partial^\mu \rhiggs \frac{\partial_\mu X}{X}
        + \frac{1}{8} \rhiggs^2 \frac{\partial_\mu X \partial^\mu X}{X^2}
        - \frac{\lambda Y}{X^2} \left(
            \frac{\rhiggs^4}{4}
             + v \sqrt{X} \rhiggs^3
        \right)
        - \frac{1}{2} \frac{Y}{X} m_\higgs^2 \rhiggs^2
    ,
    \label{eqn:scalar-abelian-higgs-action-expanded-rescaled}
\end{split}
\end{align}
where
\begin{subequations}\label{eqn:def-mathcal-F-and-G}
\begin{align}
    \mathcal{F}_{\mu \nu}
    &\equiv \partial_\mu \mathcal{A}_\nu
        - \partial_\nu \mathcal{A}_\mu
\intertext{and}
    \mathcal{G}_{\mu \nu}
    &\equiv \frac{1}{2 W}
        \left(
            \mathcal{A}_\mu \partial_\nu W
            - \mathcal{A}_\nu \partial_\mu W
        \right).
\end{align}
\end{subequations}
To treat scenarios where a homogeneous, classical background scalar field effectively modulates the
fundamental parameters of the Abelian-Higgs model through its cosmological evolution, we expand
$\phi(t, \three{x}) = \bar{\phi}(t) + \delta \phi(t, \three{x})$ where $\delta \phi$ is a small
perturbation.
The form of \cref{eqn:scalar-abelian-higgs-action-expanded-rescaled} therefore motivates absorbing
the $\bar{\phi}$ dependence of the theory into its fundamental parameters as
\begin{subequations}\label{eqn:phi-dependent-parameters}
\begin{align}
    \mA(\bar{\phi})
    &\equiv \mA \sqrt{X(\bar{\phi}) / W(\bar{\phi})}
    \label{eqn:mA-phi-def}
    ,
    \\
    v(\bar{\phi})
    &\equiv v \sqrt{X(\bar{\phi})}
    \label{eqn:higgs-vev-phi-def}
    ,
    \\
    \lambda(\bar{\phi})
    &\equiv \lambda Y(\bar{\phi}) / X(\bar{\phi})^2
    \label{eqn:lambda-phi-def}
    ,
\intertext{and, as a consequence,}
    \label{eqn:mhiggs-phi-def}
    m_\higgs(\bar{\phi})
    &= \sqrt{2 \lambda(\bar{\phi})} v(\bar{\phi})
    = m_\higgs \sqrt{Y(\bar{\phi}) / X(\bar{\phi})}
\intertext{and}
    g_D(\bar{\phi})
    &= \mA(\bar{\phi}) / v(\bar{\phi})
    = g_D / \sqrt{W(\bar{\phi})}
    \label{eqn:gD-phi-def}
    .
\end{align}
\end{subequations}
From this point on, we take the dark photon to have no homogeneous component and set the Higgs to
zero for simplicity (with results including the Higgs presented in \cref{app:SAH}).
In an FLRW spacetime [\cref{eqn:flrw-metric}], the linearized equations of motion in Fourier space
read
\begin{subequations}\label{eqn:SAH-linearized-eoms}
\begin{align}
    0
    &= \ddot{\bar{\phi}}
        + 3 H \dot{\bar{\phi}}
        + V'(\bar{\phi})
    \label{eqn:scalar-eom}
    \\
    0
    &= \ddot{\mathcal{A}}_\pm
        + H \dot{\mathcal{A}}_\pm
        +  \left[
            \frac{k^2}{a^{2}}
            + \mA(\bar{\phi})^2
            - \frac{H}{2}
            \frac{\partial_t W(\bar{\phi})}{W(\bar{\phi})}
            - \frac{\partial_t^2 \sqrt{W(\bar{\phi})}}{\sqrt{W(\bar{\phi})}}
        \right]
        \mathcal{A}_\pm
    \label{eqn:rescaled-transverse-eom}
    \\
    0
    &= \ddot{A}_\parallel
        + \frac{
            \left[
                3 H
                + \partial_t \ln X(\bar{\phi})
            \right]
            k^2
            + \left[
                H
                + \partial_t \ln W(\bar{\phi})
            \right]
            a^2 \mA(\bar{\phi})^2
        }{
            k^2 + a^2 \mA(\bar{\phi})^2
        }
        \dot{A}_\parallel
        + \left[
            \frac{k^2}{a^{2}}
            + \mA(\bar{\phi})^2
        \right]
        A_\parallel
    \label{eqn:longitudinal-eom-isolated}
    .
\end{align}
\end{subequations}
\Cref{eqn:SAH-linearized-eoms} has also decomposed the vector onto a basis of transverse
($\mathcal{A}_\pm$) and longitudinal ($A_\parallel$) polarization; see \cref{app:SAH} for full
details.
Note that the equation of motion for the longitudinal component \cref{eqn:longitudinal-eom-isolated}
is written in terms of the noncanonical field, as the rescaling applied for transverse modes does
not fully remove scalar couplings from the longitudinal mode's friction term.
\Cref{eqn:longitudinal-eom-isolated} also more transparently demonstrates that the relativistic and
nonrelativistic limits of the longitudinal dynamics coincide with those of scalars and of transverse
vectors, respectively, in accordance with the Goldstone equivalence theorem.

The structure of \cref{eqn:SAH-linearized-eoms} is quite general, allowing for a number of
qualitatively distinct dark photon production mechanisms.
Large field excursions (relative to whatever relevant energy scale), for instance, allow for large
variation in the Abelian-Higgs parameters [\cref{eqn:phi-dependent-parameters}],
in which case both the backreaction threshold and the kinematic conditions for efficient particle
production become time dependent.
We specialize to two examples that illustrate each of these possibilities.
In \cref{sec:universal-coupling} we show that a universally coupled dilaton broadens the parameter
space available to inflationary production by modulating the Higgs VEV, simultaneously weakening the
defect bound and making production more efficient.
In \cref{sec:transverse-kinetic-coupling} we consider the model presented in
Ref.~\cite{Cyncynates:2023zwj} of a scalar coupled to the dark photon kinetic term.
Such a coupling can both exponentially suppress the dark photon mass at early times and induce a
tachyonic instability that achieves extremely delayed production.
Finally, in \cref{sec:narrow} we discuss an alternative particle production mechanism that was
pointed out by Ref.~\cite{Adshead:2023qiw}: for a particular (and finely tuned) mass ratio between
the dark photon and an oscillating scalar, parametric resonance can efficiently produce dark photons
even with arbitrarily slow resonance, providing another realization of delayed production.
This scenario, in contrast to the aforementioned, only depends on the leading-order behavior of the
scalar's couplings about the minimum of its potential.
Throughout, we also discuss various unique signatures predicted by each model and their relation to
the dark photon's direct-detection parameter space.

\subsection{Universal coupling}\label{sec:universal-coupling}

In this section, we show that defect constraints on inflationary production are weakened if particle
masses (possibly including SM particles) are $\phi$ dependent.
A scalar's interactions with matter, be it the dark matter or SM matter, generate for it an
effective potential that naturally drives the field to values that minimize the free energy of the
system.
Absent a bare potential, the scalar $\phi$ rolls to values that either minimize particle masses (for
$\phi$-dependent masses) or decouple the theory (for $\phi$-dependent couplings).
For the dark Abelian-Higgs theories we consider, the latter possibility would make $\gD$, which
already must be small to avoid defect formation, decrease yet further as the Universe evolves
toward its ground state.
On the other hand, a larger dark Higgs VEV during inflation works in favor of weaker defect bounds:
defect formation is suppressed when all degrees of freedom are heavier, while cosmological evolution
makes them lighter at late times.

The particular choice of coupling functions that endows only the theory's mass scales with $\phi$
dependence is $W(\phi) = 1$ and
\begin{align}
    Y(\phi)
    &= X(\phi)^2
    \equiv Z(\phi)^2,
\end{align}
for which, using \cref{eqn:phi-dependent-parameters}, the Abelian-Higgs parameters all scale with
$Z^{d/2}$ where $d$ is the mass dimension of the particular parameter.
In \cref{sec:inflation-varying-mass} we outline the model-independent features that weaken defect
bounds from inflationary production, and in \cref{sec:damour-polyakov-dilaton} we discuss one
particularly well-motivated realization from string theory, in which all parameters in the theory
depend on the so-called dilaton~\cite{Damour:1994zq}.
While we focus on inflationary production because of its minimalism, the mechanism of
\cref{sec:damour-polyakov-dilaton} could suppress backreaction onto the Higgs for a number of other
production mechanisms discussed in this paper.

\subsubsection{Inflationary production and varying masses}\label{sec:inflation-varying-mass}

To study the impact of evolving mass scales on defect formation and the dark photon relic
abundance on a model-agnostic basis, we first work in terms of a time-dependent mass $\mA(a)$.
The energy density of the longitudinal mode [\cref{eqn:rho-perp-parallel}] scales as
\begin{align}
    \dd{\rho_\parallel}{\ln k}
    &\sim \frac{1}{2 a^2} \mA(a)^2 \dd{\vert A_\parallel \vert^2}{\ln k}
\end{align}
in both the relativistic and nonrelativistic limits.
When $\mA(a)$ evolves slowly compared to $H$, the longitudinal-mode equation of motion
[\cref{eqn:longitudinal-eom-isolated}] simplifies to
\begin{align}
    0
    &= \ddot{A}_\parallel
        + \frac{3 k^2 + a^2 \mA(a)^2}{k^2 + a^2 \mA(a)^2} H \dot{A}_\parallel
        + \left( \frac{k^2}{a^2} + \mA(a)^2 \right) A_\parallel
    .
\end{align}
In the relativistic limit $k \gg a \mA(a)$, $A_\parallel$'s dynamics are independent of $\mA(a)$,
but once $A_\parallel$ is nonrelativistic its dynamics depend nontrivially on the evolution of its
mass.
If the dark photon's mass evolves adiabatically [such that a Wentzel-Kramers-Brillouin (WKB)
solution is valid] after it becomes nonrelativistic, its amplitude then scales $\propto
1/\sqrt{\mA(a)}$.
This distinction between the relativistic and nonrelativistic regimes is crucial to determining the
dynamics of $\mA(a)$ that alleviate defect formation bounds.

\Cref{eqn:inflation-energy-spectrum} shows that the mode that dominates the abundance of dark
photons produced gravitationally during inflation is that which enters the horizon at the same time
it becomes nonrelativistic---that is, the mode with wave number
$k_\star = a_\star H(a_\star) = a_\star \mA(a_\star)$.
Assume that the peak remains at this mode, evaluated when $H(a)$ first drops below
$\mA(a)$.\footnote{
    In principle, the mass could subsequently decrease rapidly enough that this mode becomes
    temporarily relativistic and no longer dominates the energy spectrum.
    We neglect this possibility for simplicity, since it is avoided if $\mA$'s evolution
    either is sufficiently slow or begins sufficiently late.
}
At $a > a_\star$ its energy density scales as
\begin{align}
    \dd{\rho_\parallel(k_\star)}{\ln k}
    &\sim
        \frac{\mA(a)^2}{2}
        \left( \frac{k_\star H_I}{2 \pi a_\star {\mA}_I} \right)^2
        \left( \frac{a}{a_\star} \right)^{-3}
        \frac{\mA(a_\star)}{\mA(a)}
    .
\end{align}
These four factors respectively encode the longitudinal modes' (current) mass; its initial
conditions from inflation~\cite{Graham:2015rva}, during which its mass takes the value ${\mA}_I$;
its redshifting upon reentering the horizon (when it simultaneously becomes nonrelativistic); and
the adiabatic evolution induced by its evolving mass.
Assuming this mode enters the horizon during the radiation era and setting
$k_\star = a_\star \mA(a_\star)$,
\begin{align}
    \dd{\rho_\parallel(k_\star)}{\ln k}
    &\sim \frac{\mA(a)^2}{2}
        \left( \frac{\mA(a_\star)}{{\mA}_I} \right)^2
        \left( \frac{H_I}{2 \pi} \right)^2
        \left( \frac{H}{\mA(a_\star)} \right)^{3/2}
        \frac{\mA(a_\star)}{\mA(a)}
    .
    \label{eqn:rho-parallel-peak-varying-mass}
\end{align}
If $\mA$ only changes while the longitudinal mode is still relativistic, then the WKB factor in
\cref{eqn:rho-parallel-peak-varying-mass} reduces to unity and the abundance scales with the square
of its late- and early-time masses:
\begin{align}
    \dd{\rho_\parallel(k_\star)}{\ln k}
    &\sim \frac{\mA^2}{2}
        \left( \frac{H_I}{2 \pi} \right)^2
        \left( \frac{H}{\mA} \right)^{3/2}
        \left( \frac{\mA}{{\mA}_I} \right)^2
    \label{eqn:east-huang-scenario}
    .
\end{align}
Compensating for the $(\mA/{\mA}_I)^2$ suppression of the dark photon abundance (at a given present-day mass) requires increasing
$H_I \propto {\mA}_I$, which nullifies any relaxation of defect constraints
[\cref{eqn:minimal-gD-inflation-bound}], as noted by Ref.~\cite{East:2022rsi}.\footnote{
    As mentioned in \cref{sec:nonminimal-couplings}, Ref.~\cite{East:2022rsi} (in Appendix C.2)
    considers a nonminimal coupling of the dark Higgs, $\xi R \vert \Phi \vert^2$, as a means to
    drive the Higgs VEV to larger values during inflation, which then returns to its vacuum value
    shortly afterward.
    The evolution of the mass scales in the theory thus necessarily terminates long before the
    dominant dark photon mode becomes nonrelativistic.
}

Now consider the scenario where $\mA$ instead remains at its inflationary value ${\mA}_I$ until
\emph{after} the dominant dark photon mode becomes nonrelativistic.
In this case, the second factor in \cref{eqn:rho-parallel-peak-varying-mass} (which encodes the
superhorizon initial condition seeded during inflation) simplifies to unity.
As a result, $\mA$'s evolution only affects the dark photon mode amplitude through adiabatic
evolution and an earlier onset of subhorizon redshifting:
\begin{align}
    \dd{\rho_\parallel(k_\star)}{\ln k}
    &\sim \frac{\mA^2}{2}
        \left( \frac{H_I}{2 \pi} \right)^2
        \left( \frac{H}{\mA} \right)^{3/2}
        \left( \frac{\mA}{{\mA}_I} \right)^{1/2}
    .
\end{align}
Therefore, $H_I$ need only be $({\mA}_I / \mA)^{1/4}$ times larger to maintain the correct relic
abundance, while the defect bound on $H_I$ is weakened by ${\mA}_I / \mA$.
Matching $\rho_\parallel$ to the dark matter relic density requires
\begin{align}
    \label{eqn:Z_I-relic-abundance-constraint}
    \frac{{\mA}_I}{\mA}
    &= \frac{\mA}{0.1~\meV}
        \left( \frac{H_I}{5 \times 10^{13}~\GeV} \right)^{4},
\end{align}
while the defect bound limits
\begin{align}
    \label{eqn:dilaton-gD-inflation-bound}
    \gD
    \leq \frac{{\mA}_I}{H_I}
    &\lesssim
        2 \times 10^{-19}
        \left( \frac{\mA}{\eV} \right)^2
        \left( \frac{H_I}{5 \times 10^{13}~\GeV} \right)^3.
\end{align}
The requisite early-late mass ratio is
\begin{align}\label{eqn:universal-mass-hierarchy}
    \frac{{\mA}_I}{\mA}
    &= 10^{4}
        \left( \frac{\gD}{2 \times 10^{-19}} \right)^{1/2}
        \left( \frac{H_I}{5 \times 10^{13}~\GeV} \right)^{5/2}
    = 10^{4}
        \left( \frac{\gD}{2 \times 10^{-19}} \right)^{4/3}
        \left( \frac{\mA}{\eV} \right)^{-5/3}
    .
\end{align}
\Cref{eqn:dilaton-gD-inflation-bound} is parametrically weaker than in the minimal scenario
[\cref{eqn:minimal-gD-inflation-bound}] and is in fact weakest for the highest allowed inflationary
scale because the relic abundance is more sensitive to $H_I$ than to the drop in the dark photon's
mass.
Further, the bound scales with a higher power of $\mA$ than \cref{eqn:dilaton-gD-inflation-bound}
because the relic abundance no longer requires a unique inflationary scale as a function of $\mA$;
heavier (present-day) dark photons may therefore be accommodated at larger $H_I$ when simultaneously
increasing ${\mA}_I / \mA$.

\subsubsection{Damour-Polyakov dilaton}\label{sec:damour-polyakov-dilaton}

We now show that the model of the string dilaton proposed by Damour and
Polyakov~\cite{Damour:1994zq} can realize cosmological evolution of the dark photon mass that
weakens defect bounds, as described phenomenologically in the previous section.
We briefly review the mechanism of Ref.~\cite{Damour:1994zq}.
In general, the string dilaton $\phi$ is expected to couple to all Standard Model operators,
essentially promoting the constants of nature to field-dependent quantities
$\zeta \to \zeta(\varphi)$ (where $\varphi \equiv \phi/\sqrt{2}\Mpl$).
If the ``constant'' has sufficiently large linear dependence on the dilaton, then it mediates a
force between objects whose mass depends on the value of $\zeta$; such fifth forces are strongly
constrained by tests of equivalence principle violation (unless $\phi$ is sufficiently
heavy)~\cite{Schlamminger:2007ht, Wagner:2012ui,Touboul:2022yrw}.
If the dilaton only enters the action via the single function $\zeta(\phi)$ or by a collection of
functions all with the same extrema (and has a canonical kinetic term), then one may shift its value
such that it has no linear couplings, i.e.,
$\zeta(\varphi) = \zeta_0 + \kappa\varphi^2 / 2 + \mathcal{O}(\varphi^3)$.

The now-leading quadratic coupling still mediates a fifth force, but one proportional to $\varphi$'s
expectation value.
Reference~\cite{Damour:1994zq} identified a mechanism that efficiently drives $\phi$ to its minimum:
if the mass $m_X$ of a particle species $X$ is generated by the confinement of a strongly coupled
gauge group whose gauge coupling depends on $\phi$, then $m_X$ depends strongly on the dilaton:
\begin{align}
    m_X(\phi)
    &= m_X \exp \left[
            \kappa
            \ln \left( \frac{e^{1/2} \hat{\Lambda}_s}{m_X} \right)
            \frac{\varphi^2}{2}
            + \mathcal{O} \left( \varphi^3 \right)
        \right],
\end{align}
where $m_X$ is the particle mass when $\phi$ is at the minimum of its effective potential,
$\hat{\Lambda}_s$ is the string scale (chosen to be $3 \times 10^{17}~\GeV$, in line with
Ref.~\cite{Damour:1994zq}, though our results are only logarithmically sensitive to this choice),
and $\kappa$ is an $\mathcal{O}(1)$ coefficient of the Taylor expansion of the gauge coupling of the
confining sector with respect to $\phi$.
Because $\hat{\Lambda}_s\gg m_X$, the constant of proportionality
\begin{align}
    \beta_X
    \equiv \kappa \ln(e^{1/2}\hat{\Lambda}_s/ m_X)
    \approx \kappa \left( 40.75 - \ln \frac{m_X}{\GeV} \right)
\end{align}
can be quite large, making this attractor mechanism efficient (as required for the dilaton to evade
fifth force bounds).
The confining gauge group coupling changes by $\mathcal{O}(1)$ over a field range
$\Delta \phi \sim \Mpl/\sqrt{\kappa}$, ensuring perturbative control even as $m_X(\phi)$ [and,
for our purposes, $\mA(\phi)$] change by multiple orders of magnitude.
If indeed the dilaton modulates the dark photon mass in this way, then per
\cref{sec:inflation-varying-mass} these dynamics are sufficient to alleviate inflationary defect
formation bounds.

The dilaton's cosmological dynamics are governed by its interactions with the SM plasma, which
induce an effective potential that takes the form~\cite{Dolan:1973qd,Weinberg:1974hy,Kapusta:1989tk}
\begin{align}
    V(\phi)
    &\approx \sum_{\{X \vert  m_X(\phi) \lesssim T\}}\frac{g_X}{24} T^2 m_X(\phi)^2
        \cdot
        \begin{cases}
            1, & \text{boson},
            \\
            1/2, & \text{fermion}
        \end{cases}
\end{align}
at one-loop order, where $g_X$ is the number of degrees of freedom associated with species $X$ and
the sum runs over all particles whose masses are below the plasma temperature and depend on $\phi$
via the confining gauge group.
The homogeneous component of $\phi$ evolves according to
\begin{align}
    0 &= \ddot{\bar{\phi}}
        + \frac{3}{2 t} \dot{\bar{\phi}}
        + \sum_{m_X(\bar{\phi}) \lesssim T} \frac{g_X}{24} T^2 m_X^2 \beta_X
        e^{\beta_X\bar{\phi}^2 / 2 \Mpl^2}  \frac{\bar{\phi}}{\Mpl^2},
\end{align}
which may be integrated directly in the slow-roll approximation, $\ddot{\bar{\phi}}\to 0$.
The heaviest coupled species that has not annihilated away dominates the effective potential, meaning
the dilaton starts rolling from its initial condition around when the heaviest species begins to
annihilate.
The temperature at that time is
\begin{align}
\label{eqn:T_roll}
    T_\mathrm{roll}
    &\approx
        \sum_{m_X(\phi_0) \lesssim T_\mathrm{roll}}
        \sqrt{\frac{g_X}{g_\star}} m_X(\phi_0)
        \frac{\beta_X\phi_0}{\Mpl}.
\end{align}
After this point, particle masses drop in proportion to the temperature,
\begin{align}
    m_X(\phi(t))
    &\approx m_X(\phi_0) \frac{T(t)}{T_\mathrm{roll}}.
\end{align}
No particle falls out of thermal equilibrium as long as its mass continues to scale in proportion to
its temperature, effectively pausing the thermal evolution of that species (or of the Universe if
all masses are $\phi$ dependent) until the dilaton reaches the minimum of its effective
potential,\footnote{
    The energy density of the dilaton during its subsequent oscillations about its minimum dilute faster than
    matter because its mass drops with $T$, ensuring it never contributes significantly to the dark
    matter density.
} at which point all masses take on their present-day values.
These dynamics rely only on the slow-roll approximation and not the particular form of the
$\phi$ dependence of $m_X$---only that $m_X$ is a sufficiently steep function of $\phi$ and that the
scalar's bare potential is negligible.
The same phenomenon occurs in, for example, Brans-Dicke~\cite{Brans:1961sx}--inspired models (see
e.g., Ref.~\cite{Erickcek:2013dea}).

To have any impact on the available parameter space of dark photon dark matter, both $\mA$ and at
least one thermalized species must acquire a $\phi$-dependent mass in the manner described above
(though, in principle, all masses could depend on $\phi$).
Defect formation limits are thus alleviated by whatever amount the dark photon mass drops after
$H = {\mA}_I$, per \cref{sec:inflation-varying-mass}; however, the dilaton can, in principle, begin
rolling earlier, which would not contribute any relaxation of defect limits but would still require
a larger $H_I$ to set the correct relic abundance.
In order for the dark photon mass to start shrinking only after the dominant mode has become
nonrelativistic, the dilaton must start rolling at a Hubble rate
$H_\mathrm{roll} = \sqrt{\pi^2 g_\star / 90} T_\mathrm{roll}^2 / \Mpl$ below ${\mA}_I$.
Using \cref{eqn:T_roll}, the relic abundance constraint \cref{eqn:Z_I-relic-abundance-constraint},
and $m_{A, I} / \mA = \exp\left( \beta_X \phi_0^2 / 4 \Mpl^2 \right)$:
\begin{align}
\begin{split}
    \label{eqn:H_Roll/mA_I}
    \frac{H_\mathrm{roll}}{{\mA}_I}
    &\approx \frac{\beta_X}{30}
        \left( \frac{m_X}{125~\GeV} \right)^2
        \left( \frac{\mA}{0.1~\meV} \right)^{-2}
        \frac{g_X}{4} \left( \frac{g_\star}{106.75} \right)^{-1/2}
        \left( \frac{H_I}{5\times 10^{13}~\GeV} \right)^{-4}
    \\ &\hphantom{ {}={} }
        \times \ln\left[
            \left( \frac{\mA}{0.1~\meV} \right) \left( \frac{H_I}{5\times 10^{13}~\GeV} \right)^{4}
        \right].
\end{split}
\end{align}

\begin{figure}
    \centering
    \includegraphics[width=\textwidth]{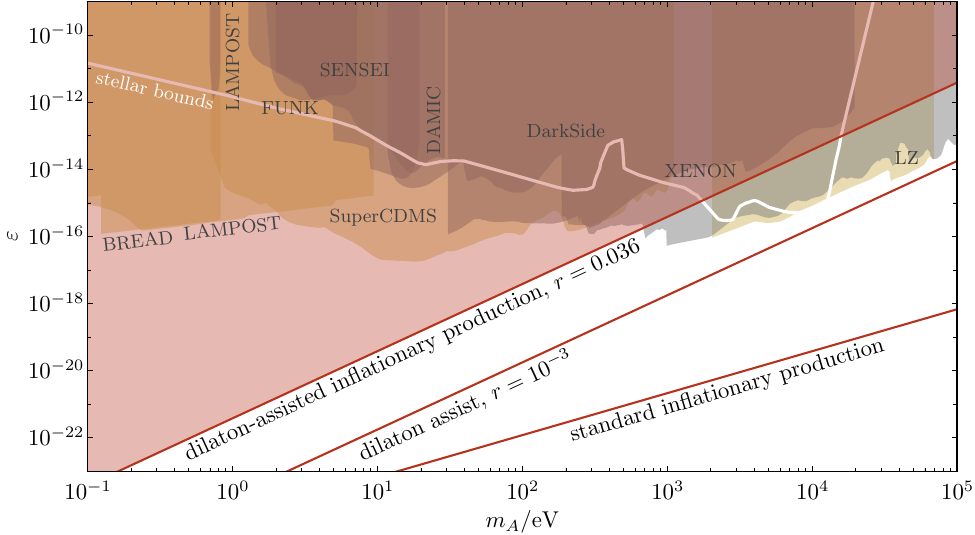}
    \caption{
        Kinetic mixing parameter space [\cref{eqn:epsilon-Benchmark}] of dark photon dark matter
        produced during inflation.
        The red region depicts parameter space that is constrained by defect formation for
        dilaton-assisted inflationary production for the maximal inflationary scale presently
        allowed (corresponding to a tensor-to-scalar ratio $r = 0.036$~\cite{BICEP:2021xfz}),
        and the upper limit for $r = 10^{-3}$ (a value that can be probed by future CMB
        observations~\cite{CMB-S4:2016ple, CMB-S4:2020lpa, SimonsObservatory:2018koc}) appears as
        labeled.
        The upper limit for standard inflationary production (without any mass evolution) is also
        plotted in red.
        Limits and projects are displayed as in \cref{fig:inflation-parameter-space}.
    }
    \label{fig:dilaton-inflation-parameter-space}
\end{figure}

The expanded parameter space available to ``dilaton-assisted'' inflationary production is depicted
in \cref{fig:dilaton-inflation-parameter-space} and could be probed by LZ~\cite{LZ:2021xov}.
This accessible parameter space would require inflationary scales so large that primordial
gravitational waves would be easily detected by future CMB observations~\cite{CMB-S4:2016ple,
CMB-S4:2020lpa, SimonsObservatory:2018koc} (but are not already excluded).
However, there are currently no direct-detection prospects at masses below a keV.
Finally, as described in Ref.~\cite{Damour:1994zq}, this model predicts violations of the weak
equivalence principle, which are stronger for $\kappa \lesssim 1$ and potentially observable in
successors to E\"ot-Wash~\cite{Schlamminger:2007ht, Wagner:2012ui} or
MICROSCOPE~\cite{Berge:2017ovy, Touboul:2022yrw}.

\subsection{Rapidly varying dark gauge coupling}\label{sec:transverse-kinetic-coupling}

Whether dark photons are produced gravitationally during inflation or via an oscillating axion
afterward, the dark sector is populated at the end of inflation (if not earlier).
What distinguishes the two scenarios is that gravitationally produced dark photons redshift even
during inflation, whereas in the latter case the dark sector density remains frozen in the
misaligned axion condensate until $H = \ma$.
In this way, the axion effectively acts as a reservoir, allowing for a lower maximum dark photon
density and reducing backreaction onto the dark Higgs once dark photons are produced.
Kinematic efficiency ultimately limits how long dark photon production from axions may be delayed
(\cref{sec:axion}), requiring larger couplings or field excursions ($\beta \theta_0 \gg 1$) and
typically also a boost of the axion--dark-photon coupling by \textit{ad hoc} means (but see
\cref{sec:slow-roll} for a possible exception).
In this section, we show that a scalar coupled to the kinetic term of the dark photon can realize
further delayed dark photon production by eliminating the kinematic barrier
(\cref{sec:resonant-kinetic-coupling-general}).
We present the dynamics in a concrete scenario (\cref{sec:rapid-kinetic-coupling-dynamics}), and we
find that such scenarios do not require parametrically large dimensionless numbers, in contrast to
axion production (\cref{sec:kinetic-coupling-quantum-corrections}); however, we do not identify a
concrete UV completion.
We then discuss unique signatures of delayed dark photon production in
\cref{sec:kinetic-coupling-signatures}.

\subsubsection{Resonant production with kinetic couplings}\label{sec:resonant-kinetic-coupling-general}

Whereas dilaton-assisted inflationary production (\cref{sec:universal-coupling}) relied on
couplings that modulate mass scales alone, the kinetic couplings $W$ and $X$ (to the dark photon and
Higgs, respectively) modulate the dark photon mass without changing the backreaction threshold
$\lambda v^4 \propto Y$ [\cref{eqn:phi-dependent-parameters}].
Thus, instead of raising the threshold for defect formation relative to the required dark photon
energy density, kinetic couplings could suppress the dark photon mass, eliminating the kinematic
barrier to resonance and delaying production to late times $H \sim m_\phi \ll \mA$.
Moreover, kinetic couplings are derivative couplings, and the requisite evolution of $X$ or $Y$ by
orders of magnitude (between their early-time and present-day values) can mechanize tachyonic
resonance that could itself source the dark photon relic abundance.

The scalar-dependent dark photon mass $m_A(\bar{\phi}) = \mA\sqrt{X(\bar{\phi})/W(\bar{\phi})}$ is
small compared to its present-day value if either $X$ is initially very small or $W$ very large.
(We normalize the values of $X$ and $Y$ today to unity without loss of generality.)
The massless limit of \cref{eqn:longitudinal-eom-isolated} is
\begin{align}
    0
    &= \ddot{A}_\parallel + \partial_t \ln X(\bar{\phi}) \dot{A}_\parallel + k^2 A_\parallel,
\end{align}
neglecting cosmic expansion for illustration's sake.
Exponential growth requires a negative friction term, but the longitudinal mode's friction term is
positive when $X$ increases with time.
The coupling $W$, on the other hand, does decrease in our hypothesized scenario.
Setting $X = Y = 1$, the canonical, transverse equation of motion \cref{eqn:rescaled-transverse-eom}
takes the simplified form
\begin{align}
    0 &= \ddot{\mathcal{A}}_\pm + H \dot{\mathcal{A}}_\pm + \omega_\pm(k)^2 \mathcal{A}_\pm,
    \label{eqn:transverse-dynamica-fine-structure-EOM}
\end{align}
where the dark photon's effective frequency squared is
\begin{align}
    \omega_\pm(k)^2
    &= \frac{k^2}{a^{2}}
        + \frac{\mA^2}{\bar{W}}
        - \frac{H}{2}
        \frac{\dot{\bar{W}}}{\bar{W}}
        - \frac{\partial_t^2 \sqrt{\bar{W}}}{\sqrt{\bar{W}}}.
    \label{eqn:transverse-omega}
\end{align}
Transverse modes thus have a negative squared effective mass if $\dot{\bar{W}}^2 < 2 \ddot{\bar{W}}
\bar{W}$; functions that are both large in magnitude and accelerating yield the fastest growth.
If the rate of change of $W$ is proportional to the scalar's effective mass $m_\phi$, i.e.,
$\dot{\bar{W}} / \bar{W} = \dot{\bar{\phi}} \partial_\phi \ln W \sim m_\phi \partial \ln W / \partial \ln \phi$,
the tachyonic growth condition $\omega_\pm^2 \leq 0$ requires the kinetic function to be larger than
the squared dark-photon--scalar mass ratio, $W(\phi_0) \gtrsim (\mA / m_\phi)^2$.
This, however, is an underestimate: efficient production requires that $\omega_\pm^2$ is negative
for a period long enough to build up all of the dark matter.
Therefore, $W$ must decrease for a parametrically long time and so by many orders of magnitude.
We next present a concrete realization of tachyonic production via a rapidly varying dark gauge
coupling, i.e., a suitable choice of coupling function $W$ and scalar potential $V$ as first
presented in Ref.~\cite{Cyncynates:2023zwj}.

\subsubsection{Dynamics}\label{sec:rapid-kinetic-coupling-dynamics}

Realizing functions $W$ that vary by orders of magnitude generally requires exponential dependence
on the scalar and that the scalar rolls over a fairly large distance (in appropriate units).
We therefore choose a potential that is compatible with large field excursions: the ``runaway,''
exponential potential
\begin{align}
    V(\phi)
    &= M^2 f^2 e^{- \phi / f}
    \equiv m_\phi^2 f^2 e^{- \left( \phi - \phi_0 \right) / f},
    \label{eqn:runaway-potential}
\end{align}
which is motivated by certain constructions in string theory~\cite{Apers:2022cyl, Revello:2023hro,
Apers:2024ffe, Seo:2024qzf} and whose self-similarity stabilizes it against quantum corrections from
self-interactions~\cite{Garny:2006wc}.
The latter equality in \cref{eqn:runaway-potential} merely defines $m_\phi$ to coincide with the
scalar's effective mass when at its homogeneous initial condition $\phi = \phi_0$ (whereas $M$ is
that when $\phi$ crosses zero).
The evolution of a homogeneous scalar in \cref{eqn:runaway-potential} is approximately
\begin{align}
    \label{eqn:runaway-approximate-sol}
    \bar{\phi}(t)
    &\approx \phi_0 + f \ln \left[ 1 + (m_\phi t)^2 \right],
\end{align}
up to oscillations that fall off with $(m_\phi t)^{-1/4}$.\footnote{
    These oscillations are logarithmic in time and change the resulting dark photon mode amplitude
    at the $\mathcal{O}(1)$ level.
    Since $\mathcal{O}(1)$ changes to the mode amplitude may be compensated by similar changes to
    parameter values, we neglect this subtlety for the sake of clarity.
}

Many possible coupling functions satisfy the tachyonic resonance condition
$\partial_t^2 \sqrt{W} / \sqrt{W} < 0$, but an exponential does so even as $\phi$'s time derivatives
become small [as for \cref{eqn:runaway-approximate-sol}] and for all field values of $\phi$.
However, were $W$ a pure exponential, nothing would prevent the dark photon's mass from continuing
to decrease to the present day.
To stabilize the vector's mass, one could augment the scalar potential with a minimum at some finite
$\phi$ or modify the coupling function $W$ to asymptote to a constant at late times.
We choose the latter option (for reasons explained at the end of this section), taking
\begin{align}
    W(\phi)
    &= 1 + e^{- \kincoup \phi / f},
    \label{eqn:kinetic-coupling}
\end{align}
normalized such that $W = 1$ as $\phi$ rolls toward $+\infty$ and with $\kincoup$ parametrizing the
coupling strength.
The initial condition $\phi_0$ must thus be negative so that $W(\phi_0) \gg 1$.

Production begins when $H \approx m_\phi$ and continues until the latter term in
\cref{eqn:kinetic-coupling} drops below unity, at which point the scalar and dark photon decouple,
the dark photon's mass reaches its present-day value, and production completes (when the Hubble rate
is $H_\star$ by definition).
Per the approximate solution \cref{eqn:runaway-approximate-sol}, the scalar crosses zero (i.e., $W$
approaches unity) at $t_\star \approx e^{- \phi_0 / 2 f} / m_\phi = 1 / M$.
While $\phi < 0$, we may take $W(\phi) \approx e^{- \kincoup \phi / f}$ and the time dependence of
the effective frequency is approximately
\begin{align}
    \label{eqn:omega-eff-approximate}
    \frac{\omega_\mathrm{eff}(k)^2}{m_\phi^2}
    &\approx \frac{k^2}{a^{2} m_\phi^2}
        + \frac{\mA^2}{W(\bar{\phi}) m_\phi^2}
        - \frac{\kincoup (2 \kincoup + 1) m_\phi^2 t^2 - 3 \kincoup}{2 (1 + m_\phi^2 t^2)^2}.
\end{align}
While $W$ remains much larger than $(\mA/m_\phi)^2$, the dark photon's bare mass is negligible;
modes with $k/a \lesssim \kincoup m_\phi (m_\phi t)^{-1}$ therefore have a negative
$\omega_\mathrm{eff}^2$ and grow by a tachyonic resonance.

For modes with negligible wave numbers and at early enough times that the bare mass term can be
neglected, the equation of motion \cref{eqn:transverse-dynamica-fine-structure-EOM} with effective
frequencies given by \cref{eqn:omega-eff-approximate} has the analytic solution (with
$\dot{\mathcal{A}}_\pm$ initially zero)
\begin{align}
    \mathcal{A}_\pm(t,0)
    &= \mathcal{A}_\pm(0,0) \left[ 1 + (m_\phi t)^2 \right]^{-\kincoup/2}.
    \label{eqn:kincoup-k-zero-sol}
\end{align}
This result may be extended to small, nonzero $k$ with a perturbative expansion in the parameter
$k/m_\phi$.
Setting $a(t) = \sqrt{m_\phi t}$, we take the ansatz
\begin{align}
    \mathcal{A}_\pm(t, k)
    &= \mathcal{A}_\pm(0, k) \left[ 1 + (m_\phi t)^2 \right]^{-\kincoup/2}
        \left[ 1 + (k/m_\phi)^2 \delta \mathcal{A}(t) \right]
    \label{eqn:kincoup-k-leading-ansatz}
\end{align}
and solve for the dimensionless perturbation $\delta \mathcal{A}(t)$.
The equation of motion then reduces to one for $\delta\mathcal{A}$ of the form
\begin{align}
    0
    &= \delta\ddot{\mathcal{A}}
        - \frac{ (m_\phi t)^2 (4 \kincoup - 1) -1}{2 t \left[ 1 + (m_\phi t)^2 \right]}
        \delta\dot{\mathcal{A}}
        + \frac{m_\phi}{t} + \frac{k^2}{m_\phi t}
        \delta\mathcal{A}.
\end{align}
The last term is negligible at leading order in $k/m_\phi$, and in this limit the equation is solved
by an integral of a hypergeometric function:
\begin{align}\label{eqn:deltaA}
    \delta\mathcal{A}(t)
    &= \int_0^{m_\phi t}\ud x (1 + x^2)^\kincoup {}_2F_1(1/4,\kincoup,5/4; -x^2) + \mathcal{O}(k^2/m_\phi^2).
\end{align}
Eventually, either the gradient or bare mass term overtakes the $W$-derivative term as the dominant
contribution to $\omega_\mathrm{eff}(k)^2$ and growth ceases.
This behavior may be approximated by cutting off the integral \cref{eqn:deltaA} at the earlier of
the two possibilities,
\begin{align}
    t_\mathrm{cut}(k)
    &\approx \min\left[
            \frac{1}{M} \left( \sqrt{\kincoup^2 + \kincoup/2} \frac{M}{\mA} \right)^{\frac{1}{\kincoup + 1}},
            \frac{(\kincoup^2 + \kincoup/2) m_\phi}{k^2}
        \right].
\end{align}
As long as $m_\phi t_\mathrm{cut}(k) \gg 1$, the integral in \cref{eqn:deltaA} is dominated by
$x \gg 1$, making it an excellent approximation to substitute $1 + x^2 \approx x^2$.
The integral may then be evaluated analytically, yielding
$\mathcal{A}_\pm(t,k) \propto (k/m_\phi)^2(m_\phi t)^{\kincoup + 1/2}$.
After this point, the mode amplitude begins to decrease as $\mathcal{A}_\pm\propto W^{1/4}/a^{1/2}$
while the dark photon mass continues to increase toward its vacuum value, per the WKB approximation.

In summary, the mode amplitude evaluated at $t_\star$ is a product of factors encoding its initial
condition $\mathcal{A}_\pm(0, k) \propto 1 / \sqrt{2 k / a}$, the $k = 0$ solution
[\cref{eqn:kincoup-k-zero-sol}] and the leading-order correction $\delta \mathcal{A}$ for nonzero
$k$ [\cref{eqn:kincoup-k-leading-ansatz}], and factors encoding the adiabatic evolution between
$t_\mathrm{cut}(k)$ and $t_\star$:
\begin{align}
    \frac{\mathcal{A}_\pm(t_\star, k)}{\mathcal{A}_\pm(0, k)}
    &=
        \frac{\mathcal{A}(t_\mathrm{cut}, 0)}{\mathcal{A}(0, 0)}
        \left(\frac{k}{m_\phi}\right)^2\delta \mathcal{A}(t_\mathrm{cut})
        \sqrt[4]{\frac{W(\bar{\phi}(t_\star))}{W(\bar{\phi}(t_\mathrm{cut}))}}
        \sqrt{\frac{a(t_\mathrm{cut}(k))}{a(t_\star)}}.
\end{align}
Inserting the preceding results in the $m_\phi t \gg 1$ limit yields
\begin{align}
    \frac{\mathcal{A}_\pm(t_\star, k)}{\mathcal{A}_\pm(0, k)}
    &= \left[ m_\phi t_\mathrm{cut}(k) \right]^{-\kincoup} \left( \frac{k}{m_\phi} \right)^2
        \left[ m_\phi t_\mathrm{cut}(k) \right]^{2\kincoup + \frac{1}{2}}
        \frac{2\Gamma(\frac{5}{4})\Gamma(\kincoup - \frac{1}{4})}{\Gamma(\kincoup)(4\kincoup + 1)}
        \left[ 2 H_\star t_\mathrm{cut}(k) \right]^{\kincoup/2 + 1/4}
    ,
    \label{eqn:transverse-scalar-mode-amplitude}
\end{align}
where the Hubble rate at production
\begin{align}
    H_\star \equiv 1 / 2 t_\star = M / 2
\end{align}
is set by the time when $\phi$ crosses $0$ and the scalar decouples.

The relic density of dark photons is obtained by a suitable integral of
\cref{{eqn:transverse-scalar-mode-amplitude}} over wave number [\cref{eqn:rho-perp-parallel}].
The mode with peak power is that for which the mass and gradient terms are equally responsible for
cutting off the resonance:
\begin{align}
    k_\star
    &\approx \sqrt{(\kincoup^2 + \kincoup/2) M m_\phi}
        \left( \sqrt{\kincoup^2 + \kincoup/2} \frac{M}{\mA} \right)^{-\frac{1}{2\kincoup + 2}}.
    \label{eqn:kin-coup-kpeak}
\end{align}
The mode amplitude is peaked at this wave number and falls off sharply at larger $k$, so the energy
density is well approximated by integrating up to $k = k_\star$.
The total dark photon energy density [\cref{eqn:rho-perp-parallel}] evaluated at $H = H_\star$ is
thus approximately
\begin{align}
    \frac{\rho_\star}{H_\star^4}
    &\approx \mathcal{N}_{\kincoup}
        \left( \frac{m_\phi}{M} \right)^{2\kincoup - 1}
        \left( \frac{M}{\mA} \right)^{ \frac{\kincoup - 7/2}{\kincoup + 1} },
\end{align}
where the $\kincoup$-dependent coefficient is
\begin{align}
    \mathcal{N}_{\kincoup}
    &= \frac{
        \left( \kincoup^2 + \kincoup/2 \right)^{\frac{9(2\kincoup + 1)}{4(\kincoup + 1)}}}{6\pi^2}
        \frac{\Gamma(5/4)^2\Gamma(\kincoup - 1/4)^2}{(1 + 4\kincoup)^2\Gamma(\kincoup)^2}.
\end{align}
In the large-$\kincoup$ limit, the power spectrum peaks at
$k_\star\sim \kincoup\sqrt{M m_\phi}\lesssim\mA$, so the dark photons are nonrelativistic when their
mass reaches its final value and growth ends.
Redshifting the energy density to matter-radiation equality, the dark photon relic abundance is
\begin{align}
    \frac{\Omega_\Ap}{\Omega_\mathrm{DM}}
    &\approx \frac{2}{3}
        \mathcal{N}_{\kincoup}\left( \frac{m_\phi}{2H_\star} \right)^{2\kincoup-1}
        \left( \frac{2H_\star}{\mA} \right)^{\frac{\kincoup - 7/2}{\kincoup + 1}}
        \frac{H_\star^2}{\Mpl^2}
        \sqrt{\frac{H_\star}{H_\mathrm{eq}}}
    .
    \label{eqn:kinetic-coupling-relic-abundance}
\end{align}
This result reproduces full numerical solutions within a couple of orders of magnitude for
$- \phi_0 / f \gtrsim 5$ [below which the errors are dominated by the approximation made in
\cref{eqn:runaway-approximate-sol}]---accuracy more than sufficient for our purposes.

For $\kincoup$ larger than a few, the relic abundance is most sensitive to $m_\phi / H_\star$, which
effectively encodes how long production takes or equivalently the scalar's initial displacement,
$\phi_0 = -2 f \ln(m_\phi/H_\star)$.
The ratio $H_\star / \mA$ may then be chosen to delay production as long as desired.
In addition, the initial dark photon mass must be far smaller than is necessitated by the
kinematic condition $\mA(\phi_0)\ll m_\phi$: neglecting the $\kincoup$-dependent prefactor in
\cref{eqn:kinetic-coupling-relic-abundance} and taking the large-$\kincoup$ limit
(for which $H_\star \approx m_\phi$) with $\Omega_\Ap = \Omega_\mathrm{DM}$ yields the estimate
\begin{align}
    \frac{\mA^2}{W(\phi_0) m_\phi^2}
    \sim \frac{H_\star^4}{m_\phi^2\Mpl^2}\frac{\mA}{m_\phi}\sqrt{\frac{H_\star}{H_\mathrm{eq}}}
    \ll 1.
    \label{eqn:kinetic-coupling-required-mass-ratio}
\end{align}
In other words, the initial condition for $\phi$ is determined primarily by requiring a sufficient
dark photon abundance rather than the kinematic requirement that $W(\phi_0) \gtrsim \mA^2/m_\phi^2$.
The dynamics are thus effectively insensitive to the scalar--dark-photon mass ratio.
This class of models thus certainly avoids backreaction onto the dark Higgs [\cref{eqn:defectBound}]
in the same parameter space as standard axion production (\cref{sec:axion}), corresponding to
$H_\star \sim m_\phi \gtrsim \mA$ [as depicted in \cref{fig:defect-limits-existing-models}].
But the suppression of the kinematic barrier also allows dark photon much heavier (today) than
$m_\phi$.
In fact, production may be delayed to at least as late as when $H_\star = 10^{-22}~\eV$ (i.e.,
roughly the earliest time observations require the dark matter to exist and behave like
matter)\footnote{
    Dark photon subcomponents of dark matter could, in principle, be produced later with even
    larger kinetic mixing, but their direct-detection signals are also suppressed by their reduced
    abundance.
    Assessing precisely what fraction of dark matter could be produced at such late times, however,
    would require a dedicated cosmological analysis.
}
and the kinetic coupling still eliminates the kinematic barrier for $\mA$ as large as
$\sim(H_\mathrm{eq}/ H_\star)^{1/2} \Mpl^2 / H_\star$.

The above calculations remain valid so long as the scalar's dynamics are well approximated by
\cref{eqn:runaway-approximate-sol}, i.e., so long as backreaction of the dark photon onto the
scalar and that of either field on the expansion history are negligible.
The energy density of a scalar in a runaway potential tracks the critical density in both the
radiation- and matter-dominated epochs:~\cite{Copeland:1997et,Steinhardt:1999nw,Copeland:2006wr}:
\begin{align}
    \Omega_\phi(t)
    \equiv \frac{\bar{\rho}_\phi(t)}{3 \Mpl^2 H^2}
    &\approx \left( \frac{2 f}{\Mpl} \right)^2.
    \label{eqn:scalar-fractional-abundance}
\end{align}
The expansion history thus only departs from the radiation solution near the onset of the matter
era, as in standard cosmology.
Comparing the scalar's energy density to $\kincoup$ times the dark photon's provides a decent proxy
for whether backreaction onto the scalar is important.
At production, the dark photon abundance (which redshifts more slowly than the radiation background
by a power of $a$) is
$\Omega_{\Ap} \approx (H_\mathrm{eq}/H_\star)^{1/2} / 2$,
so the scalar's decay constant needs to be larger than
\begin{align}
    f
    &\gtrsim 10^{-2} \sqrt{\kincoup} \left(\frac{H_\star}{10^{-22}~\eV}\right)^{-1/4} \Mpl
    \label{eqn:runaway-decay-constant-no-backreaction}
\end{align}
to ensure backreaction never occurs.

The scalar's tracking behavior, moreover, provides a natural means for the dark photon's abundance
to exceed the scalar's in the late Universe: the scalar redshifts as radiation, faster than the
nonrelativistic dark photons.
In contrast, when dark photons are produced via a harmonically oscillating axion
(\cref{sec:axion,sec:axion-enhance}) or scalar, the parent field retains a sizable energy density
at late times, as discussed in \cref{sec:axion}.
For this reason, we implemented the termination of production by modifying $W$ from a pure
exponential [\cref{eqn:kinetic-coupling}] rather than augmenting the potential
[\cref{eqn:runaway-potential}] with a minimum at a finite field value.
Barring some additional decay channel~\cite{Agrawal:2018vin, Co:2018lka, Dror:2018pdh}, production
is terminated via nonlinear backreaction that can only be studied via $3+1$-dimensional
simulations~\cite{Agrawal:2018vin} and generically leads to two-component dark matter.

\subsubsection{Quantum corrections}\label{sec:kinetic-coupling-quantum-corrections}

\begin{figure}
    \centering
    \includegraphics[width=\textwidth]{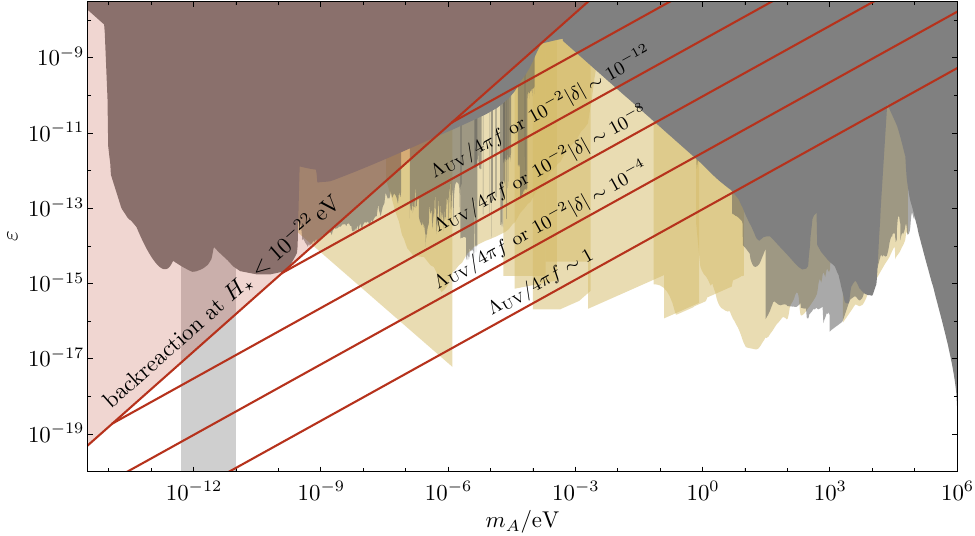}
    \caption{
        Parameter space in mass and kinetic mixing [\cref{eqn:epsilon-Benchmark}] of dark photon
        dark matter produced via kinetically coupled scalars.
        Thin red lines depict the maximum allowed kinetic mixing for a given degree of tuning
        (as labeled) of either the UV cutoff $\Lambda_\mathrm{UV} / 4 \pi f$ for which one-loop
        quantum corrections are unimportant for runaway scalars
        (\cref{sec:transverse-kinetic-coupling}) or the dark-photon--scalar mass tuning $\delta$
        [\cref{eqn:tuning}] for oscillating scalars (\cref{sec:narrow}).
        The red shaded region depicts parameter space in which backreaction or defect formation
        occurs even if the dark photon is produced as late as allowed by the CMB
        (roughly $H_\star \approx 10^{-22}~\eV$).
        Gray and yellow regions respectively depict the current limits and projected reach of
        various probes, as labeled in \cref{fig:defect-limits-existing-models}.
    }
    \label{fig:tuned-scalars-parameter-space}
\end{figure}

Unlike the axion, whose continuous shift symmetry protects its potential from perturbative quantum
corrections, the scalar theories discussed in this section are susceptible to quantum corrections.
Additional heavy degrees of freedom (that arise in a UV completion of the theory) may overwhelm any
phenomenologically viable tree-level scalar potential as it rolls over large field ranges.
Nonetheless, so long as the scalar only has self-interactions and a kinetic coupling to the dark
photon, quantum corrections can remain small because of the smallness of $\mA$ at early times, only
becoming important as $\mA$ reaches its vacuum value.
We compute the one-loop scalar effective potential in scalar--Abelian-Higgs theory in
\cref{app:quantum-corrections}, from which we find that, for the particular choices of $W$ and $V$
above, the UV cutoff $\Lambda_\mathrm{UV}$ must be bounded above by
\begin{align}
    \frac{\Lambda_\mathrm{UV}^2}{(4\pi f)^2}
    &< \frac{3 H_\star^2}{2 \mA^2}.
\end{align}
The later production occurs, the lower the scalar's potential and the smaller its quantum
corrections must be to ensure the dynamics are not disrupted.
Taking $\Lambda_\mathrm{UV} \sim 4 \pi f$ as a criterion for naturalness sets $H_\star \gtrsim \mA$
for scalar--dark-photon production---that is, we expect that below the postinflationary line in
\cref{fig:defect-limits-existing-models} the scalar models of dark photon production we discuss are
both detectable and not fine-tuned in this sense.
Larger kinetic mixing (without backreaction) is available to scalar-produced dark photons to the
extent that one allows the models to be fine-tuned, as illustrated by
\cref{fig:tuned-scalars-parameter-space}.

The above discussion, taking only quadratically divergent contributions to the effective potential,
derives from the one-loop computations in \cref{app:quantum-corrections} which show that properly
quantizing noncanonical field theories leads to a subtle cancellation of would-be quartic
divergences (that one might intuitively expect to be present).
\Cref{app:quantum-corrections} does not, however, attempt to determine whether such divergences are
present at higher-loop order.
Supposing that a quartic divergence was incurred at two loops by the kinetic coupling $W$, it would
also be suppressed by $\gD^2$.
Requiring $\gD^2 \Lambda_\mathrm{UV}^4 \lesssim H_\star^2 f^2$ imposes that
$\gD \lesssim H_\star / f \cdot (f / \Lambda_\mathrm{UV})^2$---nominally quite a constraining bound,
given that it combines with the lower bound on $f$ [\cref{eqn:runaway-decay-constant-no-backreaction}]
to set
$\gD \lesssim 1.3 \times 10^{-20} \left( H_\star / \mathrm{eV} \right)^{5/4} / \sqrt{\kincoup}$
(taking $\Lambda_\mathrm{UV} \approx f$).
This bound happens to coincide with that on standard inflationary production
[\cref{eqn:minimal-gD-inflation-bound}] in scaling (with $H_\star$ in place of $\mA$) and nearly in
magnitude and is therefore extremely constraining for delayed production.
Suppose, however, that $m_\phi \gg \mA$ (so that production is instead early) and that
$k_\star / a_\star \sim H_\star$.
The generic backreaction bound \cref{eqn:defectBound} on transverse production then scales quite
weakly with $H_\star$:
$\gD \lesssim 10^{-14} \lambda^{1/4}\left( \mA / \mu\mathrm{eV} \right)^{5/8} \left( \mA / H_\star \right)^{1/8}$.
The parameter space that is natural can be made larger than that which avoids backreaction
simply by increasing $H_\star / \mA$ by, say, eight orders of magnitude, yielding a bound only
one decade tighter than the fiducial case where $H_\star = \mA$.
We therefore conclude that, even for the worst-case (and unconfirmed) UV divergences, the runaway
scalar model allows for natural and backreaction-free parameter space in most of the fiducial
postinflationary regime.

Since this class of production mechanism relies on the dark Abelian-Higgs sector being extremely
weakly coupled at early times, one might worry that it runs afoul of weak gravity conjectures
(WGCs)~\cite{Arkani-Hamed:2006emk}.
While it is plausible but uncertain whether the WGCs constrain the gauge couplings for displaced
moduli~\cite{matt-private-comm}, were this the case, they would limit the field range of the scalar
field and possibly interfere with dark photon production.
The dark photon mass and hence the gauge coupling must be very small at early times; their sizes are
quantified by \cref{eqn:kinetic-coupling-required-mass-ratio}, which translates to an estimate for
the initial gauge coupling of
\begin{align}
    \gD(\phi_0)
    &\sim \gD\frac{H_\star}{\Mpl}\sqrt{\frac{H_\star}{\mA}}\sqrt[4]{\frac{H_\star}{H_\mathrm{eq}}}
\end{align}
in the large-$\kincoup$ limit.
The tower WGC~\cite{Heidenreich:2016aqi,Heidenreich:2017sim,Reece:2018zvv} implies a UV cutoff
of $g_D(\phi_0)^{1/3} \Mpl$.
Using the upper limit on $\gD$ implied by the backreaction bound \cref{eqn:defectBound}, we find
that the UV cutoff associated with the WGC is
\begin{align}
    \gD
    &\lesssim 3 \times 10^{-18} \cdot
        \lambda^{7/22} \left( \frac{\mA}{10^{-15}~\eV} \right)^{25/22}
        \left( \frac{H_\mathrm{max}}{10^{-15}~\eV} \right)^{-9/22},
\end{align}
where $H_\mathrm{max}$ is the maximum Hubble rate of the Universe and we have required
$T_\mathrm{max} \sim \sqrt{\Mpl H_\mathrm{max}} < \Lambda_\mathrm{UV}$.
\Cref{fig:kinetic-coupling} depicts the range of parameter space in which this model  would violate
the WGC at early times for UV cutoffs of $1~\mathrm{TeV}$ and $10^{16}~\mathrm{GeV}$ (the latter
corresponding to the energy scale of high-scale inflation).

\subsubsection{Signatures}\label{sec:kinetic-coupling-signatures}

In the regime of extremely delayed production, where the scalar potential must be protected by a low
effective UV cutoff, the signatures of production via a rapidly varying gauge coupling are
particularly striking.
Small-scale structure is significantly enhanced for sufficiently small $H_\star$ and is potentially
observable.
In principle, since power below the peak (at which the power spectrum is order unity) scales as
$(k / k_\star)^3$, cosmological scales could also be enhanced above that of the baseline adiabatic
fluctuations (whose power is $\sim 10^{-9}$) if $k_\star$ is within three decades of an observable
scale.
In our case, the peak wave number is $k_\star \approx 2 \kincoup H_\star$ in the large-$\kincoup$
limit.
As a rough guide, at matter-radiation equality
$k_\star / a_\mathrm{eq} \sim 2 \beta H_\mathrm{eq} \sqrt{H_\star / H_\mathrm{eq}}$.
If production terminates long enough before equality (say, by three decades of expansion), then
$\sqrt{H_\star / H_\mathrm{eq}} \gtrsim 10^3$; with $\beta \gtrsim 10$, enhanced power is important
on scales at least an order of magnitude smaller than the horizon at equality.
These scales are on the threshold of the sensitivity of \textit{Planck}'s CMB observations and
around those which the Lyman-$\alpha$ forest probe~\cite{Tegmark:2002cy, Planck:2018nkj,
Chabanier:2019eai}.
Further in the future, 21-cm intensity mapping could probe scales 1-2 orders of magnitude
smaller~\cite{deKruijf:2024voc}.
Ultimately, quantitative cosmological constraints require a dedicated analysis of dark matter with
such power spectra (as would a bound on the latest time the dark matter could be produced).

\begin{figure}
    \centering
    \includegraphics[width=\columnwidth]{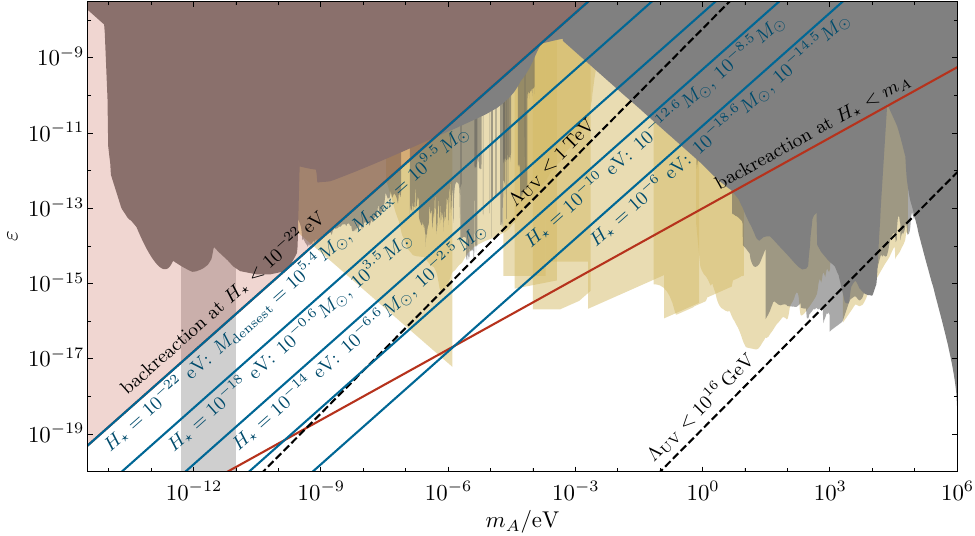}
    \caption{
        Minimum dark matter substructure sizes predicted by the kinetically coupled runaway scalar
        model in the direct-detection parameter space.
        Blue lines display substructure masses $M_\mathrm{densest}$ for structures collapsing at the
        onset of matter-radiation equality and $M_\mathrm{max}$ for structures collapsing at
        redshift $z = 250$ in scenarios where production is delayed until Hubble rates $H_\star$ as
        labeled on each line.
        Much of the expanded kinetic mixing parameter space---including that accessible to future
        direct-detection programs---features substructures that are potentially accessible to
        upcoming astrometric ($M > 10^{-3}M_\odot$)~\cite{VanTilburg:2018ykj} and photometric
        searches ($10^{-15} M_\odot < M < 10^{-8} M_\odot$)~\cite{Dai:2019lud}.
        Dashed black lines represent upper bounds on the dark gauge coupling implied by the tower
        weak gravity conjecture~\cite{Heidenreich:2017sim, Heidenreich:2016aqi, Reece:2018zvv},
        should it apply to displaced moduli.
        Results are otherwise presented as in
        \cref{fig:defect-limits-existing-models,fig:tuned-scalars-parameter-space}.
    }
    \label{fig:kinetic-coupling}
\end{figure}

Enhanced power on yet smaller scales would approach the critical overdensity $\delta_c = 1.686$
during radiation domination~\cite{Press:1973iz}, leading to the collapse of structures with radius
$\sim \pi / (2 k_\star / a_\mathrm{eq})$ around matter-radiation equality.
(The peak of the density power spectrum is at roughly twice $k_\star$.)
The density of these structures is approximately $178$ times that of the ambient matter at
matter-radiation equality~\cite{Bardeen:1985tr},
\begin{align}
    \rho_\mathrm{coll}
    &\sim 178 \cdot \frac{3}{2} \Mpl^2 H_\mathrm{eq}^2
    \approx 2.7 \times 10^{5} M_\odot / \mathrm{pc}^3,
\end{align}
and these collapsed structures contain a mass\footnote{
    Although gravitational interactions can change the subhalo's mass and density over
    time~\cite{VanTilburg:2018ykj}, we calculate their initial values at collapse for a conservative
    representation of their late-time properties.
}
\begin{align}
    M
    &\sim
        \frac{3}{2} \Mpl^2 H_\mathrm{eq}^2
        \cdot \frac{4\pi}{3} \left( \frac{\pi}{2 k_\star / a_\mathrm{eq}} \right)^3
    \approx 2.4 \times 10^5 M_\odot
        \left( \frac{\kincoup}{10} \right)^{-3}
        \left( \frac{H_\star}{10^{-22}~\eV} \right)^{-3/2}.
\end{align}
Power on larger scales is also moderately enhanced, resulting in the collapse of larger, more
diffuse structures at later times.
The structures collapsing at $z = 250$ are the largest that remain resilient against tidal
stripping~\cite{Blinov:2021axd}; their density is again $178$ times the matter density at the time
of collapse, $\rho_\mathrm{coll} \sim 9.2 \times 10^1 M_\odot / \mathrm{pc}^3$.
We may estimate their mass by extrapolating the density power spectrum from its peak to small wave
number, i.e. $\sim (2 k/k_\star)^3$, where the factor of two accounts for the broad peak of the
density power spectrum.
Since the density power spectrum is proportional the square of the fractional overdensity, it grows
with $a^2$ during the matter era, so modes of ever longer wavelength $a/k\propto a^{5/3}$ collapse
as the Universe continues to expand.
On the other hand, the matter density dilutes as $a^{-3}$.
In total, the mass of structures collapsing at a given redshift grows with $a^2$, and we estimate
that structures collapsing at $z = 250$ have mass
\begin{align}
    M
    \sim
        \frac{3}{2} \frac{\Mpl^2 H_\mathrm{eq}^2}{(a / a_\mathrm{eq})^3}
        \cdot \frac{4 \pi}{3} \left( \frac{\pi}{k_\mathrm{col} / a_\mathrm{eq}} \right)^3
    &\sim 2.9 \times 10^9 M_\odot
        \left( \frac{\kincoup}{10} \right)^{-3}
        \left( \frac{H_\star}{10^{-22}~\eV} \right)^{-3/2}.
\end{align}
\Cref{fig:kinetic-coupling} depicts the expected substructure mass for maximally delayed dark photon
production as a function of $\mA$ and $\varepsilon$, superimposed on the limits and prospective
reach of direct-detection experiments.
Note that production may well be delayed longer than strictly necessary to accommodate whatever
kinetic mixing the dark photon happens to have; the substructure masses displayed in
\cref{fig:kinetic-coupling} are thus the minimum predicted by the model (rather than a unique value
thereof).

Astrometric lensing, the correlated deflection in background stars induced by a dense, heavy halo
passing through our line of sight, is a promising indirect signature of such
substructure~\cite{Erickcek:2010fc}.
High-resolution telescopes like \textit{Gaia}~\cite{Gaia:2018ydn} and the \textit{Hubble} Space
Telescope~\cite{bellini2014hubble}, and future missions such as Theia~\cite{Theia:2017xtk},
Roman~\cite{2019JATIS...5d4005W}, the Square Kilometre Array (SKA)~\cite{FOMALONT20041473}, and the
Thirty Meter Telescope~\cite{Skidmore_2015}, could detect these effects by measuring the correlated
proper motions and accelerations of stars.
\textit{Gaia} is sensitive to subhalos heavier than $10^{5} M_\odot$ for the densest subhalos that
form around matter-radiation equality, while Theia and SKA will probe these dense substructures down
to masses around $10^{-3} M_{\odot}$ and the more dilute structures forming at $z = 250$ down to
$10^{2} M_{\odot}$ ~\cite{VanTilburg:2018ykj}.
Another signature is photometric microlensing, where changes in a star's brightness suggest an
intervening gravitational lens.
This technique is particularly effective with highly magnified stars near critical lensing caustics
of galaxy clusters.
If dark matter subhalos are present, their virial motion introduces noise in the star's position and
brightness, depending on the subhalo mass spectrum~\cite{Dai:2019lud}.
However, these projections are subject to uncertainties from galactic evolution and tidal stripping
of subhalos and require further simulation to confirm their viability~\cite{Blinov:2021axd}.

The scalar coupled to the dark photon may yield its own signatures; their inherent model dependence
could facilitate positively identifying the origin of dark photon dark matter.
In the case of the runaway scalar considered in \cref{sec:rapid-kinetic-coupling-dynamics}, delaying
production requires large enough decay constants $f$ to ensure resonant growth is not disrupted, as
well as to supply the dark photon with sufficient energy to make up the dark matter.
The model thus predicts some minimum cosmological abundance of the scalar in both the radiation and
matter eras due to its tracking behavior [\cref{eqn:scalar-fractional-abundance}], even if the dark
photon dominates the dark matter abundance.
The runaway scalar would alter cosmological dynamics to a degree controlled by $f / \Mpl$.
Recent work~\cite{Ramadan:2023ivw, Copeland:2023zqz} has quantitatively analyzed cosmology with
tracking scalars as a means to resolve the Hubble tension, the mismatch between Hubble constants
inferred from the early Universe (e.g., via CMB data~\cite{Planck:2018vyg}) and measured directly by
the calibrated distance ladder~\cite{Riess:2021jrx}.
In the scenarios studied in Refs.~\cite{Ramadan:2023ivw, Copeland:2023zqz}, current data limit the
fractional abundance of the scalar at early times at the few-percent level.
Extrapolating this result to the runaway scalars we consider estimates that $f$ must be below about
a tenth of the Planck scale.
Though these signatures are specific to runaway scalars, a scalar with a quadratic potential could
just as well mechanize delayed production, but without a natural means to ensure dark photons
dominate the dark matter density (since production terminates via nonlinear backreaction).
Such scalars could instead provide astrophysical signatures as fuzzy subcomponents of dark matter.

Finally, though production must end sufficiently early that the dark matter exists well before
matter-radiation equality, the dark photon's mass need not necessarily stop evolving at this point.
The coupling function [\cref{eqn:kinetic-coupling}], rather than asymptoting to unity, could instead
transition to a slower evolution with time.
While not necessary for the scenario we identify, a concrete realization thereof may well predict
that the dark photon mass continues to evolve to some extent after production, a possibility that
can also be tested by cosmological data~\cite{Anderson:1997un, Archidiacono:2022iuu}.

\subsection{Narrow resonance with an oscillating scalar}
\label{sec:narrow}

The final possibility we consider is, rather than a runaway scalar, a scalar oscillating in a
quadratic potential
\begin{align}
    V(\phi)
    &= \frac{1}{2} m_\phi^2 \phi^2.
\end{align}
Such a scenario was studied in Ref.~\cite{Adshead:2023qiw}, which showed that a kinetic coupling
$W(\phi) = e^{\kincoup \phi / f}$ leads not just to broad tachyonic resonance familiar from
analogous models of gauge preheating~\cite{Deskins:2013dwa, Giblin:2017wlo, Adshead:2017xll,
Adshead:2018doq} but also to a unique regime of narrow resonance when $\mA / m_\phi = 1/2$ that can,
in principle, persist for arbitrarily small scalar field amplitudes (and so an arbitrarily long
time).
As a result, conversion of the scalar into dark photons eventually becomes fast compared to the Hubble
rate without requiring large coupling~\cite{Adshead:2023qiw}---indeed, only relying on the
leading-order expansion $W(\phi) \approx 1 + \kincoup \phi / f$, which would be expected to dominate
the behavior from the perspective of bottom-up effective field theory.

Moreover, smaller couplings delay the time that dark photons are efficiently produced, which
Ref.~\cite{Adshead:2023qiw} posited as a means to alleviate defect formation bounds.
Before examining this possibility in detail, we briefly review the mechanics of this scenario,
focusing on the novel regime of persistent resonance at small scalar field amplitudes.\footnote{
    The broad resonance scenario (i.e., $\kincoup \phi_0 / f > 1$ for any mass
    $\mA \lesssim \kincoup m_\phi$) is certainly a viable production mechanism, but its phenomenology
    (in terms of allowed kinetic mixing, at least) is identical to axion production.
}
We also show that the structure of the equations of motion that gives rise to the small-amplitude
regime is not unique to the kinetic coupling considered in Ref.~\cite{Adshead:2023qiw}.

The solution to the homogeneous equation of motion \cref{eqn:scalar-eom} for a massive, misaligned
scalar that begins oscillating in the radiation era [when $a(t) \propto \sqrt{t}$] is
\begin{align}
    \bar{\phi}(t)
    &= \Gamma(5/4) \phi_0 \frac{J_{1/4}(m_\phi t)}{\sqrt[4]{m_\phi t /2}}
\end{align}
from initial conditions $\bar{\phi} = \phi_0$ and $\dot{\bar{\phi}} = 0$ at times $m_\phi t \to 0$.
Define $a_\mathrm{osc} = a(t_\mathrm{osc})$ where $H(t_\mathrm{osc}) = m_\phi$, such that
$H(t) = m_\phi [a(t) / a_\mathrm{osc}]^{-2}$.
At late times $m_\phi t \gg 1$,
\begin{align}
    \bar{\phi}(t)
    &\approx \frac{2^{3/2} \Gamma(5/4) \phi_0}{\sqrt{\pi} [a(t) / a_\mathrm{osc}]^{3/2}}
        \sin \left( m_\phi t + \pi / 8 \right)
    \equiv \Phi(t) \sin \left( m_\phi t + \pi / 8 \right),
    \label{eqn:oscillating-scalar-solution-ito-Phi}
\end{align}
where the second equality defines the time-dependent oscillation amplitude $\Phi(t)$.

Expanding coupling functions in \cref{eqn:rescaled-transverse-eom} to leading order in small
$\bar{\phi}$ and neglecting Hubble friction (as appropriate at the late times when resonance becomes
efficient),
\begin{align}
    0
    &= \ddot{\mathcal{A}}_\pm
        + \left[
            \frac{k^2}{a^{2}}
            + \mA^2
            + \mA^2 \bar{\phi}(t) \left(
                \frac{X'(\bar{\phi})}{X(\bar{\phi})}
                - \frac{W'(\bar{\phi})}{W(\bar{\phi})}
            \right)
            + \ddot{\bar{\phi}}(t) \frac{W'(\bar{\phi})}{2 W(\bar{\phi})}
        \right]
        \mathcal{A}_\pm.
    \label{eqn:rescaled-transverse-eom-small-phi}
\end{align}
Substituting \cref{eqn:oscillating-scalar-solution-ito-Phi} and taking the time coordinate
$z = m_\phi t / 2 - 3 \pi / 16$, \cref{eqn:rescaled-transverse-eom-small-phi} takes the form of the
Mathieu equation,
\begin{align}
    0
    &= \frac{\ud^2 \mathcal{A}_\pm}{\ud z^2}
        + \left[ p - 2 q \cos(2 z) \right] \mathcal{A}_\pm,
\end{align}
with $m_\phi^2 p / 4 = k^2 / a^2 + \mA^2$ and
\begin{align}
    q
    &= \Phi(t)
        \left[
            \frac{2 \mA^2}{m_\phi^2}
            \left(
                \frac{X'(\bar{\phi})}{X(\bar{\phi})}
                - \frac{W'(\bar{\phi})}{W(\bar{\phi})}
            \right)
            - \frac{W'(\bar{\phi})}{W(\bar{\phi})}
        \right].
    \label{eqn:mathieu-q-oscillating}
\end{align}
If the mass ratio $\mA / m_\phi$ is precisely $1/2$ then a narrow resonance for $k \ll m_\phi$
persists to arbitrarily small $q$ (i.e., field amplitudes)~\cite{Adshead:2023qiw}.
(Otherwise---as is typical of such scenarios---the resonance terminates once the scalar redshifts
below some nonzero $\bar{\phi}$.)

Evidently, \cref{eqn:mathieu-q-oscillating} is nonzero regardless of whether the coupling is via
the kinetic term ($W$) or the mass term ($X$) or both (i.e., a universal coupling); only coupling
functions contrived to make \cref{eqn:mathieu-q-oscillating} identically zero would exhibit no such
instability.
Furthermore, longitudinal modes experience the same instability: as discussed below
\cref{eqn:SAH-linearized-eoms}, in the $k \ll \mA(\bar{\phi})$ limit where the small-amplitude
resonance would exist if present, the longitudinal mode's dynamics
[\cref{eqn:longitudinal-eom-isolated}] are identical to the transverse modes'.

The narrow-resonance scenario allows for dark photon production parametrically later than the
standard expectation that $H_\star \gtrsim \mA$.
Modes in the narrow-resonance band have exponential growth rate
$\Gamma = \kincoup m_\phi \Phi(t) / 8 f$, which decays more slowly than the Hubble rate in the
radiation era.
Production therefore becomes efficient at a scale factor~\cite{Adshead:2023qiw}
\begin{align}\label{eqn:narrow-resonance-efficient-production-scale-factor}
    \frac{a_\star}{a_\mathrm{osc}}
    &= \left( \frac{3 \kincoup \phi_0}{4 f} \right)^{-2}
        \ln \left( \frac{m_\phi^2}{\mA^2} \frac{\phi_0^2}{m_\phi^2} \right)^2,
\end{align}
where $\phi_0$ is the initial misalignment of the scalar.
The corresponding Hubble rate at production is
\begin{align}
    \frac{H_\star}{\mA}
    &= 2
        \left( \frac{3 \kincoup \phi_0}{4 f} \right)^{4}
        \ln \left( \frac{m_\phi^2}{\mA^2} \frac{\phi_0^2}{m_\phi^2} \right)^{-4}.
    \label{eqn:narrow-H-star-ito-kincoup}
\end{align}
Since nonrelativistic dark photons are produced, the energy density in the sector always redshifts
as $a^{-3}$ (after $a_\mathrm{osc}$); the relic abundance is therefore simply that inherited from
the scalar's misalignment:
\begin{align}
    \frac{\Omega_A}{\Omega_\mathrm{DM}}
    &\approx
        \left( \frac{\mA}{2 \times 10^{-18}~\eV} \right)^{1/2}
        \left( \frac{\phi_0}{10^{16}~\mathrm{GeV}} \right)^2.
    \label{eqn:narrow-relic-abundance}
\end{align}

As suggested by Ref.~\cite{Adshead:2023qiw}, such delayed production could allow for larger $g_D$
while avoiding defect formation, akin to the mechanisms discussed above.
Moreover, since $W(\phi)$ need not be particularly large (or small) for efficient production, the
effect of the scalar on the parameters of the theory [\cref{eqn:phi-dependent-parameters}] is
negligible.
However, the persistence of narrow resonance to small amplitudes $\Phi(t)$ (and therefore at late
times) depends how precisely the mass ratio $\mA / m_\phi$ is tuned to $1/2$.
Concretely, the instability persists until $a_\star$ only if the tuning
parameter~\cite{Adshead:2023qiw}
\begin{align}\label{eqn:tuning}
    \delta
    &\equiv \left( \mA / m_\phi \right)^2 - 1/4
\end{align}
satisfies
\begin{align}\label{eqn:tuning-result}
    \abs{\delta}
    < \frac{3 \kincoup \phi_0}{16 f (a_\star / a_\mathrm{osc})^{3/2}}.
\end{align}
Inserting \cref{eqn:narrow-H-star-ito-kincoup} for $\kincoup \phi_0 / f$, production thus may only
be delayed to a Hubble rate
\begin{align}
    \frac{H_\star}{\mA}
    &> 4 \abs{\delta}
        \ln \left( \frac{m_\phi^2}{\mA^2} \frac{\phi_0^2}{m_\phi^2} \right)^{-1}
    \gtrsim 0.02 \abs{\delta},
    \label{eqn:narrow-H-star-for-tuning}
\end{align}
using the relic abundance \cref{eqn:narrow-relic-abundance} evaluated with a fiducial scalar mass
$m_\phi = 4 \times 10^{-18}~\eV$.
Using \cref{eqn:defectBound}, the maximum dark gauge coupling that evades defect formation is then
\begin{align}
    \gD
    &\lesssim
        4 \times 10^{-14} \,
        \lambda^{1/4}
        \abs{\delta}^{-3/8}
        \left( \frac{\mA}{\mu\eV} \right)^{5/8}
        \left( \frac{\Omega_\Ap}{\Omega_\mathrm{DM}} \right)^{-1/4}.
    \label{eqn:narrow-bare-mass-tuning-bound}
\end{align}
The allowed kinetic mixing parameter space as a function of $\delta$ is depicted in
\cref{fig:tuned-scalars-parameter-space}.

\subsubsection{Thermal masses and tuning}

Even assuming the existence of a UV completion that guarantees such a mass tuning that is robust to
quantum corrections, in-medium effects inevitably source a time-dependent mass for the dark photon
via its kinetic mixing with the SM photon.
To assess the impact of the SM plasma on this mechanism, we extend the scalar--Abelian-Higgs theory
to explicitly include kinetic mixing as well as a current term representing nonrelativistic electrons:
\begin{align}
\begin{split}
    \mathcal{L}
    &= \mathcal{L}_\mathrm{SAH}
        - \frac{\varepsilon}{2} F_{\mu\nu} F_\mathrm{SM}^{\mu\nu}
        - \frac{1}{4} F^\mathrm{SM}_{\mu\nu} F_\mathrm{SM}^{\mu\nu}
        - e A^\mathrm{SM}_\mu J^\mu,
\end{split}
\end{align}
where $J^\mu = n_e u_e^\mu$ and $n_e$ is the electron number density and $u_e^\mu$ is their
velocity.
We remove the explicit kinetic mixing by redefining the SM photon field to
\begin{align}
    \mathsf{A}^\mu
    &= A_\mathrm{SM}^\mu + \varepsilon A^\mu
\end{align}
and define the electromagnetic field strength tensor
$\mathsf{F}_{\mu \nu} \equiv \partial_\mu \mathsf{A}_\nu - \partial_\nu \mathsf{A}_\mu$.
In this basis, the Lagrangian takes the form
\begin{align}
\begin{split}\label{eqn:lagrangian-with-current}
    \mathcal{L}
    &=\frac{1}{2} \partial_\mu \phi \partial^\mu \phi
        - V(\phi)
        - \frac{W(\phi) - \varepsilon^2}{4} F_{\mu\nu} F^{\mu\nu}
        + \frac{X(\phi)}{2} \mA^2 A_\mu A^\mu
        - \frac{1}{4} \mathsf{F}_{\mu \nu} \mathsf{F}^{\mu \nu}
        - e J_\mu
        \left(
            \mathsf{A}^\mu
            - \varepsilon A^\mu
        \right).
\end{split}
\end{align}
The formalism developed in the absence of mixing must only be augmented by including the SM current
in the equations of motion and replacing $W(\phi)$ with $W(\phi) - \varepsilon^2$.

For simplicity we consider only transverse modes; the equations of motion for $\mathsf{A}_\pm$ and
$\mathcal{A}_\pm$ for the Lagrangian \cref{eqn:lagrangian-with-current} are
\begin{subequations}\label{eqn:eoms-with-current}
\begin{align}
    e J_\pm
    &= \ddot{\mathsf{A}}_\pm
        + H \dot{\mathsf{A}}_\pm
        + \frac{k^2}{a^{2}} \mathsf{A}_\pm
    \label{eqn:A-SM-eom-current}
    \\
    - e \varepsilon(\bar{\phi}) J_\pm
    &= \ddot{\mathcal{A}}_\pm
        + H \dot{\mathcal{A}}_\pm
        +  \left[
            \frac{k^2}{a^{2}}
            + \mA(\bar{\phi})^2
            - \frac{H}{2}
            \frac{\partial_t W(\bar{\phi})}{W(\bar{\phi}) - \varepsilon^2}
            - \frac{
                    \partial_t^2 \sqrt{W(\bar{\phi}) - \varepsilon^2}
                }{
                    \sqrt{W(\bar{\phi}) - \varepsilon^2}
            }
        \right]
        \mathcal{A}_\pm.
\end{align}
\end{subequations}
In \cref{app:plasma}, we show that the transverse modes of the current evolve approximately by
\begin{align}
    \dot{J}_\pm
    + \nu(t) J_\pm
    &= \frac{\omega_p(t)^2}{e}
        \left[
            \dot{\mathsf{A}}_\pm - \varepsilon(\bar{\phi}) \dot{\mathcal{A}}_\pm
        \right],
    \label{eqn:current-eom-main}
\end{align}
where the damping coefficient $\nu(t)$ is defined in \cref{eqn:def-nu}, and its value relative to
the plasma frequency $\omega_p \equiv \sqrt{e^2 n_e / m_e}$ is given by
\cref{eqn:resistivity-vs-plasma-frequency}.
For simplicity, we take $W = 1$ so that $\varepsilon$ is constant.
In this case, the factor of $1 - \varepsilon^2$ multiplying the kinetic term in
\cref{eqn:lagrangian-with-current} may be absorbed via a redefinition of $g_D$.

We reduce the system to a set of third-order equations of motion for $\mathsf{A}_\pm$ and
$\mathcal{A}_\pm$ alone by differentiating \cref{eqn:eoms-with-current} with respect to time,
inserting \cref{eqn:current-eom-main} for $\dot{J}_\pm$, and then substituting
\cref{eqn:A-SM-eom-current} for $J_\pm$.
The resulting equations may then be written in the form
\begin{subequations}\label{eqn:third-order-eoms-refactored}
\begin{align}
\begin{split}
    0
    &= \dd{}{t} \left(
            \ddot{\mathsf{A}}_\pm
            + \left[
                H
                + \nu
            \right]
            \dot{\mathsf{A}}_\pm
            + \left[
                \frac{k^2}{a^2}
                + \omega_p^2
                - \dot{\nu}
                + H \nu
            \right]
            \mathsf{A}_\pm
            - \varepsilon \omega_p^2 \mathcal{A}_\pm
        \right)
    \\ &\hphantom{ {}={} }
        + \left(
            \ddot{\nu}
            - H \dot{\nu}
            + \nu \frac{k^2}{a^2}
            - \nu \dot{H}
            - 2 \omega_p \dot{\omega}_p
        \right)
        \mathsf{A}_\pm
        + 2 \omega_p \dot{\omega}_p \mathcal{A}_\pm
\end{split}
    \\
\begin{split}
    0
    &= \dd{}{t} \left(
            \ddot{\mathcal{A}}_\pm
            + H \dot{\mathcal{A}}_\pm
            + \left[
                \frac{k^2}{a^{2}}
                + \mA(\bar{\phi})^2
                + \varepsilon^2 \omega_p^2
            \right]
            \mathcal{A}_\pm
            - \varepsilon \dot{\nu}
            \dot{\mathsf{A}}_\pm
            + \left[
                \dot{\nu}
                - H \nu
                - \omega_p^2
            \right]
            \varepsilon \mathsf{A}_\pm
        \right)
    \\ &\hphantom{ {}={} }
        + \varepsilon \left(
            2 \omega_p \dot{\omega}_p
            + \left[
                \dot{H}
                - \frac{k^2}{a^2}
                + H \dot{\nu}
                - \ddot{\nu}
            \right]
        \right)
        \mathsf{A}_\pm
        - 2 \varepsilon^2 \omega_p \dot{\omega}_p
        \mathcal{A}_\pm
    .
\end{split}
\end{align}
\end{subequations}
If the second lines of each equality in \cref{eqn:third-order-eoms-refactored} are negligible, then
the system reduces to second order in time with additional damping and mass contributions from the
plasma.
Since $\dot{\omega}_p / \omega_p \propto H$ and $\dot{\nu} / \nu \propto H$, this condition is met
by taking both $H$ and $k/a$ small compared to $\omega_p$ and $\mA(\bar{\phi})$.
\Cref{eqn:third-order-eoms-refactored} then simplifies to
\begin{subequations}\label{eqn:eoms-with-current-simplified}
\begin{align}
    0
    &= \ddot{\mathsf{A}}_\pm
        + \nu
        \dot{\mathsf{A}}_\pm
        + \omega_p^2
        \mathsf{A}_\pm
        - \varepsilon \omega_p^2 \mathcal{A}_\pm
    \label{eqn:A-SM-eom-with-current-simplified}
    \\
    0
    &= \ddot{\mathcal{A}}_\pm
        + \left[
            \mA(\bar{\phi})^2
            + \varepsilon^2 \omega_p^2
        \right]
        \mathcal{A}_\pm
        - \varepsilon \nu
        \dot{\mathsf{A}}_\pm
        - \varepsilon \omega_p^2
        \mathsf{A}_\pm.
    \label{eqn:rescaled-A-eom-with-current-simplified}
\end{align}
\end{subequations}

\Cref{eqn:resistivity-vs-plasma-frequency} (and the surrounding discussion) shows that the relevant regime for $T \ll m_e$ is that of negligible damping, $\nu \ll \omega_p$.\footnote{
    As a matter of interest, the system decouples in the limit of infinite damping
    ($\nu \gg \omega_p$) because the SM photon slowly rolls in this case.
    That is, $\dot{\mathsf{A}}_\pm \approx - (\omega_p^2/\nu)\left( {\mathsf{A}}_\pm -\varepsilon {\mathcal{A}}_\pm\right)$ solves \cref{eqn:A-SM-eom-with-current-simplified} when
    $\ddot{\mathsf{A}}_\pm$ is neglected at leading order in $\omega_p/\nu$.
    As a result, $\ddot{\mathcal{A}}_\pm \approx - \mA(\bar{\phi})^2 \mathcal{A}_\pm$.
    In other words, if the plasma is extremely resistive to currents, it decouples from the dark
    photon.
}
In this case, were $\mA(\bar{\phi})^2 + \varepsilon^2 \omega_p^2$ constant we could simply
diagonalize the equations and solve for the mode functions.
Though $\mA(\bar{\phi})$ is explicitly not constant, in the narrow-resonance band its relative
variations are small.
Writing this small perturbation in the form $\mA(\bar{\phi})^2 = \mA^2 + \delta\mA(\bar{\phi})^2$,
we may meaningfully define new fields
\begin{align}
    \begin{pmatrix}
        a_\pm \\ b_\pm
    \end{pmatrix}
    &= \begin{pmatrix}
            \cos \varphi & \sin \varphi
            \\
             -\sin \varphi & \cos \varphi
        \end{pmatrix}
        \begin{pmatrix}
            \mathsf{A}_\pm \\ \mathcal{A}_\pm
        \end{pmatrix}
    \approx \begin{pmatrix}
            1 & \varepsilon / [1 - \mA^2 / \omega_p^2]
            \\
            -\varepsilon / [1 - \mA^2 / \omega_p^2] & 1
        \end{pmatrix}
        \begin{pmatrix}
            \mathsf{A}_\pm \\ \mathcal{A}_\pm
        \end{pmatrix},
        \label{eqn:change-basis-plasma}
\end{align}
where
\begin{align}
    \frac{1}{2}\tan 2 \varphi
    &= \frac{\varepsilon}{1 - \varepsilon^2 - \mA^2 / \omega_p^2}
\end{align}
and the second equality in \cref{eqn:change-basis-plasma} shows the result at leading order in
$\varepsilon$.
Then to quadratic order in $\varepsilon$ and $\delta \mA(\bar{\phi})^2 / \mA^2$, these fields evolve
according to
\begin{subequations}
\begin{align}
    0
    &= \ddot{a}_\pm
        + \omega_p^2 \left(
            1 + \frac{\varepsilon^2}{1 - \mA^2 / \omega_p^2}
        \right)
        a_\pm
        - \varepsilon \frac{\delta \mA(\bar{\phi})^2}{1 -\mA^2/\omega_p^2} b_\pm
    \\
    0
    &= \ddot{b}_\pm
        + \mA(\bar{\phi})^2 \left(
            1 - \frac{\varepsilon^2}{1 - \mA^2 / \omega_p^2}
        \right)
        b_\pm
        - \varepsilon \frac{\delta \mA(\bar{\phi})^2}{1 -\mA^2/\omega_p^2} a_\pm.
\end{align}
\end{subequations}
If $\ddot{a}_\pm \sim \omega_p a_\pm$, then
$a_\pm \sim \varepsilon \delta \mA(\bar{\phi})^2 / (\omega_p^2 - \mA^2) \cdot b_\pm$.
So long as $\delta \mA(\bar{\phi})^2$ is small compared to either $\omega_p^2$ or $\mA^2$ (the
latter of which is certainly the case), the contribution of $a_\pm$ to the equation of motion for
$b_\pm$ is negligible and the two equations decouple.
Plasma effects therefore do not disrupt resonance when their contribution to the effective mass of
$b_\pm$ is smaller than the required tuning [defined in \cref{eqn:tuning}]:
\begin{align}
    \delta
    &\gtrsim \varepsilon^2
        \frac{\min\left( \omega_p^2, \mA^2 \right)}{4 \max\left( \omega_p^2, \mA^2 \right)}.
    \label{eqn:plasma-tuning-condition}
\end{align}
Note that this analysis breaks down if $\mA$ crosses $\omega_p$.

Since the thermal mass decreases monotonically with time, \cref{eqn:plasma-tuning-condition} need
only hold at the latest possible production time allowed by the degree of bare mass tuning.
Inserting \cref{eqn:plasma-frequency} for $\omega_p$ and then evaluating at $H_\star$ according to
\cref{eqn:narrow-H-star-for-tuning} yields
\begin{align}
    \varepsilon
    &\lesssim
        1.76 \times 10^{-8}
        \begin{dcases}
            \left( \frac{\mA}{6.28 \times 10^{-5}~\eV} \right)^{-3/2}
            \left( \frac{H_\star}{10^{-22}~\eV} \right)^{5/4}
            \sqrt{ \frac{\eta_B}{6 \times 10^{-10}}},
            & \mA \ll \omega_p,
            \\
            \left( \frac{\mA}{6.28 \times 10^{-5}~\eV} \right)^{1/2}
            \left( \frac{H_\star}{10^{-22}~\eV} \right)^{-1/4}
            \sqrt{ \frac{6 \times 10^{-10}}{\eta_B}},
            & \mA \gg \omega_p,
        \end{dcases}
    \label{eqn:narrow-bare-mass-tuning-bound-thermal}
\end{align}
where $\eta_B$ is the baryon-to-photon ratio.
[The plasma frequency is $6.28 \times 10^{-5}~\eV$ when $H_\star = 10^{-22}~\eV$, so
\cref{eqn:narrow-bare-mass-tuning-bound-thermal} is written relative to the most constrained dark
photon mass.]
Notably, the kinetic mixing can be much larger than the maximum that would otherwise be allowed by
backreaction [\cref{eqn:defectBound}].
The detuning due to plasma effects redshifts faster than the tuning required for efficient
resonance, and the more tuned the bare masses are, the later production may viably occur (i.e., by
reducing $\kincoup$ as much as possible).
Moreover, at a fixed bare mass tuning, \cref{eqn:narrow-bare-mass-tuning-bound-thermal} is a weaker
constraint than \cref{eqn:narrow-bare-mass-tuning-bound}.
Note also that we assumed the plasma is nonrelativistic because our interest is in maximally delayed
production; production at earlier times when electrons are relativistic is certainly allowed, but a
different analysis would be necessary.

\subsubsection{Matter power spectrum}

Although the narrow resonance is centered on $k = 0$, with time the resonance band only grows in
width over comoving wave number since~\cite{Adshead:2023qiw}
\begin{align}
    \left\vert \left(\frac{k}{a m_\phi}\right)^2 + \delta \right\vert
    &< \frac{\kincoup \phi_0}{f}\left(\frac{H}{m_\phi}\right)^{3/4}.
\end{align}
The dark photon's power spectrum thus peaks at the largest wave number that is unstable when the
scalar starts oscillating (at $H \approx m_\phi$).
As long as the detuning $\delta$ is negligible, which is certainly the case during the early phases
of the resonance, the peak wave number is approximately
$k_\star / a_\mathrm{osc} \sim 2 \mA \sqrt{\kincoup \phi_0 / f}$.
The density power spectrum is thus peaked as in \cref{sec:kinetic-coupling-signatures}, with scales
$\sim a_\mathrm{eq} / 2 k_\star$ collapsing at matter-radiation equality.
The resulting substructures have a characteristic mass of
\begin{align}
    M
    &\sim \frac{3}{2} \Mpl^2 H_\mathrm{eq}^2
        \frac{4 \pi}{3}
        \left( \frac{\pi}{2 k_\star / a_\mathrm{eq}} \right)^{3}
    \sim 1.5 \times 10^{-6} M_\odot
        \left( \frac{\mA}{10^{-12}~\eV} \right)^{-9/8}
        \left( \frac{H_\star}{10^{-22}~\eV} \right)^{-3/8}
    ,
\end{align}
where we used \cref{eqn:narrow-H-star-ito-kincoup} to solve for $\kincoup \phi_0 / f$ and then used
the relic abundance \cref{eqn:narrow-relic-abundance} to set a fiducial value for $\phi_0 / m_\phi$
in terms of $\mA$ and $H_\star$ at the benchmark parameters.
(The above result thus neglects logarithmic dependence on $\mA$.)
While these structures are smaller than those produced by the kinetic coupling in the previous
section, they may be interesting targets for photometric microlensing~\cite{Dai:2019lud}.

\section{Discussion}\label{sec:discussion}

\begin{figure}
    \centering
    \includegraphics[width=\columnwidth]{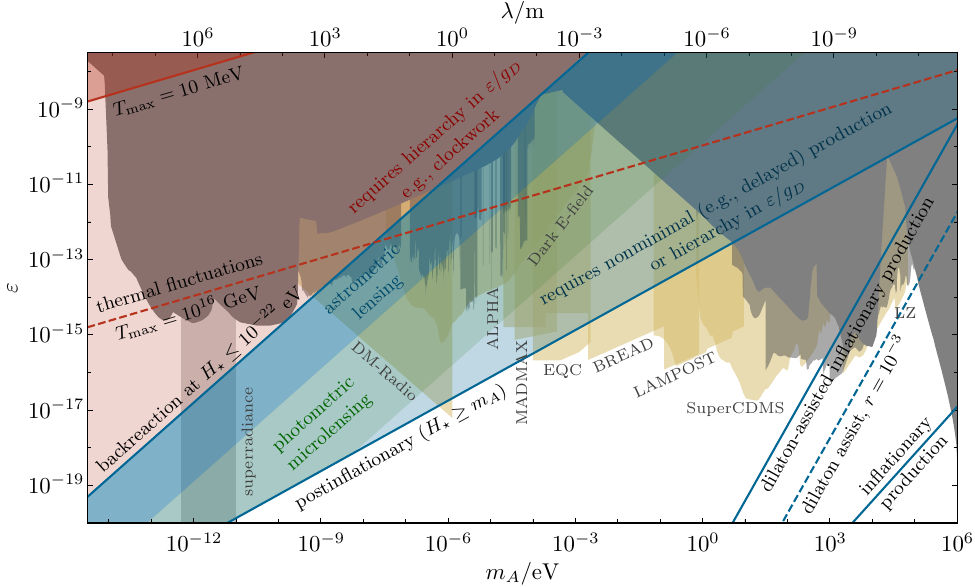}
    \caption{
        Summary of various limits on viable kinetic mixing [\cref{eqn:epsilon-Benchmark}] of dark photon dark matter.
        From right to left, blue lines correspond to the upper bounds for the various production
        scenarios outlined in the text: gravitational production during inflation
        (\cref{sec:inflation}); inflationary production assisted by a dilaton mechanizing an
        evolving dark photon mass (\cref{sec:universal-coupling}); postinflationary production
        taking place at a fiducial Hubble rate $H_\star \approx \mA$, as exemplified by tachyonic
        production via axion oscillations (\cref{sec:axion}); and maximally delayed
        postinflationary production, as realized by multiple models with rolling scalars
        (\cref{sec:axion-enhance,sec:transverse-kinetic-coupling,sec:narrow}).
        The red region above this last line cannot be realized by modified dark photon production,
        as the Proca theory breaks down even while modes observed in the CMB enter the horizon
        ($H_\star \lesssim 10^{-22}~\eV$); this parameter space instead requires a hierarchy in
        $\varepsilon$ and $\gD$ (see \cref{sec:production-independent}).
        Above the red lines, defects nucleate due solely to the Standard Model plasma's mixing with
        the dark photon, independent of whatever other mechanisms produce dark photons (see
        \cref{sec:thermal}); solid and dashed lines depict bounds assuming the minimum and maximum
        possible reheating temperatures within currently viable inflationary cosmologies.
        Signatures of the various nonminimal production scenarios become more prominent toward the
        upper boundaries of their allowed parameter space.
        For dilaton-assisted inflationary production, larger kinetic mixing requires larger
        inflationary scales and therefore could be corroborated by a detection of primordial
        gravitational waves with future CMB observations; this scenario also predicts equivalence
        principle violation.
        Dark matter substructure probed via astrometric lensing (darker blue) or photometric
        microlensing (green) could corroborate delayed production via, e.g., a runaway scalar
        (\cref{fig:kinetic-coupling}) with kinetic mixing above the fiducial postinflationary
        production line; the enhancement of small-scale structure could also be probed by future
        cosmological observations.
        Current exclusions from astrophysical~\cite{Redondo:2008aa, Zechlin:2008tj, Li:2023vpv,
        Dubovsky:2015cca, Vinyoles:2015aba, Baryakhtar:2017ngi, Wadekar:2019mpc, Linden:2024fby,
        Hong:2020bxo, Bi:2020ths, Fedderke:2021aqo} and cosmological~\cite{Arias:2012az,
        McDermott:2019lch, Caputo:2020rnx, Caputo:2020bdy, Witte:2020rvb} probes are depicted in
        dark gray and those from haloscope and other laboratory searches~\cite{Suzuki:2015sza,
        Knirck:2018ojz, Brun:2019kak, Hochberg:2019cyy, Nguyen:2019xuh, Phipps:2019cqy,
        SuperCDMS:2019jxx, XENON:2019gfn, DAMIC:2019dcn, An:2020bxd, Dixit:2020ymh,
        FUNKExperiment:2020ofv, SENSEI:2020dpa, Tomita:2020usq, XENON:2020rca, Chiles:2021gxk,
        Fedderke:2021rrm, Manenti:2021whp, XENON:2021qze, XENON:2022ltv, An:2022hhb,
        Cervantes:2022yzp, DarkSide:2022knj, DOSUE-RR:2022ise, Fan:2022uwu, Ramanathan:2022egk,
        Bajjali:2023uis, An:2023wij, BREAD:2023xhc, An:2024wmc, Levine:2024noa} in light gray.
    }
    \label{fig:final-parameter-space}
\end{figure}

In this paper, we mapped the parameter space of kinetically mixed dark photon dark matter, charting
regions that require varying degrees of assumptions and model complexity to mechanize cosmologically
consistent dark photon production.
Building off the results of Ref.~\cite{East:2022rsi}, in
\cref{sec:defect-bounds-on-existing-scenarios} we reviewed how early-Universe backreaction onto the
dark Higgs impacts the viability of established dark photon production mechanisms.
A fundamental assumption is that the dark photon's mass is UV-completed by a dark Higgs mechanism,
for which the dark gauge coupling and kinetic mixing would generally be of the same order.
This minimal scenario faces stringent constraints due to the excitation of new degrees of
freedom---namely, the dark Higgs, which disrupts production, and the formation of magnetic strings,
whose presence rules out cold dark photon dark matter~\cite{East:2022rsi}.
Our atlas of the Higgsed dark photon parameter space is summarized by
\cref{fig:final-parameter-space}; in the remainder of this section we review our main results.

\subsection{Inflationary production}

The minimal scenario of gravitational dark photon production during inflation~\cite{Graham:2015rva}
is the most strongly constrained~\cite{East:2022rsi}---no planned experiments come close to
detecting viable kinetic mixing in this case.
\Cref{sec:inflation} discussed the dependence of this result on the postinflationary expansion
history, showing that a prolonged period of postinflationary kination (rather than instantaneous
reheating) allows for kinetic mixings that are just marginally accessible to LZ~\cite{LZ:2021xov}
and XENON~\cite{XENON:2019gfn, XENON:2020rca, XENON:2021qze, XENON:2022ltv} (see
\cref{fig:inflation-parameter-space}).
A perhaps more plausible early-Universe cosmology---a prolonged matter era after inflation---leads
instead to even more stringent upper limits.

In \cref{sec:scalar} we explored a class of models extending the Abelian-Higgs sector with a singlet
scalar, whose couplings induce field-dependent modulation of the theory's parameters.
In such models, the scalar's dynamics could directly alter the backreaction bound
\cref{eqn:coupling-strength} throughout cosmological history---for instance, between inflation and
the present day.
In \cref{sec:universal-coupling} we showed that a universal coupling, for which the mass scales in
the theory depend on the scalar, can expand the parameter space available to inflationary production
if the dark photon's mass drops after it becomes nonrelativistic.
Further, we demonstrated that the dilaton model of Damour and Polyakov~\cite{Damour:1994zq} provides
a realization of such dynamics: if the dilaton is also universally coupled to Standard Model
particles, it only starts rolling at late times when Standard Model particles begin to annihilate.
Optimally weakening the backreaction bounds requires the inflationary scale to approach the upper
limit permitted by CMB constraints, $H_I \sim 5 \times 10^{13}~\GeV$~\cite{Planck:2018vyg}.
This ``dilaton assist'' model therefore predicts that if LZ detects a dark photon, future CMB
experiments will detect primordial gravitational waves from inflation (see
\cref{fig:final-parameter-space,fig:dilaton-inflation-parameter-space}).
Furthermore, the dilaton's coupling to Standard Model matter potentially violates the equivalence
principle and could be probed in future space-based experiments~\cite{Battelier:2019kmc,
Touboul:2022yrw}.

\subsection{Postinflationary and delayed production}

While gravitational production during inflation is inevitable, inflation need not occur at an energy
scale high enough for inflationary dark photon production to fully account for the dark matter
abundance.
Postinflationary mechanisms, while less minimal, allow for dark photon dark matter not only over a
much broader range of masses~\cite{Agrawal:2018vin, Co:2018lka, Dror:2018pdh, Bastero-Gil:2018uel}
but also, as explained in \cref{sec:defect-bounds-on-existing-scenarios}, a much broader range of
kinetic mixing.
The general bound of \cref{eqn:defectBound} demonstrates that backreaction on the dark Higgs and the
formation of topological defects are evaded at larger couplings the later dark photon production
occurs.
Dark matter must exist in time for the standard $\Lambda$CDM model to describe the observed
CMB; a reasonable benchmark is that production occurs no later than a Hubble rate of
$H_\star \sim 10^{-22}~\eV$ (or redshifts around $10^6$).
This bound is saturated along the ``delayed production'' contour in
\cref{fig:final-parameter-space}, and a signal near this boundary would indicate extremely delayed
production, potentially leaving a signatures in the CMB at small angular scales.

In practice, however, mechanizing such delayed production is nontrivial, as demonstrated in the case
of resonant production of dark photons from axion oscillations~\cite{Agrawal:2018vin, Co:2018lka,
Co:2021rhi, Kitajima:2023pby} discussed in \cref{sec:axion}.
Kinematic requirements (for efficient parametric resonance) generally restrict production to early
epochs when the Hubble rate exceeded the dark photon mass, setting the fiducial postinflationary
benchmark with $H_\star \approx \mA$ in \cref{fig:final-parameter-space}.
A signal from experiments such as ALPHA~\cite{Gelmini:2020kcu}, BREAD~\cite{BREAD:2021tpx}, Dark
E-Field Radio~\cite{Levine:2024noa}, DM Radio~\cite{DMRadio:2022pkf}, or
MADMAX~\cite{Gelmini:2020kcu} could therefore point to production delayed beyond this minimal
benchmark.
In contrast, a signal from LAMPOST~\cite{Baryakhtar:2018doz}, SuperCDMS~\cite{Bloch:2016sjj}, or an
extension of BREAD using a highly excited cyclotron~\cite{Fan:2024mhm} would be consistent with
these standard postinflationary production scenarios.

In \cref{sec:axion-enhance}, we showed that a number of underappreciated regimes of axion production
(see \cref{fig:axion-production-schematic}) are indeed viable in broader parameter space.
The standard regime of broad tachyonic resonance requires dimensionless axion--dark-photon coupling
$\beta \gtrsim 10$, which is parametrically larger than the typical expectation that
$\beta \sim \alpha_D$.
If axion oscillations are delayed by some means (such that $H_\star \ll \ma$), resonance becomes
more efficient even at $\beta$ much smaller than unity (see
\cref{sec:axion-enhance-narrow-resonance} and also Ref.~\cite{Kitajima:2023pby}).
This regime of narrow resonance thus both increases the efficiency of production and decreases the
degree of backreaction on the dark Higgs.
Conversely, in \cref{sec:slow-roll} we describe a slow-roll regime at very large coupling ($\beta
\gg 50$), which could delay production until $H_\star\sim \mA/\beta$ independent of the axion mass.
Although these dynamics ultimately must be corroborated with numerical simulations (see
\cref{app:axion-fragmentation-time}), it would offer a straightforward and predictive production
mechanism for dark photons at parametrically late times.
We also raised the possibility that if the axion couples to the dark photon in the magnetic basis,
then the axion--dark-photon coupling may naturally be
$\beta \sim \alpha_D^{-1} \gg 1$~\cite{Sokolov:2021eaz, Sokolov:2022fvs, Heidenreich:2023pbi},
correlating the timing of production with the coupling strength: $H_\star \sim \mA\alpha_D$ (see
\cref{fig:final-parameter-space}).
Although this scenario is rather speculative due to the theoretical difficulties outlined in
\cref{sec:slow-roll}, its potential simplicity and near-term detectability make it a compelling
avenue for further study.

Finally, a scalar field responsible for the cosmological variation of the Abelian-Higgs parameters could directly
produce dark photons.
We considered two explicit scenarios: a rapidly varying dark gauge coupling driven by a runaway
scalar~\cite{Cyncynates:2023zwj} and narrow parametric resonance induced by an oscillating
scalar~\cite{Adshead:2023qiw}.
Despite their distinct dynamics, both mechanisms can produce dark photons not just at early times
when $H_\star \gtrsim \mA$ but also---at the cost of fine-tuning (see
\cref{fig:tuned-scalars-parameter-space})---at much later times when $H_\star \ll \mA$.
In the case of a rapidly varying dark gauge coupling (\cref{sec:transverse-kinetic-coupling}), the
scalar not only sources a tachyonic resonance for the dark photon but also exponentially suppresses
its bare mass, removing the kinematic barrier.
The narrow-resonance case (\cref{sec:narrow}), in contrast, relies on a unique instability that
persists to arbitrarily late times provided that the dark-photon--scalar mass ratio is finely tuned
to $1/2$.

While these scalar production scenarios must be tuned to provide an exception to the fiducial
``postinflationary'' line in \cref{fig:final-parameter-space}, they both predict additional
signatures that could test the models.
Delayed production in both cases is accompanied by enhanced dark matter density perturbations on
larger length scales, which could be probed by high-resolution CMB observations, the Lyman-$\alpha$
forest, line intensity mapping, and searches for dark matter substructure.
Should vector dark matter be discovered at smaller masses and larger kinetic mixing, these
complementary signatures could corroborate scalar couplings as the origin of the dark photon.
The converse, however, does not hold: detecting small-scale structures or primordial gravitational
waves (for dilaton-assisted inflationary production, discussed above) of a given size does not
\emph{require} the kinetic mixing to be as large as the corresponding defect formation bounds
permit.

\subsection{Model-angostic considerations}\label{sec:model-agnostic}

A model-independent bound on the kinetic mixing derives from the possibility that the hot Standard
Model plasma restores symmetry to the dark Higgs at sufficiently high temperatures via the kinetic
mixing portal (\cref{sec:thermal}).
If symmetry is restored, strings later form through the Kibble mechanism, precluding stable, cold dark
photon dark matter at late times and leading to the bound of
\cref{eqn:irreducible-thermal-abundance-bound}.
Although this bound is much weaker than the postinflationary production bound
[\cref{eqn:defectBound}] for the lowest possible reheating temperatures, it does imply that
sufficiently light and strongly kinetically mixed dark photons (extending beyond the parameter space
depicted in \cref{fig:final-parameter-space}) are generically accompanied by a cosmic string
network---a potential signature of an Abelian-Higgs dark sector that does not constitute the dark
matter.
Signals from dark matter detected by successors to SuperMAG~\cite{Fedderke:2021aqo},
SNIPE~\cite{Sulai:2023zqw}, and AMAILS~\cite{Jiang:2023jhl}, which probe $\varepsilon \gtrsim 10^{-8}$
at masses below $10^{-12}~\eV$, would require some nondynamical explanation---specifically, a kinetic
mixing that is parametrically larger than the dark gauge coupling.

One possible resolution to this issue lies in the mechanism that generates kinetic mixing.
\Cref{sec:defect-bounds-on-existing-scenarios} assumed that kinetic mixing arises from the exchange
of heavy fermions charged under both electromagnetism and the dark gauge group, as originally
proposed in Ref.~\cite{Holdom:1985ag}.
Modifications to this scenario that allow the kinetic mixing to be parametrically larger than the
dark gauge coupling could, in principle, circumvent the bounds discussed in
\cref{sec:defect-bounds-on-existing-scenarios}.
Abelian clockwork is a well-established candidate mechanism to generate such a
hierarchy~\cite{Giudice:2016yja}.
In \cref{sec:production-independent}, we assessed whether the additional degrees of freedom (i.e.,
Higgses) introduced by clockwork would impose additional constraints on the grounds of backreaction
and defect formation~\cite{East:2022rsi}.
Our analysis indicates that clockwork indeed provides a consistent means to boost kinetic mixing: as
shown in \cref{sec:clockwork-mixing}, the dark photon's mixing with the added clockwork fields is
suppressed by the already-small dark gauge coupling, so it couples no more strongly to any clockwork
Higgs.
\Cref{sec:clockwork-axion} also indicates that any production of the clockwork gauge fields induced
by production of the dark photon itself imposes no stronger backreaction bounds.

While the clockwork mechanism does, in principle, provide an explanation for kinetic mixing larger
than expected on backreaction grounds, such a proposal is rather \textit{ad hoc} since the added
field content plays no role in the origin of the dark matter other than to make it arbitrarily
detectable.
Dark photon production from axions is an exception: clockwork can simultaneously enhance the
axion--dark-photon coupling and the kinetic mixing and indeed is likely required to ensure the
former is large enough for production to occur at all.

Finally, the dark photon's mass could originate from the St\"uckelberg mechanism rather than the
Higgs mechanism we focus on.
Like a Higgs mass, a St\"uckelberg mass in known examples from string theory entails both radial and
angular scalar degrees of freedom~\cite{Reece:2018zvv}.
The angular field is absorbed by the dark photon, while the radial mode remains as a physical degree
of freedom with a mass constrained similar to the dark Higgs, i.e., smaller than about
$4 \pi \mA / \gD$.
The phase of the St\"uckelberg field can also wind and form strings; in contrast to the Higgs case,
the symmetric point where the dark photon mass vanishes lies infinitely far away in field
space~\cite{Reece:2018zvv}.
Even if defects do not form, backreaction onto radial St\"uckelberg modes presumably occurs at the
same parametric threshold as in the Higgs case, suggesting that the low-energy Proca
theory breaks down in similar parameter space.
In this case, the excluded parameter space we report for the Abelian-Higgs case corresponds to
the regime in which the viability of dark photon production with St\"uckelberg masses has not been
established.
We leave a dedicated study of dark photon dark matter with St\"uckelberg masses to future work.

\section{Conclusion}\label{sec:conclusion}

The direct-detection parameter space for kinetically mixed dark photons has long been treated with a
tacit assumption that there exists a consistent cosmology for dark photons of any mass and kinetic
mixing---one that breaks down catastrophically for dark photons that acquire their mass through a
(dark) Higgs mechanism.
In such minimal scenarios, the magnitude of the kinetic mixing is determined by the dark gauge
coupling, which also sets the regime of validity for the low-energy effective theory of a massive
vector field.
Breaching this threshold nucleates a string network that precludes cold dark photon dark
matter~\cite{East:2022rsi}.
In exploring the implications of these constraints, one perspective is to identify the most
well-motivated regions of parameter space by examining concrete, minimal models of dark photon
production.
While the model that requires the fewest new degrees of freedom---gravitational production during
inflation---is entirely undetectable, we showed that simple, UV-inspired physics can extend its
parameter space to the reach of future electron recoil and noble gas detectors.
Additional degrees of freedom, like rolling scalars, can also resonantly produce dark photons after
inflation in parameter space that is broader but still inaccessible to most future haloscope
experiments.
We proposed a number of extensions thereof that, at the cost of some form of tuning, realize
consistent dark photon production in all of the parameter space in reach of these searches.

An alternative approach studies the general implications of direct detection by any particular
upcoming experiment, whose search space is already determined by its design.
Many of these dark photon searches are simply byproducts of experiments specifically designed to
search for axions; much kinetic mixing parameter space will therefore be probed at little added
cost, regardless of theoretical motivation.
Our analysis demonstrates that the full dark photon parameter space can indeed be made viable,
either through dynamical mechanisms or more \emph{ad hoc} model building such as clockwork.
The reach of modified production dynamics, however, is limited by the requirement that dark photons
behave like cold dark matter at observed cosmological epochs, imposing a model-independent threshold
beyond which mechanisms like clockwork are required.
Notably, this regime encompasses almost all existing cosmological constraints, exemplifying the
importance of theoretical input to infer the nature of the dark sector from direct-detection
searches.

One motif of the various nonminimal models of dark photon production we studied is a correlation
between laboratory measurements and cosmological and astrophysical signatures.
Namely, the dynamics that allow for experimentally detectable kinetic mixing also predict unique
phenomenology that could be probed by, for instance, future observations of the cosmic microwave
background or searches for dark matter substructure.
Ultralight dark photon dark matter scenarios also place strong requirements on cosmology more
broadly.
For the dilaton assist model discussed in \cref{sec:universal-coupling}, the dark photon's
kinetic mixing is only detectable if the energy scale of inflation is large enough to be observed
via gravitational waves in the CMB.
On the other hand, even if dark photon dark matter originates by some other mechanism---as it must
if lighter than an meV, regardless of its kinetic mixing---avoiding the nucleation of defects during
inflation places a strong bound on its energy scale [\cref{eqn:minimal-gD-inflation-bound}].
Detectable, postinflationary production scenarios therefore generally require the energy scale of
inflation be too low to be observed by future CMB experiments.
By contrapositive, a detection thereof would place stringent limits on viable dark photon dark
matter scenarios.
The reheating temperature also cannot be too high, lest the Standard Model plasma restore the dark
U(1) symmetry, seeding a string network via the Kibble mechanism.

Among the nonminimal models we studied, the scenarios with the most promising detection prospects
frequently require some degree of parameter tuning.
Nevertheless, the dynamical mechanisms we exploit---and our more model-agnostic analyses---provide a
blueprint for pursuing viable constructions that may be more elegant or motivated.
The potential for additional unique signatures, moreover, offers an avenue to falsify models or
identify the physics underlying the dark sector.
Then again, another reasonable response to the challenges in constructing dark photon models with
large kinetic mixing is to pursue purely gravitational signatures of vector dark matter as a means
of detection~\cite{Adshead:2021kvl, Zhang:2021xxa, Jain:2021pnk, Gorghetto:2022sue, Amin:2022pzv,
Amin:2022nlh, Jain:2023ojg, Jain:2023tsr, Zhang:2024bjo, Liu:2024pjg, Ling:2024qfv, An:2024axz}.
A deeper, synergistic understanding of the interplay between particle models and cosmological
dynamics stands to further advance the theoretical status of detectable dark photon dark matter.

\acknowledgments
We are grateful to Peter Adshead, Isabel Garcia Garcia, Anson Hook, Justin Kaidi, Andrew Long,
Matthew Reece, Katelin Schutz, Olivier Simon, Lorenzo Sorbo, and especially Masha Baryakhtar and
Junwu Huang for discussions that have been invaluable in the preparation of this manuscript.
D.C.\ is supported through the Department of Physics and College of Arts and Science at the
University of Washington and by the U.S. Department of Energy Office of Science under Award No.
DE-SC0024375.
Z.J.W.\ is supported in part by the Department of Physics and College of Arts and Science at the
University of Washington and the Dr. Ann Nelson Endowed Professorship.
D.C.\ is grateful for the hospitality of Perimeter Institute where part of this work was carried
out.
Research at Perimeter Institute is supported in part by the Government of Canada through the
Department of Innovation, Science and Economic Development Canada and by the Province of Ontario
through the Ministry of Colleges and Universities.
This material is partially supported by a grant from the Simons Foundation and the hospitality of
the Aspen Center for Physics.
This research was supported in part by Grant No. NSF PHY-2309135 to the Kavli Institute for Theoretical
Physics (KITP).
Dark photon parameter space limits and projections are compiled in Ref.~\cite{Caputo:2021eaa,
AxionLimits}.

\appendix

\section{Scalar--Abelian-Higgs model}\label{app:SAH}

We first review the standard formulation of the spontaneously broken Abelian-Higgs model, extended
to include arbitrary couplings to a scalar field.
The classical Lagrangian describing a scalar--Abelian-Higgs theory is
\cref{eqn:scalar-abelian-higgs-action-unbroken}, quoted here:
\begin{align}
    \mathcal{L}_\mathrm{SAH}
    &= \frac{1}{2} \partial_\mu \phi \partial^\mu \phi
        - V(\phi)
        - \frac{W(\phi)}{4} F_{\mu\nu} F^{\mu\nu}
        + \frac{X(\phi)}{2} D_\mu \Phi \left( D^\mu \Phi \right)^\ast
        - \frac{\lambda Y(\phi)}{4} \left( \abs{\Phi}^2 - v^2 \right)^2
    \label{eqn:scalar-abelian-higgs-action-unbroken-app}
    .
\end{align}
The Higgs $\Phi$ is a complex scalar that we decompose into a phase and radial displacement from its
vacuum expectation value, $\Phi = \left( v + \higgs \right) e^{\I \goldstone / v}$.
The covariant derivative is $D_\mu \Phi = \partial_\mu \Phi - \I q_\Phi g_D A_\mu \Phi$, with $g_D$
the dark gauge coupling and $q_\Phi$ the (integer-valued) Higgs charge.
In the broken phase and written in terms of radial fluctuations $\higgs$ about the VEV and the
Goldstone boson $\Pi$, \cref{eqn:scalar-abelian-higgs-action-unbroken-app} expands to
\begin{align}
\begin{split}
    \mathcal{L}_\mathrm{SAH}
    &= \frac{1}{2} \partial_\mu \phi \partial^\mu \phi
        - V(\phi)
        - \frac{W(\phi)}{4} F_{\mu\nu} F^{\mu\nu}
        + \frac{X(\phi)}{2} \mA^2 A_\mu A^\mu
    \\ &\hphantom{{}={}}
        + \frac{X(\phi)}{2} \partial_\mu \higgs \partial^\mu \higgs
        - Y(\phi) \left(
            \frac{\lambda}{4} \higgs^4
             + \lambda v \higgs^3
             + \frac{1}{2} m_\higgs^2 \higgs^2
        \right)
    \\ &\hphantom{{}={}}
        + \frac{X(\phi)}{2} \partial_\mu \goldstone \partial^\mu \goldstone
        + \frac{X(\phi)}{2} \left( \frac{\higgs^2}{v^2} + \frac{2 \higgs}{v} \right)
        \left( \partial_\mu \goldstone - \mA A_\mu \right)
        \left( \partial^\mu \goldstone - \mA A^\mu \right),
    \label{eqn:scalar-abelian-higgs-action-expanded}
\end{split}
\end{align}
where $\mA = q_\Phi g_D v$ and $m_\higgs = \sqrt{2 \lambda} v$.

We fix unitary gauge, in which $\goldstone = 0$.
In a general spacetime, the Euler-Lagrange equation for the vector is
\begin{align}
    - \nabla_\mu \left[
        W(\phi) F^{\mu \nu}
    \right]
    &= X(\phi) (1 + h / v)^2 \mA^2 A^\nu
    \equiv \mathcal{X}(\phi, h) \mA^2 A^\nu
    .
    \label{eqn:vector-ele-ito-field-tensor}
\end{align}
The covariant divergence thereof sets
\begin{align}
    0
    &= \nabla^\alpha \left[ \mathcal{X}(\phi, h) A_\alpha \right]
    ,
    \label{eqn:covariant-lorenz}
\end{align}
which for $\mathcal{X}(\phi, h) = 1$ reduces to the Lorenz ``gauge'' condition
$\nabla^\alpha A_\alpha = 0$ as applicable for massive vectors (i.e., Proca fields).
The equations of motion are therefore
\begin{subequations}\label{eqn:general-eoms}
\begin{align}
\begin{split}
    0
    &= \nabla_\mu \nabla^\mu \phi
        + V'(\phi)
        + \frac{W'(\phi)}{4} F_{\mu \nu} F^{\mu \nu}
        - \frac{X'(\phi)}{2} \mA^2 A_\mu A^\mu
    \\ &\hphantom{ {}={} }
        - \frac{X'(\phi)}{2} \nabla_\mu \higgs \nabla^\mu \higgs
        + \frac{\lambda Y'(\phi)}{4} \left( \higgs^2 + 2 \higgs v \right)^2
        \label{eqn:scalar-ele}
\end{split}
    \\
    0
    &= \nabla_\mu \nabla^\mu \higgs
        + \frac{X'(\phi)}{X(\phi)} \nabla_\mu \phi \nabla^\mu \higgs
        + \frac{Y(\phi)}{X(\phi)}
        \left[ \lambda \higgs^3 + 3 \lambda v \higgs^2 + m_\higgs^2 \higgs \right]
        + \left( 1 + \higgs / v \right) \mA^2 A_\mu A^\mu / v
        \label{eqn:higgs-ele}
    \\
    0
    &= \nabla^\alpha \nabla_\alpha A_\beta
        - R_{\sigma \beta} A^\sigma
        + \nabla_\nu \left[ A_\mu \nabla^\mu \ln \mathcal{X}(\phi, h) \right]
        + \frac{\nabla^\alpha W(\phi)}{W(\phi)} F_{\alpha\beta}
        + \frac{\mathcal{X}(\phi, h)}{W(\phi)} \mA^2 A_\beta
        \label{eqn:vector-ele}
    ,
\end{align}
\end{subequations}
where $R_{\sigma \beta}$ is the Ricci tensor and the Higgs mass is
$m_\higgs = \sqrt{2 \lambda} v$.

As in the main text, we restrict our discussion to the FLRW spacetime with metric
\begin{align}\label{eqn:flrw-metric}
    \ud s^2
    &= \ud t^2
        - a(t)^2 \delta_{i j} \ud x^i \ud x^j
\end{align}
and take the scalar to be homogeneous: $\phi(t, \three{x}) = \bar{\phi}(t)$.
The form of \cref{eqn:general-eoms} motivates absorbing the
$\bar{\phi}$ dependence of the theory into its fundamental parameters as in
\cref{eqn:phi-dependent-parameters}.
We expand the Higgs and vector components in Fourier space as
\begin{subequations}\label{eqn:fourier-expansions}
\begin{align}
    \higgs(t, \three{x})
    &= \bar{\higgs}(t)
        + \int \frac{\ud^3 k}{(2\pi)^3}
        \delta \higgs(t, \three{k})
        e^{\I \three{k} \cdot \three{x}}
    \label{eqn:higgs-fourier-expansion}
    \\
    A_0(t, \three{x})
    &= \int \frac{\ud^3 k}{(2\pi)^3}
        A_{0}(t, \three{k})
        e^{\I \three{k} \cdot \three{x}}
    \label{eqn:A0-fourier-expansion}
    \\
    A_i(t, \three{x})
    &= \sum_{\lambda}^{\{\pm, \parallel\}} \int \frac{\ud^3 k}{(2\pi)^3}
        A_{\lambda}(t, \three{k})
        \varepsilon^\lambda_i(\three{k})
        e^{\I \three{k} \cdot \three{x}},
    \label{eqn:Ai-polarization-fourier-expansion}
\end{align}
\end{subequations}
where the polarization vectors $\bm{\varepsilon}^\pm(\three{k})$ and
$\bm{\varepsilon}^\parallel(\three{k})$ form an orthogonal basis of polarizations
[i.e., $\varepsilon^\lambda_m(\three{k}) \varepsilon^{\lambda'}_m(\three{k})^\ast = \delta^{\lambda \lambda'}$]
that are respectively transverse
[$k_m \varepsilon^\pm_m(\three{k})$] and longitudinal
[$\I k_m \varepsilon^\parallel_m(\three{k}) = k$] to the
wave number $\three{k}$.
They are Hermitian [$\varepsilon^\lambda_m(-\three{k}) = \varepsilon^\lambda_m(\three{k})^\ast$]
and the circular polarization vectors satisfy
$\varepsilon^\pm_m(\three{k})^\ast = \varepsilon^\mp_m(\three{k})$.
The linearized equations of motion decomposed onto this basis (and written in terms of the
$\bar{\phi}$-dependent parameters) evaluate to
\begin{subequations}\label{eqn:linearized-eoms-app}
\begin{align}
    0
    &= \ddot{\bar{\phi}}
        + 3 H \dot{\bar{\phi}}
        + V'(\bar{\phi})
    \\
    0
    &= \ddot{\bar{\higgs}}
        + \left[
            3 H
            + \frac{\partial_t X(\bar{\phi})}{X(\bar{\phi})}
        \right]
        \dot{\bar{\higgs}}
        + \left[
            X(\bar{\phi}) \lambda(\bar{\phi}) \bar{\higgs}^2
            + 3 \sqrt{X(\bar{\phi})} \lambda(\bar{\phi}) v(\bar{\phi}) \bar{\higgs}
            + m_\higgs(\bar{\phi})^2
        \right]
        \higgs
    \label{eqn:higgs-eom-linearized}
    \\
    0
    &= \delta \ddot{\higgs}
        + \left[
            3 H
            + \frac{\partial_t X(\bar{\phi})}{X(\bar{\phi})}
        \right]
        \delta \dot{\higgs}
        + \left[
            \frac{k^2}{a^{2}}
            + 3 X(\bar{\phi}) \lambda(\bar{\phi}) \bar{\higgs}^2
            + 6 \sqrt{X(\bar{\phi})} \lambda(\bar{\phi}) v(\bar{\phi}) \bar{\higgs}
            + m_\higgs(\bar{\phi})^2
        \right]
        \delta \higgs
    \label{eqn:delta-higgs-eom-linearized}
    \\
    0
    &= \ddot{A}_\pm
        + \left[
            H
            + \frac{\partial_t W(\bar{\phi})}{W(\bar{\phi})}
        \right]
        \dot{A}_\pm
        + \left[
            \frac{k^2}{a^{2}}
            + \mA(\bar{\phi}, \bar{h})^2
        \right]
        A_\pm
    \label{eqn:Ai-eom-linearized-transverse}
    \\
\begin{split}
    0
    &= \ddot{A}_\parallel
        + \left[
            H
            + \frac{\partial_t W(\bar{\phi})}{W(\bar{\phi})}
        \right]
        \dot{A}_\parallel
        + \left[
            \frac{k^2}{a^{2}}
            + \mA(\bar{\phi}, \bar{h})^2
        \right]
        A_\parallel
        - \left[
            2 H
            + \frac{\partial_t \mathcal{X}(\bar{\phi}, \bar{h})}{\mathcal{X}(\bar{\phi}, \bar{h})}
            - \frac{\partial_t W(\bar{\phi})}{W(\bar{\phi})}
        \right]
        k A_0.
    \label{eqn:Ai-eom-linearized-longitudinal}
\end{split}
\end{align}
\end{subequations}
Here we use the shorthand $\mA(\phi, h) = \mA(\phi) (1 + \higgs / v)^2$.
In addition, \cref{eqn:covariant-lorenz} reduces to
\begin{align}
    0
    &= - \dot{A}_0
        - \left(
            3 H
            + \frac{\partial_t \mathcal{X}(\bar{\phi}, \bar{h})}{\mathcal{X}(\bar{\phi}, \bar{h})}
        \right)
        A_0
        + \frac{\partial_i A_i}{a^{2}}
    \label{eqn:lorenz-flrw}
\end{align}
and the $\nu = 0$ component of \cref{eqn:vector-ele-ito-field-tensor} (i.e., Gauss's law) reads
\begin{align}
    0
    &= k \dot{A}_\parallel
        + \left[
            k^2
            + a^2 \mA(\bar{\phi}, \bar{h})
        \right]
        A_0;
    \label{eqn:gauss-law-linearized-longitudinal}
\end{align}
solving \cref{eqn:gauss-law-linearized-longitudinal} for $A_0$ and substituting into
\cref{eqn:Ai-eom-linearized-longitudinal} yields
\begin{align}
    0
    &= \ddot{A}_\parallel
        + \frac{
            \left[
                3 H
                + \partial_t \ln \mathcal{X}(\bar{\phi}, \bar{h})
            \right]
            k^2
            + \left[
                H
                + \partial_t \ln W(\bar{\phi})
            \right]
            a^2 \mA(\bar{\phi}, \bar{h})^2
        }{
            k^2 + a^2 \mA(\bar{\phi}, \bar{h})^2
        }
        \dot{A}_\parallel
        + \left[
            \frac{k^2}{a^{2}}
            + \mA(\bar{\phi}, \bar{h})^2
        \right]
        A_\parallel.
\end{align}
For the rescaled fields $\rhiggs = \sqrt{X(\bar{\phi})} \higgs$
and $\mathcal{A}_\pm = \sqrt{W(\bar{\phi})} A_\pm$, the equations of motion reduce to
\begin{align}
    0
    &= \ddot{\bar{\rhiggs}}
        + 3 H \dot{\bar{\rhiggs}}
        + \left[
            \lambda(\bar{\phi}) \bar{\rhiggs}^2
            + 3 \lambda(\bar{\phi}) v(\bar{\phi}) \bar{\rhiggs}
            + m_\higgs(\bar{\phi})^2
            - \frac{3 H}{2} \frac{\partial_t X(\bar{\phi})}{X(\bar{\phi})}
            - \frac{\partial_t^2 \sqrt{X(\bar{\phi})}}{\sqrt{X(\bar{\phi})}}
        \right]
        \bar{\rhiggs}
    \\
    0
    &= \delta \ddot{\rhiggs}
        + 3 H \delta \dot{\rhiggs}
        + \left[
            \frac{k^2}{a^{2}}
            + 3 \lambda(\bar{\phi}) \bar{\rhiggs}^2
            + 6 \lambda(\bar{\phi}) v(\bar{\phi}) \bar{\rhiggs}
            + m_\higgs(\bar{\phi})^2
            - \frac{3 H}{2} \frac{\partial_t X(\bar{\phi})}{X(\bar{\phi})}
            - \frac{\partial_t^2 \sqrt{X(\bar{\phi})}}{\sqrt{X(\bar{\phi})}}
        \right]
        \delta \rhiggs
    \\
    0
    &= \ddot{\mathcal{A}}_\pm
        + H \dot{\mathcal{A}}_\pm
        +  \left[
            \frac{k^2}{a^{2}}
            + \mA(\bar{\phi}, \bar{h})^2
            - \frac{H}{2}
            \frac{\partial_t W(\bar{\phi})}{W(\bar{\phi})}
            - \frac{\partial_t^2 \sqrt{W(\bar{\phi})}}{\sqrt{W(\bar{\phi})}}
        \right]
        \mathcal{A}_\pm
    .
\end{align}
Observe that the only couplings to $\bar{\phi}$ not embedded in the parameters defined in
\cref{eqn:phi-dependent-parameters} are derivative couplings.

The stress-energy tensor for \cref{eqn:scalar-abelian-higgs-action-expanded} (in unitary gauge) is
\begin{align}
\begin{split}
    T^{\mu}_{\hphantom{\mu} \nu}
    &= W(\phi) \left[
            F^{\mu \alpha} F_{\nu \alpha}
            - \mA(\phi, h)^2 A^\mu A_\nu
            + \delta^{\mu}_{\hphantom{\mu} \nu} \left(
                - \frac{1}{4} F_{\alpha \beta} F^{\alpha \beta}
                + \frac{1}{2} \mA(\phi, h)^2 A_\alpha A^\alpha
            \right)
        \right]
    \\ &\hphantom{ {}={} }
        + X(\phi) \left[
            - \partial^\mu \higgs \partial_\nu \higgs
            + \delta^{\mu}_{\hphantom{\mu} \nu} \left(
                \frac{1}{2} \partial_\alpha \higgs \partial^\alpha \higgs
                - \frac{Y(\phi)}{X(\phi)} \left[
                    \frac{\lambda}{4} \higgs^4
                     + \lambda v \higgs^3
                     + \frac{1}{2} m_\higgs^2 \higgs^2
                \right]
            \right)
        \right]
    \\ &\hphantom{ {}={} }
        - \partial^\mu \phi \partial_\nu \phi
            + \delta^{\mu}_{\hphantom{\mu} \nu} \left(
                \frac{1}{2} \partial^\alpha \phi \partial_\alpha \phi
                - V(\phi)
            \right).
\end{split}
\end{align}
Using \cref{eqn:gauss-law-linearized-longitudinal} to solve for $A_0$, we may express the dark photon's
average energy density $\bar{\rho}_A$ and pressure $\bar{P}_A$ in terms of the polarization decomposition
\cref{eqn:Ai-polarization-fourier-expansion}.
To the same level of approximation as \cref{eqn:linearized-eoms-app} (and assuming isotropic initial conditions),
the contributions from the transverse and longitudinal components to the energy density $\bar{\rho}_A$ are respectively
\begin{subequations}\label{eqn:rho-perp-parallel}
\begin{align}
    \bar{\rho}_\pm(t)
    &= \frac{W(\bar{\phi})}{4 \pi^2 a^2}
        \int \ud k \, k^2
        \left[
            \abs{ \dot{A}_\pm }^2
            + \frac{k^2}{a^2} \abs{A_\pm}^2
            + \mA(\bar{\phi}, \bar{h})^2 \abs{ A_\pm }^2
        \right]
\intertext{and}
    \bar{\rho}_\parallel(t)
    &= \frac{W(\bar{\phi}) \mA(\bar{\phi}, \bar{h})^2}{4 \pi^2 a^2}
        \int \ud k \, k^2
        \left[
            \frac{1}{k^2 + a^2 \mA(\bar{\phi}, \bar{h})^2}
            \abs{ a \dot{A}_\parallel }^2
            + \abs{ A_\parallel }^2
        \right],
\end{align}
\end{subequations}
and those to the pressure $P_A$ are
\begin{subequations}
\begin{align}
    \bar{P}_\pm(t)
    &= \frac{W(\bar{\phi})}{12 \pi^2 a^2}
        \int \ud k \, k^2
        \left[
            \abs{ \dot{A}_\pm(\tau, k) }^2
            + \frac{k^2}{a^2} \abs{A_\pm(\tau, k)}^2
            - \mA(\bar{\phi}, \bar{h})^2 \abs{ A_\pm(\tau, k) }^2
        \right]
\intertext{and}
    \bar{P}_\parallel(t)
    &= \frac{W(\bar{\phi}) \mA(\bar{\phi}, \bar{h})^2}{12 \pi^2 a^2}
        \int \ud k \, k^2
        \left[
            \frac{3 k^2 + a^2 \mA(\bar{\phi}, \bar{h})^2}{\left[ k^2 + a^2 \mA(\bar{\phi}, \bar{h})^2 \right]^2}
            \abs{ a \dot{A}_\parallel(\tau, k) }^2
            - \abs{ A_\parallel(\tau, k) }^2
        \right].
\end{align}
\end{subequations}

\section{Quantum corrections}\label{app:quantum-corrections}

We seek to compute the one-loop quantum corrections to the theory of a scalar coupled to the
standard Abelian-Higgs model, \cref{eqn:scalar-abelian-higgs-action-unbroken}.
The quantization of theories with noncanonical kinetic terms entails subtleties whose neglect leads
to incorrect scattering amplitudes and can, in our case, severely alter the ultraviolet (UV)
divergences of radiative corrections.
We first review the appropriate formulation before proceeding to compute one-loop effects in the
scalar--Abelian-Higgs model.

\subsection{Quantization of noncanonical field theories}\label{sec:quantization-noncanonical}

To compute quantum corrections, we require the Hamiltonian $\mathcal{H}$ (specifically as a
functional of fields and their four-gradients and \textit{not} the conjugate momenta as well)
corresponding to \cref{eqn:scalar-abelian-higgs-action-expanded}, which generates the time evolution
and interactions of the theory.
In classical field theory, the Hamiltonian and Lagrangian are related by a Legendre transform, and the only complication in
theories with noncanonical kinetic terms is that fields' conjugate momenta are not trivially related
to their time derivatives.
In quantum field theory, the appropriate Hamiltonian coincides with the Legendre transform of the
Lagrangian \textit{only} if the kinetic terms are canonical.

Deriving the effect of kinetic couplings is most straightforward in the path-integral
formalism (see Ref.~\cite{Weinberg:1995mt} section 9.3, whose notation we partly adopt below), in which the
generating functional for a general multiplet of bosons $\phi_a$ with conjugate momenta $\pi_a$ is
\begin{align}
    \mathcal{Z}
    &= \int \prod_a \left[ \mathcal{D} \phi_a \frac{\mathcal{D} \pi_a}{2 \pi} \right]
        \exp \left\{
            \frac{\I}{\hbar} \int \ud^4 x \,
            \left[
                \sum_a \dot{\phi}_a(x) \pi_a(x)
                - \mathcal{H}[\phi_a(x), \pi_a(x)]
            \right]
        \right\}.
\end{align}
For ease of illustration and generality, parametrize the Hamiltonian as
\begin{align}
    \mathcal{H}[\phi_a, \pi_a]
    &= \frac{1}{2} \sum_{a, b} \pi_a A_{a b}[\phi] \pi_b
        + \sum_a B_a[\phi] \pi_a
        + C[\phi].
    \label{eqn:general-hamiltonian}
\end{align}
In a canonical theory, $A_{a b}[\phi] = \delta_{a b}$.
The functional $C[\phi]$ includes (spatial) gradient and potential terms for a scalar theory and
mass, current-coupling, and magnetic-field terms for a gauge theory.
The salient effects we aim to illustrate depend only on $A_{a b}[\phi]$; we retain a general
parametrization for succinctness and provide an explicit mapping of \cref{eqn:general-hamiltonian}
to the scalar--Abelian-Higgs theory later on.

Because the kinetic term in \cref{eqn:general-hamiltonian} is still quadratic in the conjugate
momenta, the path integral over $\pi_a$ may be computed exactly.
The argument of the exponential in the path integral indeed coincides with the
(inverse) Legendre transform of the Hamiltonian \cref{eqn:general-hamiltonian}, as a consequence of
the integration over $\pi_a$ instead of Hamilton's equations~\cite{Weinberg:1995mt}:
\begin{align}
    \mathcal{Z}
    &= \int \prod_a \left[ \mathcal{D} \phi_a \right]
        \det \left( 2 \pi \I A[\phi] / \hbar \right)^{-1/2}
        \exp \left\{
            \frac{\I}{\hbar} \int \ud^4 x \,
            \mathcal{L}[\phi, \dot{\phi}]
        \right\},
    \label{eqn:generating-functional}
\end{align}
where
\begin{align}
    \mathcal{L}[\phi_a, \dot{\phi}_a]
    &= \sum_{a, b}
        \left[
            \frac{1}{2}
            \dot{\phi}_a
            A[\phi]^{-1}_{a b} \dot{\phi}_b
            + \frac{1}{2}
            B_a[\phi]
            A[\phi]^{-1}_{a b} B_b[\phi]
            - B_a[\phi] A[\phi]^{-1}_{a b} \dot{\phi}_b
        \right]
        - C[\phi].
    \label{eqn:lagrangian-from-path-integral}
\end{align}
While \cref{eqn:lagrangian-from-path-integral} is superficially identical to the corresponding
classical result, any field dependence in the determinant factor of \cref{eqn:generating-functional}
(via the kinetic term in $A[\phi]$) \textit{does} modify the path integral, effectively introducing
to the Lagrangian an additional term
\begin{align}
    \Delta \mathcal{L}
    &= \frac{\I \hbar}{2} \int \frac{\ud^4 k}{(2 \pi)^4} \trace \ln A[\phi].
    \label{eqn:delta-L-from-noncanonical-kinetic-term}
\end{align}

Instead of performing the path integral over $\pi_a$, we could shift the conjugate momentum to
\begin{align}
    \pi_a'
    &= \pi_a - A[\phi]^{-1}_{a b} \left( B_b[\phi] - \dot{\phi}_b \right)
\end{align}
and retain it as an auxiliary degree of freedom in the Lagrangian that interacts with $\phi$:
\begin{align}
    \mathcal{L}[\phi_a, \dot{\phi}_a, \pi_a']
    &= - \frac{1}{2} \sum_{a, b} A_{a b}[\phi] \pi_a' \pi_b'
        + \mathcal{L}[\phi_a, \dot{\phi}_a].
    \label{eqn:delta-L-from-noncanonical-kinetic-term-auxiliary}
\end{align}
Though $\pi_a'$ is nonpropagating, it modifies correlation functions of $\phi_a$ through loops.
That the leading contributions from the new terms in the Lagrangian should be
counted at one-loop level is less obvious in the form
\cref{eqn:delta-L-from-noncanonical-kinetic-term}, but is affirmed by the relative factor of $\hbar$
acquired by $\Delta \mathcal{L}$ when the determinant factor is reabsorbed into the
Lagrangian.\footnote{
    Related observations were made in previous studies of noncanonical field theories: the first
    exposition on the additional interactions arising in noncanonical field
    theories~\cite{Gerstein:1971fm} found in chiral perturbation theory that their contribution to
    the pion propagator appears at next-to-leading order in the pion decay constant, and when
    considering loop corrections during inflation their contribution is suppressed in the slow-roll
    expansion~\cite{Seery:2007we, Adshead:2008gk}.
}

We now map the preceding general results to a massive vector field kinetically coupled to a scalar
field.
We replace $\phi_a \to A_i$ and $\pi_a \to E_i$.
Recovering the covariant Lagrangian
\begin{align}
    \mathcal{L}
    &= - \frac{W}{4} F_{\mu \nu} F^{\mu \nu}
        + \frac{1}{2} \mA^2 A_\mu A^\mu
        + J_\mu A^\mu
\end{align}
(with $J_\mu$ standing in for the Higgs interactions) requires choosing
\begin{subequations}
\begin{align}
    A_{i j}
    &= W^{-1} \delta_{i j}
    \\
    B_i
    &= \partial_i A_0
    \\
    C
    &= \frac{1}{2} \mA^2 \left( A_i \right)^2
        + \frac{W}{2}
        \left( \epsilon^{i j k} \partial_j A_k \right)^2
        + J_i A_i
\end{align}
\end{subequations}
in \cref{eqn:general-hamiltonian} and, just as for a canonical vector field, introducing $A_0$ as an
auxiliary field (also integrated over in $\mathcal{Z}$) and adding to the Hamiltonian
\begin{align}
    \Delta \mathcal{H}
    &= - \frac{1}{2} \mA^2
        \left( A_0 - \frac{\partial_i E_i - J_0}{\mA^2} \right)^2.
\end{align}
The Higgs and the Goldstone, whose kinetic terms in the Lagrangian are multiplied by $X$, each
contribute a copy of \cref{eqn:delta-L-from-noncanonical-kinetic-term} with $A$ set to $X^{-1}$.
The effective Lagrangian therefore includes the contribution
\begin{align}
    \Delta \mathcal{L}
    &= - \frac{3 \I \hbar}{2} \int \frac{\ud^4 k}{(2 \pi)^4} \ln W
        - \I \hbar \int \frac{\ud^4 k}{(2 \pi)^4} \ln X
    \label{eqn:delta-L-for-vector}
\end{align}
in addition to the classical Lagrangian.

\subsection{Background-field expansion}

With the full quantum theory in tow, we may compute radiative corrections.
We regulate with a finite momentum cutoff $\Lambda$ under the usual interpretation that it
corresponds to the scale of new physics in a given UV completion of the effective theory described
by \cref{eqn:scalar-abelian-higgs-action-expanded}.
We seek to assess what limitations on $\Lambda$ are imposed to ensure consistency of the scalar
production scenario---principally, avoiding large corrections to the scalar's classical potential.
Since we consider scalar field configurations with classical homogeneous components, we perform a
standard background-field expansion to compute the effective potential due to quantum fluctuations.

We first expand $\phi \to f \varphi_b + \phi$ in \cref{eqn:scalar-abelian-higgs-action-expanded},
where $\varphi_b = \phi_b / f$ is treated as a spacetime constant.
We then drop constant terms and tadpoles (i.e., terms with fewer than two powers of the fluctuating fields).
We gauge fix according to the standard Faddeev-Popov procedure, choosing the gauge fixing functional
\begin{align}
    G(A_\mu, \goldstone)
    &= \sqrt{X(\phi)} \left(
            \partial_\mu A^\mu - \xi \mA \goldstone
        \right);
\end{align}
the factor of $\sqrt{X(\phi)}$ introduced relative to the textbook choice ensures that terms that
mix $A_\mu$ and $\goldstone$ in the quadratic action still cancel when
$- G(A_\mu, \goldstone)^2 / 2 \xi$ is added to the action.
The Faddeev-Popov determinant also picks up this factor of $\sqrt{X(\phi)}$, which we account for by
including it in the action for the ghosts $c$ and $\bar{c}$.
Gauge fixing then adds to the Lagrangian
\begin{align}
    \mathcal{L}_\mathrm{FP}
    &= - \frac{X(\phi)}{2 \xi} \left(
            \partial_\mu A^\mu - \xi \mA \goldstone
        \right)^2
        + \sqrt{X(\phi)}
        \left(
            \partial_\mu \bar{c} \partial^\mu c
            - \xi \mA^2 c \bar{c}
        \right).
\end{align}
Finally, we canonically
normalize\footnote{
    ``Canonically normalize'' in the sense that propagators take their standard
    form---no interactions are altered by doing so, only the point at which we rearrange factors of
    $W(\varphi_b)$ and $X(\varphi_b)$ to write results in canonical form.
} fields by replacing $A_\mu$ with $A_\mu / \sqrt{W(\varphi_b)}$, $\higgs$ with
$\higgs / \sqrt{X(\varphi_b)}$, $\goldstone$ with $\goldstone / \sqrt{X(\varphi_b)}$, and
$c$ with $c / \sqrt[4]{X(\varphi_b)}$.
To compute the effective potential we require only the quadratic action (including background-field
dependence in effective masses).
The part of $\mathcal{L}_\mathrm{SAH} + \mathcal{L}_\mathrm{FP}$ quadratic in fields is
\begin{align}
    \mathcal{L}_{(2)}
    &= - \frac{1}{4} F_{\mu \nu} F^{\mu \nu}
        - \frac{1}{2 \xi(\varphi_b)} \left( \partial_\mu A^\mu \right)^2
        + \frac{1}{2} \mA(\varphi_b)^2 A_\mu A^\mu
        + \frac{1}{2} \partial_\mu \phi \partial^\mu \phi
        - \frac{1}{2} m(\varphi_b)^2 \phi^2
    \\ &\hphantom{{}={}}
        + \frac{1}{2} \partial_\mu \higgs \partial^\mu \higgs
        - \frac{1}{2} m_\higgs(\varphi_b)^2 \higgs^2
        + \frac{1}{2} \partial_\mu \goldstone \partial^\mu \goldstone
        - \frac{\xi \mA^2}{2} \goldstone^2
        + \partial_\mu \bar{c} \partial^\mu c
        - \xi \mA^2 c \bar{c},
    \label{eqn:scalar-abelian-higgs-quadratic-lagrangian}
\end{align}
where $\mA(\varphi_b)$ and $m_\higgs(\varphi_b)$ are defined in analogy to
\cref{eqn:phi-dependent-parameters} and
\begin{align}
    m(\varphi_b)^2
    &\equiv V''(f \varphi_b) / f^2
    \label{eqn:m-varphi-b-def}
    \\
    \xi(\varphi_b)
    &\equiv \xi W(\varphi_b) / X(\varphi_b).
    \label{eqn:xi-varphi-b-def}
\end{align}

For completeness, we also report the interaction Lagrangian (in unitary gauge,
$\xi \to \infty$, so that we can neglect the Goldstone boson and ghosts), as would be needed to
compute radiative corrections:
\begin{align}
\begin{split}
    \mathcal{L}_\mathrm{int}
    &= - \frac{1}{4} F_{\mu \nu} F^{\mu \nu}
        \sum_{n = 1}^\infty \frac{1}{n!} \frac{W^{(n)}(\varphi_b)}{W(\varphi_b)} \frac{\phi^n}{f^n}
        + \left[
            \frac{1}{2} \partial_\mu \higgs \partial^\mu \higgs
            + \frac{1}{2} \mA(\varphi_b)^2 A_\mu A^\mu
        \right]
        \sum_{n = 1}^\infty \frac{1}{n!} \frac{X^{(n)}(\varphi_b)}{X(\varphi_b)} \frac{\phi^n}{f^n}
    \\ &\hphantom{{}={}}
        + \mA(\varphi_b)^2 \left(
            \frac{\higgs}{v(\varphi_b)}
            + \frac{\higgs^2}{2 v(\varphi_b)^2}
        \right)
        A_\mu A^\mu
        \sum_{n = 0}^\infty \frac{1}{n!} \frac{X^{(n)}(\varphi_b)}{X(\varphi_b)} \frac{\phi^n}{f^n}
        - \sum_{n = 3}^\infty \frac{\tilde{V}^{(n)}(\varphi_b)}{n! f^n} \phi^n
    \\ &\hphantom{{}={}}
        - \frac{1}{2} m_\higgs(\varphi_b)^2 \higgs^2
        \sum_{n = 1}^\infty \frac{1}{n!} \frac{Y^{(n)}(\varphi_b)}{Y(\varphi_b)} \frac{\phi^n}{f^n}
        - \left[
            \frac{\lambda(\varphi_b)}{4} \higgs^4
            + \lambda(\varphi_b) v(\varphi_b) \higgs^3
        \right]
        \sum_{n = 0}^\infty \frac{1}{n!} \frac{Y^{(n)}(\varphi_b)}{Y(\varphi_b)} \frac{\phi^n}{f^n}
        .
    \label{eqn:scalar-abelian-higgs-interaction-lagrangian}
\end{split}
\end{align}
The additional interactions incurred in the quantum theory [\cref{eqn:delta-L-for-vector}] expand to
\begin{align}
    \Delta \mathcal{L}
    &= - \frac{\I}{2} \int \frac{\ud^4 k}{(2 \pi)^4}
        \sum_{n = 0}^\infty
        \frac{1}{n!}
        \left( \frac{\phi}{f} \right)^n
        \left[
            3 \frac{\partial^{n} \ln W(\varphi)}{\partial \varphi^n}
            + 2 \frac{\partial^{n} \ln X(\varphi)}{\partial \varphi^n}
        \right]_{\varphi = \varphi_b}
        .
\end{align}
Both the $\phi$-independent term and the quadratic term are required explicitly for our purposes;
they are
\begin{align}
    \Delta \mathcal{L}_{(0)}
    &= - \frac{\I}{2} \int \frac{\ud^4 k}{(2 \pi)^4}
        \left[ 3 \ln W(\varphi_b) + 2 \ln X(\varphi_b) \right]
    \label{eqn:measure-contribution-to-lagrangian-constant-term}
    \\
\begin{split}
    \Delta \mathcal{L}_{(2)}
    &= - \frac{3 \I}{4 f^2}
        \int \frac{\ud^4 k}{(2 \pi)^4}
        \left[
            \frac{W^{(2)}(\varphi_b)}{W(\varphi_b)}
            - \left(
                \frac{W^{(1)}(\varphi_b)}{W(\varphi_b)}
            \right)^2
        \right]
        \phi^2
    \\ &\hphantom{{}={}}
        - \frac{\I}{2 f^2}
        \int \frac{\ud^4 k}{(2 \pi)^4}
        \left[
            \frac{X^{(2)}(\varphi_b)}{X(\varphi_b)}
            - \left(
                \frac{X^{(1)}(\varphi_b)}{X(\varphi_b)}
            \right)^2
        \right]
        \phi^2
    .
    \label{eqn:measure-contribution-to-lagrangian-mass-term}
\end{split}
\end{align}
The coefficient of $- \phi^2 / 2$ is an effective one-loop contribution to the scalar's squared
mass.
In practice, \cref{eqn:measure-contribution-to-lagrangian-constant-term,eqn:measure-contribution-to-lagrangian-mass-term}
act like counterterms to photon and Higgs loops and cancel leading divergences.

\subsection{Effective potential}\label{sec:effective-potential}

The one-loop effective potential may be computed by integrating the quadratic action over the
fields, \cref{eqn:scalar-abelian-higgs-quadratic-lagrangian}~\cite{Jackiw:1974cv}.
The contribution from scalar fluctuations is
\begin{align}
    e^{\I \Delta \Gamma[\phi_b]^{(1), \phi}}
    &= \int \mathcal{D} \phi
        \exp \left\{
            \I
            \int \ud^4 x \,
            \left[
                \frac{1}{2} \partial_\mu \phi \partial^\mu \phi
                - \frac{1}{2} m(\varphi_b)^2 \phi^2
            \right]
        \right\},
\end{align}
which, integrated over $\phi$, adds to the effective action
\begin{align}
    \I \Delta \Gamma[\phi_b]^{(1), \phi}
    &= - \frac{1}{2}
        \int \ud^4 x
        \int \frac{\ud^4 k}{(2 \pi)^4}
        \ln \left( 1 - \frac{m(\varphi_b)^2}{k^2} \right).
\end{align}
At next-to-leading order in $\Lambda \gg m(\varphi_b)$, the resulting effective potential is
\begin{align}
    \Delta V[\phi_b]^{(1), \phi}
    &= \frac{\Lambda^2 m(\varphi_b)^2}{2 (4 \pi)^2}
        - \frac{m(\varphi_b)^4}{8 (4 \pi)^2}
        \left[ 1 + 2 \ln \left( 1 + \frac{\Lambda^2}{m(\varphi_b)^2} \right) \right],
    \label{eqn:one-loop-effective-potential-scalar}
\end{align}
which is subdominant to the bare potential when $\Lambda^2 / 2 (4 \pi f)^2 < 1$.
In the special case that $V(\varphi_b)$ is a pure exponential function, the quantum corrections to
the scalar effective potential arising from scalar loops organize into an expansion in the number of
vertices~\cite{Garny:2006wc}.
At one-vertex level, the corrections are directly proportional to $V(\varphi_b)$ and can therefore
be absorbed into the overall normalization of the potential.

The Higgs contribution is
\begin{align}
    e^{\I \Delta \Gamma[\phi_b]^{(1), \higgs}}
    &= \int \frac{\mathcal{D} \higgs}{\sqrt{X(\varphi_b)}}
        \exp \left\{
            \I \int \ud^4 x \,
            \left[
                \frac{1}{2} \partial_\mu \higgs \partial^\mu \higgs
                - \frac{1}{2} m_\higgs(\varphi_b)^2 \higgs^2
            \right]
        \right\}
\end{align}
Observe that the measure is rescaled in accordance with the field redefinition made previously.
The resulting effective potential contribution is
\begin{align}
    \Delta V[\phi_b]^{(1), \higgs}
    &= - \frac{\I}{2}
        \int \frac{\ud^4 k}{(2 \pi)^4}
        \left[
            \ln X(\varphi_b)
            + \ln \left( 1 - \frac{m_\higgs(\varphi_b)^2}{k^2} \right)
        \right].
    \label{eqn:higgs-one-loop-effective-potential}
\end{align}
The calculation for the Goldstone proceeds identically:
\begin{align}
    \Delta V[\phi_b]^{(1), \goldstone}
    &= - \frac{\I}{2}
        \int \frac{\ud^4 k}{(2 \pi)^4}
        \left[
            \ln X(\varphi_b)
            + \ln \left( 1 - \frac{\xi \mA^2}{k^2} \right)
        \right].
    \label{eqn:goldstone-one-loop-effective-potential}
\end{align}

Recall that the ghosts were rescaled by a factor of $\sqrt[4]{X(\varphi_b)}$; however, as Grassmann
numbers their measure changes by the inverse of the applied scaling transformation.
Therefore,
\begin{align}
    e^{\I \Delta \Gamma[\phi_b]^{(1), c}}
    &= \int \mathcal{D} c \mathcal{D} \bar{c} \, \sqrt{X(\varphi_b)}
        \exp \left\{
            \I \int \ud^4 x \,
            \left(
                \partial_\mu \bar{c} \partial^\mu c
                - \xi \mA^2 c \bar{c}
            \right)
        \right\},
\end{align}
so
\begin{align}
    \Delta V[\phi_b]^{(1), c}
    &= \frac{\I}{2}
        \int \frac{\ud^4 k}{(2 \pi)^4}
        \left[
            \ln X(\varphi_b)
            + 2 \ln \left( 1 - \frac{\xi \mA^2}{k^2} \right)
        \right].
    \label{eqn:ghost-one-loop-effective-potential}
\end{align}

Finally, the quadratic action for the photon may be written as
\begin{align}
    e^{\I \Delta \Gamma[\phi_b]^{(1), A}}
    &= \int \frac{\mathcal{D} A_\mu}{W(\varphi_b)^{2}}
        \exp \left\{
            - \frac{\I}{2} \int \ud^4 x \, A^\mu \prop^{-1}_{A, \mu \nu} A^\nu
        \right\}
    = W(\varphi_b)^{-2} \det\left[ \prop_A^{-1} \right]^{- 1/2}
    \label{eqn:effective-action-from-photon}
\end{align}
where
\begin{align}
    \prop^{-1}_{A, \mu \nu}
    &\equiv
        - g^{\mu \nu} \left[
            \partial_\alpha \partial^\alpha
            + \mA(\varphi_b)^2
        \right]
        + \left( 1 - \frac{1}{\xi(\varphi_b)} \right) \partial^\mu \partial^\nu.
\end{align}
The functional determinant may be computed in Euclidean time ($\tau = t / \I$) as an integral over
its (Fourier-space) eigenvalues,
\begin{align}
    \ln \det \prop^{-1}_{A, \mu \nu}
    &= \I
        \int \frac{\ud^4 k_E}{(2 \pi)^4}
        \trace \ln \left\{
            \left[
                - k_E^2
                - \mA(\varphi_b)^2
            \right]
            \left[
                - \delta^{\mu \nu}
                + \left( 1 - \frac{1}{\xi(\varphi_b)} \right)
                \frac{k_E^\mu k_E^\nu}{k_E^2 + \mA(\varphi_b)^2}
            \right]
        \right\}.
    \label{eqn:effective-action-from-photon-ito-integral}
\end{align}
Series expanding the matrix logarithm, simplifying each term, resumming, and shuffling constant
factors (which are dropped from the effective potential) yields
\begin{align}
    \ln \det \prop^{-1}_{A, \mu \nu}
    &= \I
        \int \frac{\ud^4 k_E}{(2 \pi)^4}
        \left\{
            3
            \ln
            \left(
                1
                + \frac{\mA(\varphi_b)^2}{k_E^2}
            \right)
            + \ln \left(
                1
                + \frac{\xi \mA^2}{k_E^2}
            \right)
            - \ln \xi(\varphi_b)
        \right\}.
    \label{eqn:trace-log-photon-propagator}
\end{align}
The first term is the contribution from the three physical dark photon polarizations.
Substituting \cref{eqn:xi-varphi-b-def} to compute $\Delta V[\phi_b]^{(1), A}$ via
\cref{eqn:effective-action-from-photon} and summing with
\cref{eqn:higgs-one-loop-effective-potential,eqn:goldstone-one-loop-effective-potential,eqn:ghost-one-loop-effective-potential}
yields a net contribution from the Abelian-Higgs sector of
\begin{align}
    \Delta V[\phi_b]^{(1)}
    &= \frac{1}{2}
        \int \frac{\ud^4 k_E}{(2 \pi)^4}
        \left[
            3 \ln \left( 1 + \frac{\mA(\varphi_b)^2}{k_E^2} \right)
            + \ln \left( 1 + \frac{m_\higgs(\varphi_b)^2}{k_E^2} \right)
            + 3 \ln W(\varphi_b)
            + 2 \ln X(\varphi_b)
        \right].
    \label{eqn:abelian-higgs-one-loop-effective-potential-integral}
\end{align}
Observe that the latter two terms, which diverge quartically, precisely cancel the contributions
incurred from quantizing the theory, $\Delta \mathcal{L}_{(0)}$
[\cref{eqn:measure-contribution-to-lagrangian-constant-term}], once
\cref{eqn:abelian-higgs-one-loop-effective-potential-integral} is subtracted from the Lagrangian.
(In addition, the $\xi$-dependent contributions from the Goldstone, ghosts, and photon all cancel,
though they are independent of the background field anyway.)
Absorbing $\Delta \mathcal{L}_{(0)}$ into the effective potential, at next-to-leading order in large
$\Lambda$ we have
\begin{align}
\begin{split}
    \Delta V[\phi_b]^{(1)}
    &= \frac{3 \Lambda^2 \mA(\varphi_b)^2}{2 (4 \pi)^2}
        - \frac{3 \mA(\varphi_b)^4}{8 (4 \pi)^2}
        \left[ 1 + 2 \ln \left( 1 + \frac{\Lambda^2}{\mA(\varphi_b)^2} \right) \right]
    \\ &\hphantom{{}={}}
        + \frac{\Lambda^2 m_\higgs(\varphi_b)^2}{2 (4 \pi)^2}
        - \frac{m_\higgs(\varphi_b)^4}{8 (4 \pi)^2}
        \left[ 1 + 2 \ln \left( 1 + \frac{\Lambda^2}{m_\higgs(\varphi_b)^2} \right) \right].
\end{split}
\end{align}
Since we expect $\Lambda$ to be much larger than any mass scale in the problem, the one-loop
contributions to the effective potential are subdominant to the classical potential when
\begin{align}
    \frac{\Lambda^2}{(4 \pi f)^2}
    &\lesssim \frac{m(\varphi_b)^2}{\max\left[ m(\varphi_b)^2, \mA(\varphi_b)^2, m_\higgs(\varphi_b)^2 \right]}.
    \label{eqn:cutoff-bound-from-effective-potential}
\end{align}
If coupling functions are chosen such that any of the fields' masses are \textit{not} dependent on
$\varphi_b$, then they do not contribute to the one-loop effective potential and are not considered in
\cref{eqn:cutoff-bound-from-effective-potential}.

At two-loop order, it is unclear whether quartic divergences cancel in a manner similar to those
that arise at one loop [via \cref{eqn:measure-contribution-to-lagrangian-mass-term}].
Though addressing this question is important to assess the naturalness of coupled scalar theories in
general, the two-loop calculations are substantially more involved and we leave them to future work.

\section{Axion to dark photon conversion at large coupling}\label{app:axionOscillations}

In this section, we detail the dynamics of dark photon production from axions in the large-$\beta$
regime up until perturbation theory breaks down.
We show that the breakdown of perturbation theory coincides with the time when friction sourced by
the dark photon becomes important.
As a result, the full study of this production mechanism ultimately requires nonlinear simulations.

The equations of motion that we consider are \cref{eqn:dark-photon-axion-eom} for the dark photon
and
\begin{align}
    0
    &= \ddot{\phi} + 3 H \dot{\phi} + \left( \frac{k^2}{a^2} + \ma^2 \right) \phi
        - \frac{\beta}{\fa} \left[ \three{E}\cdot\three{B} \right] (k)
\end{align}
for the axion, neglecting possible higher-order self-interactions in the axion potential.
The backreaction of the dark photon onto the axion's equation of motion is given by
(see, e.g., Ref.~\cite{Anber:2009ua})
\begin{align}
    \gen{\three{E}\cdot \three{B}}
    &= -\frac{1}{2a^3} \sum_{\lambda = \pm} \lambda
        \int \frac{\ud^3 k}{(2\pi)^3 } \dd{}{t} \left\vert A_\lambda \right\vert^2
\end{align}
at the background level.
We also require
\begin{align}
\begin{split}
    \label{eqn:EBEB}
    &\gen{
        \left\{
            [\three{E}\cdot\three{B}](\xi_1, \three{k})
            - \gen{\three{E}\cdot\three{B}}(\xi_1)
        \right\}
        \left\{
            [\three{E}\cdot\three{B}](\xi_2, \three{0})
            - \gen{\three{E}\cdot\three{B}}(\xi_2)
        \right\}
    }
    \\
    &= \frac{1}{a^3(\xi_1) a^3(\xi_2)}
        \int\frac{\ud^3 p}{(2\pi)^3}
        \abs{\three{p}}
        \abs{\bm{\varepsilon}^+(\three{p})^\ast \cdot \bm{\varepsilon}^+(\three{p} + \three{k})}^2
    \\ &\hphantom{ {}={} \frac{1}{a^3(\xi_1) a^3(\xi_2)}
            \int\frac{\ud^3 p}{(2\pi)^3}
        }
        \times
        \Bigg[
            \abs{\three{p} + \three{k}}
            \dot{A}_+(\xi_1, \three{p})
            A_+(\xi_1, \three{p} + \three{k})^\ast
            \dot{A}_+(\xi_2, \three{p} + \three{k})^\ast
            A_+(\xi_2, \three{p})
    \\ &\hphantom{ {}={} \frac{1}{a^3(\xi_1) a^3(\xi_2)}
            \int\frac{\ud^3 p}{(2\pi)^3}
            \times
            \Bigg[
        }
            + \abs{\three{p}}
            \dot{A}_+(\xi_1, \three{p} + \three{k})
            A_+(\xi_1, \three{p})^\ast
            \dot{A}_+(\xi_2, \three{p} + \three{k})^\ast
            A_+(\xi_2, \three{p})
        \Bigg].
\end{split}
\end{align}
where the polarization vectors are defined below \cref{eqn:fourier-expansions}, $\xi_1$ and $\xi_2$
denote two different times, and we keep only the $+$ polarization in the four-point function for
simplicity.
Note the identity
\begin{align}
    \bm{\varepsilon}^{\lambda}(\three{p})^\ast \cdot \bm{\varepsilon}^{\lambda'}(\three{p}+\three{k})
    &= \frac{1}{2}
        + \frac{\lambda\lambda'}{2}
        \frac{
            \left\vert \three{p} \cdot (\three{p} + \three{k})\right\vert
        }{
            \left\vert \three{p} \right\vert \left\vert \three{p} + \three{k} \right\vert
        }
    .
\end{align}
The initial condition of the dark photon is set by the Bunch-Davies vacuum:
\begin{subequations}
\begin{align}
    A_\lambda(\three{k})
    &= \frac{1}{\sqrt{2} \sqrt[4]{k^2 + a^2 \mA^2}}
    \\
    \dot{A}_\lambda(\three{k})
    &= \I \frac{k}{a} A_\lambda(\three{k}).
\end{align}
\end{subequations}
We decompose the axion into a homogeneous part $\bar{\phi}(t)$ and a perturbation
$\delta\phi(t,\three{k})$ which satisfy
\begin{align}
\label{eqn:HomogeneousEOM}
    \frac{\beta}{\fa}\gen{\three{E}\cdot\three{B}}&=\ddot{\bar{\phi}} + 3H\dot{\bar{\phi}} + \ma^2\bar{\phi},\\
\label{eqn:PerturbationEOM}
    \frac{\beta}{\fa}\p{\three{E}\cdot\three{B} - \gen{\three{E}\cdot\three{B}}}&=\delta\ddot{\phi} + 3H\delta\dot{\phi} + \p{\frac{k^2}{a^2} + \ma^2}\delta\phi.
\end{align}
We take a homogeneous initial condition for the axion,
\begin{align}\label{eqn:perturbationBC}
    \delta\phi = \delta\dot{\phi} = 0.
\end{align}
Finally, we enumerate a number of useful identities.
If
\begin{align}
    A(t) = A_0 \bar{A}(t) \exp\pc{ - \frac{1}{2}\p{\frac{k - k_0(t)}{a(t)\epsilon(t)}}^2},
\end{align}
then
\begin{align}\label{eqn:simplifiedA}
    A(t)A(t')
    \xrightarrow[\epsilon\to 0]{} A_0^2 \bar{A}^2(t)\pi^{3/4}2^{1/4} [a(t)\epsilon(t)]^{3/2}
        \delta(k - k_0(t))\sqrt{\delta(k_0(t) - k_0(t'))}.
\end{align}
Note that we have represented the temporal delta function inside a square root, anticipating the
appearance of another factor of the delta function coming from the other pair of $A$s.
One can solve \cref{eqn:PerturbationEOM} for the variance of the axion field in terms of the dark
photon as
\begin{align}\label{eqn:phiVariance}
\begin{split}
    \gen{\delta\phi^2}
    &= \frac{\beta^2}{\fa^2} \int_0^t\ud\xi_1\ud\xi_2\int\frac{\ud^3 k}{(2\pi)^3}
        G_k(t,\xi_1) G_k(t,\xi_2)
    \\ &\hphantom{ {}={} \frac{\beta^2}{\fa^2} \int_0^t\ud\xi_1\ud\xi_2\int\frac{\ud^3 k}{(2\pi)^3}}
        \times
        \gen{
        \left\{
            [\three{E}\cdot\three{B}](\xi_1, \three{k})
            - \gen{\three{E}\cdot\three{B}}(\xi_1)
        \right\}
        \left\{
            [\three{E}\cdot\three{B}](\xi_2, \three{0})
            - \gen{\three{E}\cdot\three{B}}(\xi_2)
        \right\}
    }
    .
\end{split}
\end{align}
If we assume that time derivatives can be replaced a factor of some scale $\Gamma$ and that $A_{\three{k}}(\xi)$ is of the form \cref{eqn:simplifiedA} with $A_0 = 1/\sqrt{k}$, then
\begin{align}
    \label{eqn:axionVarianceSimplified}
    \gen{\delta\phi^2}
    &= \frac{1}{2^{7/2}\pi^{5/2}}\frac{\beta^2 \mA^4}{\fa^2}\int_0^\tau\frac{\ud\xi}{\xi^{3/2}}\abs{\frac{d \kappa_0}{d\xi}}^{-1} \kappa_0^2\Gamma^2 \bar{A}^4\epsilon^3
        \int_0^{2 k_0} \kappa\ud \kappa G_\kappa^2(\xi,\tau)\p{1 - \p{\frac{\kappa}{2 \kappa_0}}^2}^2
\end{align}
with $t = \tau/\mA$, $a = a_0 \sqrt{\tau}$ and $\kappa = a_0 \mA\tau$, with all other quantities
implicitly rewritten in units of $\mA$.

Having enumerated the important formulas, we now discuss the dynamics.
The dark photon is not produced so long as $\dot{\bar{\phi}}$ remains below the production threshold velocity $\dot{\bar{\phi}}_\mathrm{P}$, during which time
the axion evolves as an entirely free field:
\begin{align}
    \bar{\phi} = \bar{\phi}(0)\p{1 - \frac{1}{5}(\ma t)^2}
\end{align}
and $\delta\phi = 0$.
The threshold velocity $\dot{\bar{\phi}}_\mathrm{P}$ is the velocity beyond which at least one dark photon mode becomes tachyonic, i.e.,
\begin{align}
    \frac{k_\mathrm{P}^2}{a^2} + \frac{\beta}{\fa}\dot{\bar{\phi}}_\mathrm{P} \frac{k_\mathrm{P}}{a} + \mA^2 = 0,
\end{align}
which picks out the velocity
\begin{align}
    \dot{\bar{\phi}}_\mathrm{P} = -\frac{2 \mA \fa}{\beta}
\end{align}
and wave number $k_\mathrm{P} / a = \mA$.
Once the velocity of the axion exceeds this threshold one of the dark photon polarizations (in our
convention, the $+$ polarization) undergoes a tachyonic resonance.
The characteristic dark photon wave number produced is roughly $k/a\sim \beta\dot{\Phi}/\fa$, so very shortly after resonance starts we may neglect the dark photon's mass. In order to simplify our discussion, we introduce the following set of variables:
\begin{align}\label{eqn:dimensionlessVariables}
    a = a_0\tau^{1/2},\hspace{0.5cm}
    t = \tau/\mA,\hspace{0.5cm}
    k = \mA a_0 \kappa,\hspace{0.5cm}
    \theta_0\beta = \frac{\mA^2}{\ma^2}B.
\end{align}
and the dimensionless initial axion displacement is $\theta_0 \equiv \bar{\phi}(0)/\fa$. In terms of these dimensionless variables, the dark photon equation of motion becomes
\begin{align}
    0&=\frac{1}{\tau^{1/4}} \ddd{}{\tau} (\tau^{1/4}A) + \p{\frac{\kappa^2}{\tau} + 1 + \frac{3}{16\tau^2} - \frac{2}{5}B\kappa\tau^{1/2}}A.
\end{align}
We may solve this equation in the WKB approximation to find
\begin{align}
    A^+\approx A^+(0)\ps{\frac{\tau_\mathrm{P}}{\tau}}^{1/4}\exp\ps{\int_{\tau_\mathrm{P}}^\tau\ud \tau'\sqrt{\frac{2}{5}B\kappa\tau'^{1/2} - \frac{\kappa^2}{\tau'} - 1 - \frac{3}{16\tau'^2}}},
\end{align}
where $\tau_\mathrm{P}$ is the dimensionless time at which the exponent becomes real. For modes that start growing some time after $\tau_\mathrm{P}$, it is a decent approximation to drop both the third and fourth terms (bare mass and Hubble respectively). In this approximation, we have
\begin{align}
    \tau_\mathrm{P} = \p{\frac{5\kappa}{2 B}}^{2/3},
\end{align}
and the integral reduces to
\begin{align}\label{eqn:A+Full-Expression}
     A^+\approx A_0\ps{\frac{\tau_\mathrm{P}}{\tau}}^{1/4}\exp\ps{\frac{\kappa^{4/3}}{B^{1/3}}\int_{(5/2)^{2/3}}^{\tau (B/\kappa)^{2/3}}\ud x\sqrt{\frac{2}{5}x^{1/2} - \frac{1}{x}}},
\end{align}
which may be expressed in terms of a Gaussian hypergeometric function.
However, we simply approximate $A^+$ by a Gaussian centered on its local maximum---an excellent
approximation because the dark photon transfer function is sharply peaked.

The local maximum of the argument of the exponent in \cref{eqn:A+Full-Expression} is located at
\begin{align}
    \kappa =\kappa_0\equiv C_0 B\tau^{3/2}
\end{align}
with $C_0 = 0.077082$
and is locally approximated by a parabola
\begin{align}
    C_1 B\tau^2\p{1 - C_2\p{\frac{\kappa}{\kappa_0} - 1}^2}
\end{align}
with $C_1=0.0788847$ and $C_2=0.253765$.
With this approximation, it is straightforward to calculate
\begin{align}
    \gen{\three{E}\cdot \three{B}}
    &= -\frac{\mA^5 a_0}{2\tau^{3/2}}\int\frac{\ud^3 \kappa}{(2\pi)^3}\kappa\dd{}{\tau} \left\vert A \right\vert^2
    \\
    &\approx -\frac{\mA^4}{(2\pi)^2}\sqrt{\frac{2\pi C_1 B}{ C_2 }} e^{2 C_1 B \tau^2}\p{\frac{5\kappa_0}{2B}}^{1/3}\frac{\kappa_0^3}{\tau^{2}},
\end{align}
where we have approximated the Gaussian by a $\delta$-function
\begin{align}
    e^{ - 2 C_1 B \tau^2 C_2\ps{\frac{\kappa}{\kappa_0} - 1}^2}
    \approx \sqrt{\frac{\pi}{2 C_1 C_2 B \tau^2}}\delta\p{\frac{\kappa}{\kappa_0} - 1}.
\end{align}

Using this estimate for $\gen{\three{E}\cdot\three{B}}$, we can compute the backreaction onto the homogeneous mode $\bar{\phi}$ using the Green's function for \cref{eqn:HomogeneousEOM} with appropriate boundary conditions:
\begin{align} \label{eqn:HomogeneousGreensFN}
    G(t,\xi) = \frac{\pi}{2^{1/2}}\frac{ \xi^{5/4}}{t^{1/4}} \left(J_{-\frac{1}{4}}(\ma \xi ) J_{\frac{1}{4}}(\ma t)-J_{\frac{1}{4}}(\ma \xi ) J_{-\frac{1}{4}}(\ma t)\right).
\end{align}
We note that for large $\beta$, backreaction kicks in well before $H = m_{a}$, and therefore the axion may be approximated as massless [or equivalently, we may take the small $t,\xi$ limit of \cref{eqn:HomogeneousGreensFN}], and so we may approximate
\begin{align}
\label{eqn:HomogeneousGreensFNApp}
    G(t,\xi)\approx 2\xi\p{1 - \sqrt{\frac{\xi}{t}}}.
\end{align}
Thus, we find
\begin{align}
    \bar{\phi} &= \bar{\phi}(0)\p{1-\frac{1}{5}(\ma t)^2} + \frac{\beta}{\fa}\int_0^t\ud\xi G(t,\xi)\gen{\three{E}\cdot\three{B}}
    \\
    &\approx \fa\theta_0\p{1-\frac{1}{5}(\ma t)^2} + \frac{\beta}{\fa}N_\mathrm{FR,0}\ma^2 (\beta\theta_0)^{3/2}(\ma t)e^{2 C_1\beta\theta_0 (\ma t)^2},
\end{align}
where the numerical factor is
\begin{align}
    N_\mathrm{FR,0} = \sqrt{\frac{C_1}{(2\pi)^3 C_2 }} \frac{5^{1/3} }{2^{1/3}} \frac{C_0^{10/3}}{16 C_1^2} = 9.4058\times 10^{-5}.
\end{align}
The time of backreaction is roughly determined by $\dot{\bar{\phi}} = 0$, which occurs at
\begin{align}
    t_\mathrm{BR} = \frac{1}{\ma}\sqrt{\frac{1}{4 C_1\beta \theta_0}W\ps{\frac{ \fa^4}{25 C_1 N_\mathrm{FR,0}^2 \ma^4\beta^6 \theta_0^2}}},
\end{align}
where $W$ is the Lambert $W$ function evaluated on its principal branch.
At this time, the energy density stored in the dark photon is approximately $\rho_\Ap\sim \rho_a/\beta$.

Now that we have determined the evolution of the homogeneous mode, we compute the perturbations to the axion to determine whether they become large prior to $t_\mathrm{BR}$. We apply \cref{eqn:phiVariance},
where $G_k$ is the Green's function of the Klein-Gordon equation with appropriate boundary
conditions [\cref{eqn:perturbationBC}],
\begin{align}\label{eqn:GreenKG}
    G_k(t,\xi)
    &=2 \xi e^{-\I \ma (\xi +t)} \ps{{_1F_1}\left(\frac{\I k^2 t}{2 \ma a^2}+\frac{1}{4};\frac{1}{2};2 \I \ma \xi \right) {_1F_1}\left(\frac{\I k^2 t}{2 \ma a^2}+\frac{3}{4};\frac{3}{2};2 \I \ma t\right)\right.\\&\hspace{2cm}\left.-\sqrt{\frac{\xi}{t}} {_1F_1}\left(\frac{\I k^2 t}{2 \ma a^2}+\frac{3}{4};\frac{3}{2};2 \I \ma\xi \right) {_1F_1}\left(\frac{\I k^2 t}{2 \ma a^2}+\frac{1}{4};\frac{1}{2};2 \I \ma t\right)}.
\end{align}
In order to evaluate the integral in \cref{eqn:phiVariance}, we make use of the $\delta$-function approximation for $A$. In this case, we are dealing with the product of two $A$s evaluated at different momenta. Since each $A$ on its own is essentially a $\delta$-function, the product of four $A$'s of different arguments imposes an additional condition: not only must $k$ and $k + p$ match $\mA a_0 \kappa_0$, but we must also have $\xi_1 = \xi_2$ because $\kappa_0$ is a function of $\xi_1$ or $\xi_2$. To see this explicitly, we complete the square as in \cref{eqn:simplifiedA}, with
\begin{align}
    A = A_0\ps{\frac{\tau_\mathrm{P}}{\tau_\xi}}^{1/4}e^{C_1B\tau_\xi^2}\exp\pc{-\frac{1}{2}\p{\frac{\kappa - \kappa_0}{\kappa_0/\sqrt{2C_1C_2 B\tau_\xi^2}}}^2}
\end{align}
where $\tau_\xi$ is $\tau$ in \cref{eqn:dimensionlessVariables} evaluated at $t = \xi$. Upon multiplying $A(\tau_{\xi_1})A(\tau_{\xi_2})$, we find a pair of Gaussians, one of which imposes $\kappa = \kappa_0$, and the other which imposes $\tau_{\xi_1} = \tau_{\xi_2}$. This is the simplification that allows us to reduce \cref{eqn:phiVariance} to \cref{eqn:axionVarianceSimplified}.

To further simplify the integral, we can approximate the Green's function \cref{eqn:GreenKG} in precisely the same way as we approximated it for the homogeneous mode, i.e., by dropping the axion's bare mass, leading to
\begin{align}
\label{eqn:inhomogeneousGreensFNApp}
    G_k(t,\xi)\approx \frac{\tau_\xi}{\mA\kappa\sqrt{\tau}}\sin\ps{2\kappa\p{\sqrt{\tau} - \sqrt{\tau_\xi}}}.
\end{align}
Further, it turns out that the dominant contribution to the integral comes from $\kappa$ and $\tau_\xi$ such that the argument of $\sin$ is small, so \cref{eqn:HomogeneousGreensFNApp} is an excellent approximation to \cref{eqn:inhomogeneousGreensFNApp} when evaluating \cref{eqn:axionVarianceSimplified}.
Finally, we replace time derivatives of $A$ with $\Gamma = 2 C_1 B\tau$, which is valid once modes start growing rapidly, i.e. $C_1 B\tau^2\gg 1$. With these approximations, evaluating $\gen{\delta\phi^2}$ becomes trivial. We apply \cref{eqn:axionVarianceSimplified} with
\begin{align}
\begin{split}
    \kappa_0 = C_0 B \tau^{3/2},
    \hspace{0.5cm}
    \epsilon = \frac{\kappa_0}{\sqrt{2 C_1 C_2 B \tau^3}},
    \hspace{0.5cm}
    \Gamma = 2 C_1 B \tau,
    \hspace{0.5cm}
    \abs{\frac{d\kappa_0}{d\tau}}^{-1} = \frac{2}{3 C_0 B \tau^{1/2}},
    \\
    G_\kappa = 2\xi\p{1 - \sqrt{\frac{\xi}{\tau}}},
    \hspace{0.5cm}
    \bar{A}^2 = \p{\frac{\tau_\mathrm{P}}{\tau}}^{1/2}e^{2C_1 B\tau^2},
    \hspace{0.5cm}
    \text{and }
    \tau_\mathrm{P} = \p{\frac{5\kappa_0}{2B}}^{2/3}
    ,
\end{split}
\end{align}
yielding
\begin{align}
    \gen{\delta\phi^2}
    &\approx N_\mathrm{FR,1}\frac{\ma^4\beta^2}{\fa^2}(\beta\theta_0)^{7/2}(\ma t)^{3}e^{4 C_1 \beta\theta_0 (\ma t)^2},
\end{align}
where
\begin{align}
    N_\mathrm{FR,1} = \frac{5^{2/3}C_0^{20/3}}{2^{2/3}\pi^{5/2}4608 C_1^{5/2}C_2^{3/2}} = 3.88565\times 10^{-9}.
\end{align}
Having computed the perturbations, we may now compare them to the backreaction onto the homogeneous mode
\begin{align}
    \frac{\gen{\delta\phi^2}}{\ps{\bar{\phi}(t) -\bar{\phi}(0)\p{1 - \frac{1}{5}(\ma t)^2}}^2}\approx 0.43921 \sqrt{\beta\theta_0} \ma t.
\end{align}
Evaluating this ratio at $t = t_\mathrm{BR}$, we find that the perturbations become order 1 at roughly the same time backreaction becomes important. This coincidence could have been anticipated, given that it is a common process $\three{E}\cdot\three{B}$ that drives both perturbations and backreaction.

\section{Delayed axion oscillations}
\label{app:Delayed-Axion-Oscillations}

\subsection{Efficient growth condition}
\label{app:Delayed-Axion-Oscillations-Condition}
As discussed in \ref{sec:axion-enhance-narrow-resonance}, dark photon production from axions can be efficient with $\beta\theta_0 \ll 1$ if oscillations are substantially
delayed relative to the standard time, $H \sim \ma$.
In this appendix, we derive the modified condition for efficient resonance, \cref{eqn:axion-DP-Production-Threshold-Efficient-Delay}.

The growth rate of the transverse modes may be calculated from the equation of motion \cref{eqn:dark-photon-axion-eom} by mapping it onto an equivalent Mathieu-type equation,
\begin{align}\label{eqn:matheiu-type}
    0
    &= A''(\tau) + (a + 2 q \cos 2\tau) A(\tau),
\end{align}
with $2\tau = \ma t$, primes representing $\tau$ derivatives, and
\begin{align}
    \tilde{a} &= 4 \left( \frac{k^2}{a^2 \ma^2} + \frac{\mA^2}{\ma^2} \right)
    \\
    \tilde{q} &= \frac{2 k\beta\theta_0}{a \ma} \left( 2 H_\mathrm{osc} t \right)^{-3/4}.
\end{align}
\Cref{eqn:matheiu-type} is not technically a Mathieu-type equation since the coefficients $\tilde{a}$ and $\tilde{q}$ depend on $t$.
However, so long as $\ma t\gg1$, this time dependence is slow and can be treated adiabatically.
The instantaneous growth rate for the transverse dark photon modes is therefore
\begin{align}
    \Gamma
    &= \pm \frac{\ma}{4}\sqrt{\tilde{q}^2 - (1 - \tilde{a})^2}.
\end{align}
The total growth of a given mode is then
\begin{align}
    A(t) / A(0)
    \approx \exp\pc{\int_{1/2H_\mathrm{osc}}^t\ud t' \, \Re\left[ \Gamma \right]}.
\end{align}
Efficient production requires growth rates exceeding $H$.
Near threshold, the resonance takes place almost entirely at a single wave number; as long as this
single wave number remains inside the resonance band (i.e., has not redshifted out of it), the
energy density growth is approximated by\footnote{
    Once this mode has redshifted outside of the resonance band, the energy density growth is linear
    instead of exponential.
}
\begin{align}\label{eqn:delayed-oscillation-growth-rate}
    \frac{\dot{\rho}_{\Ap}}{\rho_\Ap}
    &\approx \max_k 2 \Gamma(k)
    = \frac{1}{2} \beta \theta_0 \ma
        \sqrt{
            \left(1-\frac{4 \mA^2}{\ma^2}\right) \left( \frac{H}{H_\mathrm{osc}} \right)^{3/2}
            +\frac{1}{4} \beta^2 \theta_0^2 \left( \frac{H}{H_\mathrm{osc}} \right)^{3}
        }.
\end{align}
There are two limits of this equation to consider.
First, if $\ma \leq 2\mA$, then the first term inside the root is negative, and must be compensated
for by the second term, requiring $\beta\theta_0\gg 1$.
In this case, resonance starts almost immediately and quickly becomes nonperturbative (corresponding
to the case discussed in \cref{app:axionOscillations}).
On the other hand, if $\beta\theta_0$ is small, then resonance requires $\ma \geq 2 \mA$.
In this limit, modes redshift from one edge of the resonance band to the other in a time interval
\begin{align}
    \Delta t = \frac{\beta\theta_0}{H_\mathrm{osc}},
\end{align}
where we have taken the $2\mA\ll \ma$, $t\sim 1/2H_\mathrm{osc}$, and $\beta\theta_0\ll 1$ limits. The resulting total growth is
\begin{align}
    \rho_\Ap\p{[2H_\mathrm{osc}]^{-1} + \Delta t}
    \sim \rho_\Ap\p{ [2H_\mathrm{osc}]^{-1}}
        \exp\pc{\frac{\ma}{2 H_\mathrm{osc}}\beta^2\theta_0^2}.
\end{align}
Thus, we arrive at the efficient resonance condition \cref{eqn:axion-DP-Production-Threshold-Efficient-Delay},
\begin{align}
    \beta\theta_0\sqrt{\frac{\ma}{2 H_\mathrm{osc}}}\gg 1.
\end{align}

\subsection{Axion fragmentation time}\label{app:axion-fragmentation-time}

As discussed in \cref{sec:axion-enhance-narrow-resonance}, attractive self-interactions can fragment
the axion before it transfers all of its energy to the dark photon.
For a general axion potential approximated as
\begin{align}
    V(\phi_a) = \ma^2\fa^2\ps{\frac{1}{2}\p{\frac{\phi_a}{\fa}}^2 +\frac{1}{3!}A\p{\frac{\phi_a}{\fa}}^3 + \frac{1}{4!}B\p{\frac{\phi_a}{\fa}}^4},
\end{align}
the growth rate of axion perturbations (derived in Appendix C of Ref.~\cite{Cyncynates:2021xzw}) is
\begin{align}
    \Gamma = \left\vert \delta\omega \right\vert
        \Re\sqrt{1 - \p{1 + \frac{k^2}{2\ma\delta\omega a^2}}^2},
\end{align}
where $\delta\omega$ is the change in the oscillation frequency of the axion relative to its bare mass.
At leading order in self-interactions,
\begin{align}
    \delta\omega = \ma\frac{3 B - 5 A^2}{24}\gen{\Theta^2}.
\end{align}
We compare this to the expression \cref{eqn:delayed-oscillation-growth-rate} in the limit $\mA\to 0$, $t\sim 1/2H_\mathrm{osc}$, and $\beta\theta_0\ll 1$: the dark photon mode growth rate is $\Gamma = \ma\beta\theta_0/4$.
Thus, the axion fragments before the dark photon absorbs $\mathcal{O}(1)$ of the axion energy density unless
\begin{align}
    \left\vert \ma\beta\theta_0/4 \right\vert > -\delta\omega.
\end{align}
Implicit in this inequality is the statement that if axion self-interactions are repulsive then the
axion perturbations do not grow.
To achieve efficient resonance at the natural expectation $\beta\theta_0\sim\alpha_D$ without
backreaction, attractive self-interactions are practically excluded.

In summary, though delayed axion oscillations can significantly reduce the requisite dimensionless
axion--dark-photon coupling, even very weak attractive self-interactions cause the axion to fragment
before all the dark photon dark matter is produced.
Viable dark photon production essentially requires repulsive self-interactions.
As the axion is a periodic field, oscillations of large enough amplitude are guaranteed to induce
attractive self-interactions.
Mechanizing repulsive self-interactions requires quite a bit of machinery (see the extensive
discussion in Ref.~\cite{Fan:2016rda}), which even then only extend over a limited field range.

``Axion friendship''~\cite{Cyncynates:2021xzw,Cyncynates:2022wlq} is a phenomenon whereby pairs of
axions with nearby masses exchange energy through nonlinear
autoresonance~\cite{landau2013course,bogoliubov1961asymptotic,fajans2001autoresonant,rajasekar2016autoresonance,glebov2017oscillations}.
If the axions in the friendly pair have hierarchical decay constants $f_{j+1}/f_j = \mathcal{F}\gg
1$, the axion with the smaller decay constant can siphon nearly all the energy from its friend,
parametrically delaying the onset of decaying oscillations to $H_\mathrm{osc}\sim \ma
\mathcal{F}^{4/3}$.
Although the requirement of nearby masses constitutes a coincidence, such pairs may be common in
axiverse-type scenarios~\cite{Arvanitaki:2009fg}.
In fact, it has been found that in realistic string compactifications the friendship parameter $\mu\equiv m_j/m_{j+1}$, which measures the ratio of neighboring axion masses, lies in the requisite range $\mu\in[0.75,1]$ for a growing fraction of pairs as the number of axions increases~\cite{Gendler:2023kjt}.\footnote{The range $\mu\in[0.75,1]$ is for a cosine potential which has attractive self-interactions. In the case of repulsive self-interactions the range switches to $\mu\in[1,1.25]$ or so.}

\section{Plasma dynamics}\label{app:plasma}

Between electron-positron annihilation and recombination, the SM plasma comprises photons,
nonrelativistic electrons, and nonrelativistic nuclei.
The Universe is charge neutral, so the electron and proton number densities are both
\begin{align}
    n_e
    = \eta_B \frac{2 \zeta(3)}{\pi^2} T^3,
    \label{eqn:ne-eta}
\end{align}
where $\eta_B \approx 6 \times 10^{-10}$ is the baryon-to-photon ratio and $T$ the plasma temperature.
The rate of Thomson scattering between photons and electrons and of Coulomb scattering between
electrons and protons are respectively $\dot{\kappa}_T = n_e \sigma_T$ and
$\dot{\kappa}_C = n_e \sigma_C$, where the corresponding cross sections are~\cite{Dubovsky:2015cca}
\begin{align}
    \sigma_T
    &= \frac{8 \pi \alpha^2}{3 m_e^2}
    \\
    \sigma_C
    &= \frac{4 \sqrt{2 \pi} \alpha^2}{3 \sqrt{m_e T^3}}
        \ln \left( \frac{4 \pi T^3}{\alpha^3 n_e} \right),
\end{align}
In terms of the proton-to-electron mass ratio $\mu_p \equiv m_p / m_e$, the Thomson cross section
for protons is $\mu_p^{-2} \sigma_T$.
In the epoch of interest, the Thomson and Coulomb rates are quite rapid relative to the Hubble rate:
\begin{subequations}\label{eqn:thomson-coulomb-rates-relative-to-H}
\begin{align}
    \frac{\dot{\kappa}_T}{H}
    &\approx 10^3 \frac{T}{\eV} \frac{\eta_B}{6 \times 10^{-10}}
    \\
    \frac{\dot{\kappa}_C}{H}
    &\approx 5.8 \times 10^{12}
        \left( \frac{T}{\eV} \right)^{1/2}
        \frac{\eta_B}{6 \times 10^{-10}}
\end{align}
\end{subequations}
taking $H^2 = g_\star \pi^2 T^4 / 90 \Mpl^2$ where $g_\star \approx 3.38$.
Treating the plasma as three coupled, perfect fluids that jointly satisfy the energy-momentum
conservation equations is therefore an excellent approximation.
For simplicity of exposition, we ignore the fact that about a quarter of the protons are bound in
helium nuclei, considering a single fluid of protons.
Strictly speaking, this description is only applicable after electron-positron annihilation
completes, which occurs around a temperature of
$T \approx 20~\mathrm{keV}$~\cite{Grayson:2023flr}; at higher temperatures the positrons and
electrons are equally abundant, and $e^+ e^-$ scattering processes are also important.

Large-scale photons (i.e., with wave number much smaller than the plasma temperature) couple to the
bulk motion of the fluid via Lorentz forces.
Longitudinal modes (or scalar degrees of freedom in cosmological perturbation theory) couple to
density perturbations and gravitational potentials, making for a large, complex set of coupled
differential equations.
Transverse photon modes, on the other hand, induce vorticity in the plasma.
Vector perturbations of the metric, however, do not appear explicitly in the momentum conservation
equations and in fact do not couple directly to a fluid's velocity if that fluid supports no
anisotropic stress~\cite{Weinberg:2008zzc}.
As already argued, the photons, electrons, and protons interact rapidly enough to suppress any
anisotropic stress in the epoch of interest.
The equations of motion for the transverse (vortical) modes [decomposed onto the same basis as
\cref{eqn:Ai-polarization-fourier-expansion}] in the plasma are
\begin{subequations}\label{eqn:vector-momentum-equations}
\begin{align}
    \dot{u}_{\gamma, \pm}
    &= \dot{\kappa}_T
        \left( u_{e, \pm} - u_{\gamma, \pm} \right)
        + \mu_p^{-2} \dot{\kappa}_T
        \left( u_{p, \pm} - u_{\gamma, \pm} \right)
    \label{eqn:photon-momentum-eqn}
    \\
    \dot{u}_{e, \pm}
    &= - H u_{e, \pm}
        + R_e \dot{\kappa}_T
        \left( u_{\gamma, \pm} - u_{e, \pm} \right)
        + \dot{\kappa}_C \left( u_{p, \pm} - u_{e, \pm} \right)
        - \frac{e}{m_e} \dot{A}_\pm
    \label{eqn:u-e-pm-eom}
    \\
    \dot{u}_{p, \pm}
    &= - H u_{p, \pm}
        + R_e \mu_p^{-3} \dot{\kappa}_T
        \left( u_{\gamma, \pm} - u_{p, \pm} \right)
        + \mu_p^{-1} \dot{\kappa}_C \left( u_{e, \pm} - u_{p, \pm} \right)
        + \frac{e}{m_p} \dot{A}_\pm.
    \label{eqn:u-p-pm-eom}
\end{align}
\end{subequations}
Here $u_{a, \pm}$ are the transverse polarizations of the fluid velocity for the photons, electrons,
and protons ($a = \gamma$, $e$, and $p$, respectively).
The photon-to-electron density ratio
\begin{align}\label{eqn:def-Re}
    R_e
    &\equiv \frac{4 \bar{\rho}_\gamma}{3 \bar{\rho}_e}
    = 1.175 \times 10^{4} \,
        \frac{T}{\eV}
        \left( \frac{\eta_B}{6 \times 10^{-10}} \right)^{-1}
\end{align}
encodes the relative rate of momentum transfer for electrons and photons due to Thomson scattering.

The photon couples to the SM 3-current, which is given terms of the proton-electron slip by
\begin{align}
    J_\pm
    &= n_e \left( u_{p, \pm} - u_{e, \pm} \right)
\end{align}
and, combining \cref{eqn:u-e-pm-eom,eqn:u-p-pm-eom}, evolves according to
\begin{align}
\begin{split}
    \dot{J}_\pm
    + \left[
        4 H
        + \left( 1 + \mu_p^{-1} \right) \dot{\kappa}_C
    \right]
    J_\pm
    &= - R_e \dot{\kappa}_T n_e
        \left[
            u_{\gamma, \pm} - u_{e, \pm}
            + \mu_p^{-3} \left( u_{\gamma, \pm} - u_{p, \pm} \right)
        \right]
    \\ &\hphantom{ {}={} }
        + n_e \frac{e}{m_e} \left( 1 + \mu_p^{-1} \right) \dot{A}_\pm
    .
    \label{eqn:current-eom}
\end{split}
\end{align}
From here on out we drop the term arising from Thomson scattering of protons,
$\mu_p^{-3} \left( u_{\gamma, \pm} - u_{p, \pm} \right)$, and also neglect Hubble friction when it
is directly added with $\dot{\kappa}_T$ or $\dot{\kappa}_C$.
Taking a time derivative of \cref{eqn:current-eom} and simplifying with \cref{eqn:u-e-pm-eom},
the current equation of motion may be written in the form
\begin{align}
    \ddot{J}_\pm
    + \nu(t) \dot{J}_\pm
    + \omega(t)^2 J_\pm
    &=
        - H R_e \dot{\kappa}_T n_e u_{e, \pm}
        + n_e \frac{e}{m_e}
        \ddot{A}_\pm
        + \gamma(t)
        n_e \frac{e}{m_e} \dot{A}_\pm
    \label{eqn:current-eom-def-rates}
\end{align}
where
\begin{subequations}\label{eqn:def-current-rates}
\begin{align}
    \nu(t)
    &= \dot{\kappa}_C
        + \left( 1 + R_e \right) \dot{\kappa}_T
    \label{eqn:def-nu}
    \\
    \omega(t)^2
    &= \left( 1 + \mu_p^{-1} R_e \right)
        \dot{\kappa}_C \dot{\kappa}_T
    \\
    \gamma(t)
    &= \left( 1 + \mu_p^{-1} R_e \right)
        \dot{\kappa}_T.
\end{align}
\end{subequations}
In \cref{eqn:current-eom-def-rates,eqn:def-current-rates} we took $1 + \mu_p^{-1} \approx 1$, but
$\mu_p^{-1} R_e$ is not necessarily small.
However, consultation with \cref{eqn:thomson-coulomb-rates-relative-to-H,eqn:def-Re} shows that
$\omega(t)$ and $\gamma(t)$ are always negligible relative to $\nu(t)$.
On the same basis, the term proportional to the electron velocity in
\cref{eqn:current-eom-def-rates} is also negligible unless the electron velocity is drastically
larger than $u_{p, \pm} - u_{e, \pm}$.
With these approximations, \cref{eqn:current-eom-def-rates} depends only $\dot{J}_\pm$,
$\ddot{J}_\pm$, and $\ddot{A}_\pm$.
Since all time-dependent coefficients vary only on the Hubble timescale, which we have already
taken to be negligible, we may approximate
\begin{align}
    \dot{J}_\pm
    + \nu(t) J_\pm
    &= \frac{\omega_p(t)^2}{e}
        \dot{A}_\pm
    \label{eqn:current-eom-approx}
\end{align}
with the plasma damping rate $\nu(t)$ defined in \cref{eqn:def-nu} and the plasma frequency defined by
\begin{align}\label{eqn:plasma-frequency}
    \omega_p^2
    &= \frac{e^2 n_e}{m_e}
    = \frac{2 \zeta(3) e^2 \eta_B}{\pi^2} \frac{T^3}{m_e}
    = \left[
        5.121 \times 10^{-9}~\eV
        \sqrt{ \frac{\eta_B}{6 \times 10^{-10}}}
        \left( \frac{T}{\eV} \right)^{3/2}
    \right]^2.
\end{align}
Relative to Ref.~\cite{Dubovsky:2015cca}, which considered the coupling of kinetically mixed dark
photons to the electron-ion plasma in the interstellar medium, the Thomson scattering of photons
with electrons relevant at early times provides an additional resistance of the plasma to bulk
charge separation.
In fact, momentum transfer due to Thomson scattering becomes more important than that due to Coulomb
scattering at temperatures above $T \sim 190~\eV$, coincident with the time period of
interest (redshifts $z \sim 10^6$).

Finally, the Thomson scattering and momentum exchange rates and the Coulomb rate relative to the
plasma frequency are
\begin{subequations}\label{eqn:resistivity-vs-plasma-frequency}
\begin{align}
    \frac{R_e \dot{\kappa}_T}{\omega_p}
    &= \frac{8 \alpha^2 \sqrt{2 \zeta(3) \eta_B}}{3 e}
        \frac{2 \pi^4}{45 \zeta(3) \eta_B}
        \left( \frac{T}{m_e} \right)^{5/2}
    \approx 106.9 \left( \frac{T}{m_e} \right)^{5/2}
    \\
    \frac{\dot{\kappa}_C}{\omega_p}
    &= \frac{4 \sqrt{2 \pi} \alpha^2 \sqrt{m_e n_e}}{3 e \sqrt{m_e T_e^3}}
        \ln \Lambda_C
    = \frac{8 \alpha^2 \sqrt{\zeta(3) \eta_B / \pi}}{3 e}
        \ln \Lambda_C
    \approx 2.84 \times 10^{-7}.
\end{align}
\end{subequations}
At temperatures $T \lesssim 10^{-2} m_e$ (but still before recombination), the resistivity of the
plasma is therefore negligible compared to the plasma frequency.

\bibliography{manual,bibliography}

\end{document}